\documentclass[12pt]{article}
\pdfoutput=1

\newlength{\abstractwidth}
\usepackage{epsf}
\usepackage{color}
\usepackage{graphicx}
\usepackage{tikz}
\usepackage{hyperref}

\usepackage{amsmath}
\usepackage{amssymb}
\usepackage{latexsym}

\renewcommand{\thefootnote}{\fnsymbol{footnote}}
\renewcommand{\thanks}[1]{\footnote{#1}}
\newcommand{\starttext}{
\setcounter{footnote}{0}
\renewcommand{\thefootnote}{\arabic{footnote}}}

\newcommand{\bea}{\begin{eqnarray}}
\newcommand{\eea}{\end{eqnarray}}
\newcommand{\be}{\begin{eqnarray}}
\newcommand{\ee}{\end{eqnarray}}

\def\cA{{\cal A}}
\def\cB{{\cal B}}
\def\cC{{\cal C}}
\def\cD{{\cal D}}

\def\cF{{\cal F}}
\def\cG{{\cal G}}
\def\cH{{\cal H}}

\def\cK{{\cal K}}
\def\cL{{\cal L}}
\def\cM{{\cal M}}
\def\cN{{\cal N}}
\def\cO{{\cal O}}

\def\cR{{\cal R}}
\def\cS{{\cal S}}

\def\bA{{\bf A}}

\def\bn{{\bf n}}
\def\bo{{\bf o}}

\def\mA{\mathfrak{A}}
\def\mB{\mathfrak{B}}
\def\mC{\mathfrak{C}}
\def\mD{\mathfrak{D}}

\def\mI{\mathfrak{I}}

\def\mM{\mathfrak{M}}
\def\mN{\mathfrak{N}}

\def\mb{\mathfrak{b}}
\def\mc{\mathfrak{c}}

\def\ZZ{{\mathbb Z}}
\def\RR{{\mathbb R}}
\def\NN{{\mathbb N}}
\def\CC{{\mathbb C}}

\def\QQ{{\mathbb Q}}

\def\Re{{\rm Re \,}}
\def\Im{{\rm Im \,}}

\def\half{{1\over 2}}
\def\thalf{{\tfrac{1}{2}}}
\def\p{\partial}

\def\a{\alpha}
\def\b{\beta}

\def\ep{\varepsilon}

\def\pbz{\p _{\bar z}}

\def\Li{{\rm Li}}

\def\multi{multiple }

\def\no{\nonumber}
\def\sm{\smallskip}

\newtheorem{thm}{Theorem}[section]
\newtheorem{lem}[thm]{Lemma}

\newtheorem{conj}[thm]{Conjecture}

\begin{document}
\starttext
\setcounter{footnote}{0}

\begin{flushright}
2019 June 4 \\
revised 2019 July 10
\end{flushright}

\vskip 0.3in

\begin{center}

{\Large \bf Exploring transcendentality in superstring amplitudes}

\vskip 0.2in

{\Large Eric D'Hoker$^{(a)}$ and Michael B. Green$^{(b)}$} 

\vskip 0.15in

{ \sl  (a) Mani L. Bhaumik Institute for Theoretical Physics}\\
{\sl  Department of Physics and Astronomy}\\
{\sl University of California, Los Angeles, CA 90095, USA}

\vskip 0.1in

{ \sl (b) Department of Applied Mathematics and Theoretical Physics }\\
{\sl Wilberforce Road, Cambridge CB3 0WA, UK}, and \\
{\sl Centre for Research in String Theory, School of Physics, }\\
{\sl Queen Mary University of London, Mile End Road, London, E1 4NS, England}

\vskip 0.15in

{\tt \small dhoker@physics.ucla.edu, M.B.Green@damtp.cam.ac.uk}

\vskip 0.5in

\begin{abstract}
\vskip 0.1in
It is well known that the low energy expansion of tree-level superstring scattering amplitudes satisfies a suitably defined version of uniform transcendentality.  In this paper it is argued that there is a natural extension of this definition that applies to the  genus-one four-graviton Type II superstring amplitude to all orders in the low-energy expansion. To obtain this result, the integral over the genus-one moduli space is partitioned into a region $\cM_R$ surrounding the cusp and its complement $\cM_L$, and  an exact expression is obtained for the contribution to the amplitude from $\cM_R$. The low-energy expansion of the $\cM_R$ contribution is proven to be free of irreducible \multi zeta-values to all orders. The contribution to the amplitude from $\cM_L$ is computed in terms of modular graph functions  up to order $D^{12} \cR^4$ in the low-energy expansion, and general arguments are used beyond this order to conjecture the transcendentality properties of the $\cM_L$ contributions. Uniform transcendentality of the full amplitude  holds provided we assign a non-zero weight to certain harmonic sum functions, an assumption which is  familiar from transcendentality assignments in quantum field theory amplitudes.

\end{abstract}
\end{center}

\newpage
 
\setcounter{tocdepth}{2} 
\newpage

\baselineskip=15pt
\setcounter{equation}{0}
\setcounter{footnote}{0}

\newpage

\section{Introduction}
 \label{sec:1}
\setcounter{equation}{0}

Many of the mathematical structures that arise in  quantum field theory, such as multiple polylogarithms,  elliptic multiple polylogarithms and various generalizations, also arise in string perturbation theory,  but they arise in a qualitatively different fashion.  This is already clear from the low-energy expansion  of $N$-particle tree-level amplitudes. For an open string amplitude, which is associated with a worldsheet disk, the coefficients  of the low-energy expansion are \multi zeta-values (MZVs).  For a closed-string amplitude,  which is associated with a worldsheet sphere, the coefficients of the low-energy expansion are single-valued MZVs. 
Thus, superstring tree-level amplitudes are generating functions for MZVs and  single-valued MZVs, while tree-level amplitudes in quantum field theory exhibit no analogously interesting mathematical structure.

\sm

The rich algebraic structure of MZVs may be traced back to the algebraic structure of the vertex operators in the conformal field theory construction of string amplitudes.  This leads to a grading by  transcendental weight in  the low-energy expansion of string theory already at tree-level, whereas in quantum field theory transcendentality arises only at loop orders via dimensional regularization and renormalization.  The distinction between string theory and quantum field theory becomes even more striking when considering the low-energy expansion of  loop amplitudes.  Most notably, superstring loop amplitudes have no ultraviolet divergences and are also infrared finite in consistent backgrounds of space-time dimension greater than four.   This raises the obvious question of how the notion of transcendentality generalizes to the low-energy expansion of superstring loop amplitudes.

\sm

Genus-one and genus-two  closed superstring amplitudes can be reduced to integrals of modular invariant integrands over their respective (bosonic) moduli spaces, whereas such explicit expressions are not available beyond genus two.   For genus one, the low-energy expansion of the integrand is a power series in the kinematic variables with coefficients that are linear combinations over $\QQ$ of modular graph functions \cite{DHoker:2015gmr,DHoker:2015wxz}.  Each modular graph function corresponds to a Feynman graph for a massless scalar field on a torus with modulus $\tau$, which is invariant under the action of $SL(2,\ZZ)$ on $\tau$,   and can be represented by a Kronecker-Eisenstein series whose  transcendental weight $w$ equals the number of Green functions in the Feynman graph. For genus two and beyond, natural generalizations of modular graph functions were introduced in \cite{DHoker:2017pvk}, which include the Kawazumi-Zhang invariants \cite{Kawazumi,Zhang}. 

\sm

Genus-one modular graph functions, and their generalization to modular graph forms,  are of mathematical interest because they  generate elliptic generalizations of single-valued multiple polylogarithms, and hence of  single-valued MZVs.  Indeed, near the cusp $\tau \to i \infty$, a modular graph function of weight $w$ reduces to a Laurent polynomial  $a_w y^w + \cdots + a_{1-w} y^{1-w}$ with $y = \pi \Im (\tau)$, up to exponentially suppressed terms, where the coefficients $a_r$ are linear combinations over $\QQ$ of single-valued MZVs \cite{DHoker:2015wxz}. Therefore, the algebraic structure of shuffle and stuffle relations satisfied by MZVs points to the algebraic structure satisfied by modular graph forms  \cite{DHoker:2016mwo,DHoker:2016quv}. Similarly, the integrands of one-loop amplitudes of open-string theory have coefficients that are closely related to holomorphic elliptic polylogarithms. 
While the integral representation of the genus-two closed-string amplitudes is explicitly known \cite{DHoker:2002hof,DHoker:2005vch, Berkovits:2005df}, only the first few terms in its low-energy expansion have been analyzed \cite{DHoker:2005jhf,DHoker:2013fcx,DHoker:2014oxd}.

\sm 

However, although it is mathematically interesting to consider modular graph functions, string amplitudes require integration over the moduli space of all Riemann surfaces, which is  challenging, even at  genus one. The integral representation of the amplitude is absolutely convergent only for purely imaginary values of the kinematic variables, and the construction of the physical amplitude requires suitable analytic continuation. Among other things, great care needs to be taken to account for the resulting non-analytic behavior in the kinematic variables which arises from the integration region in moduli space where the surface degenerates, and is expected from unitarity. For the  genus-one four-graviton amplitude this  analytic continuation was shown to exist, and used to calculate the decay rates and mass-shifts of massive string states accessible in the four-graviton amplitude in \cite{DHoker:1994gnm}.

\sm 
 
This paper is concerned with establishing  further properties of the genus-one four-graviton amplitude  of Type II superstring theory, in the low-energy expansion. This builds upon earlier work \cite{Green:1999pv, Green:2008uj},  which considered low order terms in the low-energy expansion.
A major objective here is the construction of  the exact non-analytic contribution to the amplitude, to all orders in the low-energy expansion. We will restrict our study to the case of ten-dimensional Minkowski space-time, although our analysis can be readily extended to the case of toroidal compactification for which the infrared behavior of the amplitude, including  the non-analytic threshold dependence, will be changed in line with the predictions of unitarity. 

\sm

In order to separate the analytic and non-analytic contributions to the amplitude, as in  \cite{Green:1999pv}, we partition the moduli space of the torus $\cM= \cM_R \cup \cM_L$ into a neighborhood $\cM_R= \cM \cap \{ \Im(\tau)>L \}$ of the cusp, and its complementary set. The parameter $L$ is arbitrary  and cancels in the complete integral.  Throughout we shall choose $L \gg1$.

\sm

The contribution from $\cM_L$ is analytic in the standard kinematic variables $s,t,u$ and,  at any order in the expansion in powers of $s,t,u$, is given by a sum of integrals  of modular graph functions times rational coefficients.  These integrals evaluate to a Laurent polynomial in $L$, plus a term proportional to $\log L$, and terms exponentially suppressed for large $L$.  With our present understanding the integrals over $\cM_L$ are particularly difficult to evaluate. The contribution from $\cM_L$ up to order $D^{10}\cR^4$ was derived in earlier work \cite{Green:1999pv,Green:2008uj, DHoker:2015gmr}, while the evaluation to order $D^{12}\cR^4$ will be carried out in section~\ref{sec:3} of this paper.  The bulk of this calculation will be relegated to appendix \ref{sec:A}.

\sm

Integration over $\cM_R$ will be considered in  section~\ref{sec:4}.   This gives non-analytic contributions that have logarithmic branch points in the kinematic variables $s,t,u$, as required by unitarity, together with further analytic terms.  We will determine both types of terms to all orders in the low-energy expansion, demonstrating  that the expansion coefficients are free of irreducible MZVs, and  are given by polynomials in odd zeta-values only, a property reminiscent of the results established in \cite{DHoker:2019xef,Zagier:2019eus}. The details of this calculation will be presented in appendix \ref{sec:B}, while in  appendix~\ref{sec:C} we will check explicitly that the coefficients of the logarithmic terms are consistent with two-particle unitarity.

\sm

 The low-energy expansion of the full genus-one amplitude  up to order $D^{12}\cR^4$ is obtained by combining the exact result from the integration over $\cM_R$ with the direct integration over $\cM_L$ of the modular graph functions which contribute up to this order.   

\sm

In  section~\ref{sec:5} we will argue that transcendental weight may be assigned in such a way that it provides a consistent grading (also referred to as uniform transcendentality) for the low-energy expansion of the tree-level amplitude and the genus-one amplitude up to order $D^{12}R^4$, under the following  assumptions. 

\bigskip

\noindent
{\large \bf Transcendentality assignments}
\begin{enumerate}
\itemsep=0in
\item 
The Riemann zeta-value $\zeta (n)$ for $n\geq 2$ has weight $n$, and therefore $\pi$ has weight  1;
\item 
The kinematic variables $s,t,u$ have weight $-1$;
\item 
For $n\ge 3$ the threshold contributions only arise in the combination $\ln(-2\pi s)-Z_n$ where $Z_n$ involves Euler's constant $\gamma$ and is defined by,
\bea
Z_n = { \zeta '(n) \over \zeta (n)} - { \zeta '(n-1) \over \zeta (n-1)} - \gamma
\eea
It therefore makes sense to  assign weight 1 to the combination $\ln(-2\pi s)-Z_n$.  This  is the minimal requirement to have a grading for the genus-one amplitude. Alternatively,  one might replace this with the stronger assumptions that $\ln (-2 \pi s)$ and $Z_n $ with $n \geq 3$ separately have weight 1.
\item 
Finite harmonic sum functions $H_k = \sum _{\ell=1}^k \ell^{-1}$ have weight 1. 
\end{enumerate}
Several comments on these assignments are in order.  Assumptions 1. and 2. are required already  to provide a grading for the tree-level amplitude. Assumption 3. implies that the differences $Z_m-Z_n$ for $m>n\geq 3$ have weight 1, an implication for which we shall present convincing evidence in section \ref{sec:52}. 

\sm

Assumption 4 is perhaps the most surprising one since it assigns a non-zero weight to a function given by a finite harmonic sum whose individual values are, after all, just rational numbers. Actually, one should think of $H_k$ as a function of $k$ to be inserted into an infinite series, and not as an individual rational number whose weight, of course, would be zero. For example, $H_k$ occurs in this manner in a double zeta-value, 
\bea
\zeta (a,1) = \sum _{k > \ell \geq 1} { 1 \over k^a \ell} = \sum _{k=2} ^\infty { 1 \over k^a} \, H_{k-1}
\eea 
The weights of $\zeta(a)$ and $\zeta(a,1)$ are respectively $a$ and $a+1$, thereby justifying the assignment of weight 1 to the function $H_k$.  Actually, assumption 4. is  familiar in considerations of transcendentality in $\cN=4$ quantum field theory amplitudes, going back to the work of \cite{Kotikov:2002ab,Beccaria:2009vt}, and we also find that it is required in order to provide string amplitudes with a consistent grading, and uniform transcendentality.  

\sm

Even without such an assignment of transcendentality to finite harmonic sum functions  the results of this paper would still put very strong restrictions on transcendentality.  The failure of uniform transcendentality would be concentrated in the terms that involve the harmonic sums, and the violation would be universally by one unit of the weight. 

 \sm

In order to make general arguments beyond order $D^{12}R^4$, we shall use a combination of exact results for the analytic contributions arising from two-loop modular graph functions \cite{DHoker:2019mib}, and general arguments rather than explicit formulas, to motivate the conjecture that transcendentality provides a grading to all orders in the low-energy expansion of the genus-one four-graviton amplitude. 

\sm

It is important to stress that, while each individual modular graph function  is well-defined and modular invariant, the string amplitude involves a special linear combination of several modular graph functions. The results of this paper are sensitive to this special combination in two respects. First, the cancellation of all dependence on $L$ between the contributions from the integrals of modular graph functions on $\cM_L$ against the exact all-orders results from $\cM_R$ requires a conspiracy between the modular graph functions in the special combination at every order in the low-energy expansion.  Second, the consistent grading by transcendental weight similarly can emerge only through this  special combination.

\vskip 0.2in 

\noindent
{\bf \large Acknowledgments}

\medskip

The authors are happy to thank Pierre Vanhove for collaboration at an early stage of this work, David Broadhurst, Justin Kaidi, Julio Parra-Martinez, Oliver Schlotterer, and Don Zagier  for useful conversations, and Justin Kaidi for helpful comments on the manuscript. ED is happy to acknowledge DAMTP in Cambridge, LPTHE at Jussieu, and LPTENS at the Ecole Normale Sup\'erieure in Paris, and both authors gratefully acknowledge  the Galileo Galilei Institute in Florence and the Niels Bohr International Academy in Copenhagen for the warm hospitality extended to them during part of this work.  

\sm

The research of ED  is supported in part by the National Science Foundation under research grant PHY-16-19926, and was supported by a Fellowship from the Simons Foundation.  MBG has been partially supported by STFC consolidated grant ST/L000385/1, by a Leverhulme Emeritus Fellowship, and by a  Simons Visiting Professorship at the NBIA.

\newpage

\section{The four-graviton Type II superstring amplitude }
 \label{sec:2}
\setcounter{equation}{0}

In this section, we shall review the tree-level amplitude and the integral representation of the genus-one four-graviton amplitudes in Type II superstring theory in ten-dimensional flat Minkowski space-time. We will show how a consistent assignment of transcendental weight may be introduced under which the tree-level amplitude has uniform transcendentality, so  the weight provides a grading of the low-energy expansion. 

\sm

Next, we consider the genus-one amplitude, review the partitioning of its moduli space $\cM$ into a neighborhood $\cM_R$ of the cusp and its complementary set $\cM_L$, expand the string integrand in powers of the kinematic variables $s,t,u$ multiplying modular graph functions, and calculate their integral over $\cM_L$ up to (and including) order $D^{12}\cR^4$. Various technical parts of the calculation are deferred to Appendix~\ref{sec:A}.

\subsection{Kinematics of the four-graviton amplitude}

The four-graviton Type II superstring amplitude $\bA (k_i, \ep_i) $ is given in string perturbation theory by
the following expression,
\bea
\bA (k_i, \ep_i) = \kappa_{10}^2 \,\cR^4 \sum _{h=0}^\infty g_s ^{2h-2} \cA^{(h)}(s_{ij})
\label{ampdef}
\eea
where $k_i$ and $\ep_i$ are respectively the momentum vector  and the polarization tensor of the external graviton $i=1,2,3,4$, satisfying $k_i^2=k_i \cdot \ep_i=0$ and momentum conservation $\sum_{i=1}^4 k_i=0$. The string coupling constant is denoted by $g_s$  and the ten-dimensional gravitational constant $\kappa _{10}^2$ is related to the string length scale by $\kappa_{10}^2 =   2^6 \pi^7 (\alpha')^4  g_s^2$.   The genus of the surface is denoted by $h$,  and the Lorentz-invariant kinematic variables $s_{ij} $ are defined to be dimensionless $s_{ij}=-\alpha '(k_i+k_j)^2/4$. All dependence on the polarization tensors is contained in the combination $\cR^4$ which stands for a particular scalar contraction of four linearized Riemann tensors built out of $\ep_i$, whose precise structure is dictated by maximal supersymmetry. Finally, the coefficient functions $\cA^{(h)}(s_{ij})$ are scalar functions which depend only on the kinematic variables $s_{ij}$.  Throughout we shall set $s=s_{12}=s_{34}$, $t=s_{14}=s_{23}$, and $u=s_{13}=s_{24}$, which satisfy $s+t+u=0$.

\subsection{The tree-level four-graviton amplitude}

The tree-level contribution $\cA^{(0)}(s_{ij})$ is given as follows, 
\bea
\label{A0}
\cA^{(0)} (s_{ij}) = { 1 \over stu} { \Gamma (1-s) \Gamma (1-t) \Gamma (1-u) \over \Gamma (1+s) \Gamma (1+t) \Gamma (1+u)}
\eea
which is the superstring version of the ``Virasoro'' amplitude \cite{Virasoro:1969me}. The dependence on $s,t,u$ is invariant under permutations of these variables, and is  through the symmetric polynomials $\sigma _k = s^k + t^k + u^k$ only, of which $\sigma_1=0$ and $\sigma _2$ and $\sigma_3$ are the remaining independent symmetric polynomials.\footnote{In view of $\sigma_1=s+t+u=0$ the polynomial $\sigma_k$ is a linear combination of $\sigma_2^l\sigma_3^m$ with $2l+3m=k$.}  The ratio of $\Gamma$-functions in (\ref{A0})  is analytic near $s=t=u=0$ and admits a Taylor series expansion whose radius of convergence is given by $|s|, |t|, |u| <1$. The amplitude may be recast in terms of the following series, 
\bea
\label{A0a}
\cA^{(0)} (s_{ij}) = { 1 \over stu} \, \exp \left \{ 2 \sum _{m=1}^\infty {\zeta (2m+1)  \over 2m+1}  \,  \sigma_{2m+1} \right \}
\eea
The Riemann zeta-function $\zeta(a)$ is defined for $\Re(a) >1$ by the following infinite sum,
\bea
\zeta(a) = \sum _{n=1}^\infty { 1 \over n^a}
\eea
and may be analytically continued in $a$ to a meromorphic function in $\CC$ with a simple pole at $a=1$. It is useful to record the first few terms in the low-energy expansion of $\cA^{(0)} (s_{ij})$ in powers of $s,t,u$, or alternatively in powers of $\sigma _2 $ and $\sigma _3$, 
\bea
\label{A0b}
\cA^{(0)} (s_{ij}) = { 3 \over \sigma _3} + 2 \zeta (3) + \zeta(5) \sigma _2 + {2\over 3}  \zeta (3)^2 \sigma _3 + \cO(\sigma _2^2)
\eea
The first term corresponds to the tree-level exchange of massless states. It is non-analytic and represents a  non-local interaction. All subsequent terms are analytic and produce local effective interactions, which we may represent schematically by $\cR^4$, $D^4 \cR^4$, and $D^6 \cR^4$.

\sm

It will be important for our subsequent analysis to note that the coefficients of the low-energy expansion of (\ref{A0a}) are powers of  odd Riemann zeta-values with rational coefficients.  Irreducible MZVs do not arise in this  expansion, although they do arise in the  expansion of $N$-particle tree amplitudes with $N>4$ \cite{SCHLotterer:2012ny}.

\sm

We shall now define a grading by transcendental weight $w[x]$ on the algebra of zeta-values and kinematic variables. The weight of a sum of two elements of the same weight equals that weight, while the weight of the  product of elements $x$ and $y$ is the sum of their weights, $w[xy] = w[x] + w[y]$. The weights assigned to the generators of the algebra are, 
\bea 
w [\zeta(a)] = a,  \hskip 0.5in  w[s] = w[t]=w[u]=-1
\eea
By construction, the transcendental  weight provides a  {\sl grading}. The argument of the exponential in (\ref{A0a}) has vanishing weight, giving weight 3 to the full amplitude.

\subsection{Multiple zeta-values and single-valued multiple zeta-values}
 
The coefficients of terms in the  low-energy expansion of $N$-particle tree-level superstring amplitudes with $N>4$ involve not only Riemann zeta-values, but also MZVs \cite{SCHLotterer:2012ny}.    The \multi zeta-function of {\sl depth} $\ell$ is defined by the following $\ell$-fold infinite sum (for overviews, see for example  \cite{ZagierMZV,Blumlein:2009cf}), 
\bea
\zeta (a_1, \cdots, a_\ell) = \sum _{n_1 > n_2> \cdots > n_\ell \geq 1} { 1 \over n_1^{a_1} \cdots n_\ell ^{a_\ell}}
\eea
The series is convergent for $\Re(a_1) >1$ and $\Re(a_i) \geq 1$ for $i \geq 2$, and may be analytically continued in $\CC^\ell$.  When  $a_i \in \NN$, we assign it the following transcendental weight,
\bea
w[ \zeta (a_1, \cdots, a_\ell) ] = a_1 + \cdots + a_\ell
\eea 
For $\ell=1$, we recover the Riemann zeta-value $\zeta(a)$ which has depth 1 and weight $a$. MZVs arise as special values of multiple polylogarithms. Multiple polylogarithms of depth $\ell$ and weight $a_1 + \cdots + a_\ell$ are defined by,
\bea
\Li _{a_1, \cdots, a_\ell} (z_1, \cdots , z_\ell) 
= \sum _{n_1 > n_2> \cdots > n_\ell \geq 1} { z_1^{n_1} \cdots z_\ell ^{n_\ell}  \over n_1^{a_1} \cdots n_\ell ^{a_\ell}}
\eea
and we have $\zeta (a_1, \cdots, a_\ell) = \Li _{a_1, \cdots, a_\ell} (1,\cdots, 1) $. Polylogarithms and MZVs obey shuffle and stuffle relations, of which we shall just exhibit the simplest stuffle relations,
\bea
\label{Lzeta}
L_a(y)L_b(z) & = & L_{a,b}(y,z) + L_{b,a}(z,y) + L_{a+b}(yz)
\no \\
\zeta (a) \zeta (b) & = & \zeta (a,b)+\zeta(b,a) + \zeta (a+b)
\eea
Shuffle and stuffle relations for polylogarithms and MZVs preserve the weight assignments  introduced above, as may be checked explicitly on the example given in (\ref{Lzeta}). The second relation in (\ref{Lzeta}) was derived by Euler almost 300 years. For $a,b\in \NN$ and $a+b$ odd, Euler also showed that the relation may be inverted and that every double zeta-value $\zeta (a,b)$ with $a+b \geq 5$ an odd integer may be reduced to a polynomial in zeta-values. For even $a+b$, and for MZVs with $\ell \geq 3$, the system is not  invertible and  MZVs are not generally reducible to a polynomial in zeta-values with rational coefficients. A great deal has been discovered concerning the basis of MZVs assuming that zeta values are transcendental, although this has not  been proved.  For example, stuffle and shuffle relations demonstrate that MZVs with weight less than eight are reducible to polynomials in ordinary Riemann Zeta-values with rational coefficients.  However,  the weight-eight MZVs  $\zeta(5,3)$ and $\zeta (6,2)$ are irreducible -- only the combination $2\zeta(5,3)+5 \zeta (6,2)$ can be expressed as a polynomial in  ordinary zeta-values with rational coefficients. The fact that the transcendentality of MZVs has not been proved may be  circumvented by adopting the motivic approach, where motivic MZVs are introduced as  abstract elements that satisfy the same stuffle and shuffle relations as MZVs but are otherwise unconstrained \cite{brown4, SCHLotterer:2012ny,Brown:2013gia}.   

\sm

The coefficients of terms in the low-energy expansion of closed-string  $N$-particle tree-level amplitudes are actually {\sl single-valued} MZVs  \cite{SCHLotterer:2012ny}.  These are identified with special values of single-valued multiple polylogarithms and form a subspace of the space of MZVs. The single-valued MZVs of weight $< 11$  are reducible to  powers of {\sl odd} Riemann zeta-values.  The lowest-weight example of an irreducible single-valued MZV arises at weight $11$ and depth $3$.  The extensive mathematical literature on single-valued MZVs originated with \cite{{Brown2004,Brown:2013gia,Stieberger:2014hba}}.

\sm

\subsection{The genus-one four-graviton amplitude}

The genus-one contribution $\cA^{(1)}(s_{ij})$  is given by the  integral \cite{Green:1981yb},  
\bea
\label{2a1}
\cA^{(1)}(s_{ij}) = 2\pi \, \int _{\cM} { d^2\tau \over \tau_2^2} \, \cB(s_{ij}|\tau)
\eea
where the overall normalization factor was determined in \cite{DHoker:2005jhf} by unitarity (see also \cite{Sakai:1986bi})\footnote{Note that the normalizations of the volume forms on $\Sigma$ and on $\cM$ used in \cite{DHoker:2005jhf} differ, by having an extra factor of 2 each, from the standard normalization adopted here. As a result, the overall factor of $\pi/16$ for the genus-one amplitude given in \cite{DHoker:2005jhf} becomes the overall factor of $2 \pi$ here.} The volume form is $d^2 \tau = { i \over 2} d\tau \wedge d \bar \tau$. The integrand $\cB(s_{ij}|\tau)$ is invariant under the action of $SL(2,\ZZ)$ on $\tau$ by M\"obius transformations. The integration is over a fundamental domain for the quotient $\cM=\cH/PSL(2,\ZZ)$ of the Poincar\'e upper half plane $\cH$ by the modular group $SL(2,\ZZ)$. We  choose for $\cM$ the standard fundamental domain given  by, 
\bea
\cM= \left \{ \tau \in \cH, ~ |\tau| \geq 1, ~ |\Re (\tau) | \leq \half \right \}
\eea
The integrand $\cB(s_{ij}|\tau)$  is given by an integral over four copies of the torus $\Sigma_\tau$ of complex structure  modulus $\tau$, which may be represented as the quotient  $\Sigma _\tau = \CC/\Lambda _\tau$ where  the lattice is given by $\Lambda _\tau = \ZZ+\tau \ZZ$,
\bea
\label{cB}
\cB (s_{ij} |\tau) = 
\prod _{k=1}^4 \int _{\Sigma_\tau } { d^2 z_k \over \tau_2} \, \exp \left \{ \sum _{1 \leq i<j \leq 4} s_{ij} \, G(z_i-z_j|\tau) \right \}
\eea
The scalar Green function $G(z|\tau)$ is defined as follows,
\bea
\tau_2 \pbz \p_z G(z|\tau) = - \pi \delta ^2(z) + \pi
\hskip 1in 
\int _{\Sigma _\tau} d^2 z \, G(z|\tau)=0
\eea
where $d^2 z = { i \over 2 } dz \wedge d\bar z$ and $\int _{\Sigma _\tau} d^2z \, \delta (z)=1$. The Green function may be represented as a Fourier series in the variable $z$,
\bea
\label{GFourier}
G(z|\tau) = \sum _{p \in \Lambda _\tau'} { \tau_2 \over \pi |p|^2} \, e^{2 \pi i (m \beta - n \alpha)} 
\eea
where $\Lambda ' _\tau = \Lambda _\tau \setminus \{ 0 \}$,  the real variables $\alpha, \beta$ are related to $z$ by $z = \alpha + \beta \tau$, and the integers $m,n$ are related to $p$ by $p = m + n \tau$. One verifies that the Green function is invariant under $SL(2,\ZZ)$ acting as follows, 
\bea
G\left ({ z \over c \tau +d} \bigg |{ a \tau + b \over c\tau+d} \right ) = G(z|\tau) 
\hskip 1in 
\left ( \begin{matrix} a & b \cr c & d \cr \end{matrix} \right ) \in SL(2,\ZZ)
\eea
As a result, the string amplitude  integrand $\cB(s_{ij} |\tau)$ is invariant under $SL(2,\ZZ)$ as well.

\sm

It will be convenient to partition $\cM$ into two complementary sets $\cM_L$ and $\cM_R$, 
\bea
\cM  = \cM_{L} \cup \cM_{R}, \qquad  \cM_{L}  = \cM \cap \{ \tau_2 \leq L \}, \qquad \cM_{R} = \cM \cap \{ \tau_2 >  L \}
\eea
where $L >1$. Our subsequent analysis will be simplified considerably by choosing $L \gg 1$. 
The reduced amplitude $\cA^{(1)}(s_{ij})$ decomposes accordingly, 
\begin{align}
\label{A1ALAR}
\cA^{(1)}  (s_{ij})  = 2\pi \Big ( \cA_L (L; s_{ij}) + \cA_R (L; s_{ij}) \Big ) 
\end{align}
The contributions $\cA_L$ and $\cA_R$ are given by the integral of (\ref{2a1}) in which the integration domain $\cM$ is replaced respectively by $\cM_L$ and $\cM_R$. Clearly, by construction, all dependence on $L$ must cancel  in the sum of $\cA_L$ and $\cA_R$.

\subsection{Low-energy expansion of the analytic contribution $\cA_L$}

The contribution $\cA_L(L; s_{ij}) $ is obtained by integrating $\cB(s_{ij}|\tau)$ over the bounded domain $\cM_L$. 
The integrand $\cB(s_{ij} |\tau)$ has simple poles in $s,t,u$ at positive integer values of $s,t,u$, which arise from the operator product expansion of the vertex operators of the massless gravitons, and physically correspond to the exchange of massive string states. Away from these poles, $\cB(s_{ij}|\tau)$ is analytic and may be expanded in a Taylor series in $s,t,u$ at $s=t=u=0$.  Since $\cB(s_{ij}|\tau)$  is a symmetric function in $s,t,u$ the  Taylor series may be organized in terms of symmetric polynomials $\sigma _2 = s^2+t^2+u^2$ and $\sigma _3 = s^3 + t^3 + u^3=3stu$, and we have,
\bea
\label{Bpq}
\cB (s_{ij}|\tau) = \sum _{p, q =0}^\infty  \cB _{(p,q)} (\tau) \,  \frac{\sigma _2^p \sigma _3^q}{p!\, q!}
\eea
 The  coefficients $\cB _{(p,q)} (\tau)$ are modular graph functions  of weight  $w=2p+3q$. The convergence of this Taylor series  in the domain  $|s|,|t|,|u|<1$  is uniform in $\tau$ throughout $\cM_L$. Since the domain $\cM_L$ is bounded, the integral $\cA_L(L; s_{ij})$ of $\cB(s_{ij}|\tau)$ over $\cM_L$ admits a Taylor series  with the same domain of convergence.  The coefficients $\cA _{(p,q)} (L)$ of the Taylor series  of $\cA_L(L; s_{ij})$ in powers of $\sigma_2$ and $\sigma_3$,
\bea
\label{Apq}
\cA_L(L;  s_{ij} ) = \sum _{p, q = 0}^\infty  \cA _{(p,q)} (L) \,  \frac{\sigma _2^p \, \sigma _3^q}{p!\, q!}
\eea 
are given by integrals over $\cM_L$ of the modular graph functions $\cB_{(p,q)} (\tau)$,
\bea
\label{Apq1}
\cA _{(p,q)} (L) = \int _{\cM_L} { d^2\tau \over \tau_2^2} \, \cB_{(p,q)} (\tau)
\eea
The integrals giving $\cA_{(p,q)}(L)$ are all absolutely convergent for arbitrary fixed $L$.

\sm

Making use of our assumption that $L \gg 1$, the dependence of $\cA_{(p,q)}(L)$ on $L$ will be governed by the $\tau_2$-dependence of $\cB(s_{ij}|\tau)$ for large $\tau_2$, which was obtained in \cite{DHoker:2015wxz}, 
\bea
\label{bpq}
\cB_{(p,q)}(\tau) = \sum _{k=1-w}^w \mb_{(p,q)}^{(k)} \tau _2 ^k + \cO(e^{-2 \pi \tau_2}) \hskip 0.6in w=2p+3q
\eea  
The resulting $L$-dependence of $\cA_{(p,q)} (L)$ is given by, 
\bea
\cA_{(p,q)} (L) = \cA_{(p,q)}^{(0)} + \mb_{(p,q)}^{(1)} \ln L + \sum _{{k=1-w \atop  k \not=1}} ^w \mb_{(p,q)}^{(k)} \,  { L^{k-1}  \over k-1} + \cO(e^{-2\pi L})
\eea
The quantity $\cA_{(p,q)}^{(0)}$ is independent of $L$, and is the only contribution that will survive the cancellation of all $L$-dependence when we add $\cA_L$ and $\cA_R$. It must be obtained by full integration of the modular graph function $\cB$ over $\cM_L$, and its calculation is the most difficult part of the evaluation of the genus-one contribution $\cA^{(1)}$. Henceforth, in evaluating $\cA_L$ and $\cA_R$, we shall omit all positive and negative powers of $L$, as well as all exponentially suppressed terms, and only retain the $L^0$ and $\ln L$ contributions.

\subsection{The modular graph functions $\cB_{(p,q)}$ up to weight 6}

The modular graph functions $\cB_{(p,q)}$ up to weight six were obtained in \cite{Green:2008uj}. For completeness,  their derivation will be reviewed in Appendix \ref{sec:A}. To facilitate their integration over $\cM_L$, the expressions for $\cB_{(p,q)}$  will be simplified dramatically by using the many identities obeyed by modular graph functions \cite{DHoker:2016mwo,DHoker:2016quv,Broedel:2018izr}. The goal of the simplifications is to expose Eisenstein series whenever possible in favor of more complicated modular graph functions, and to factor out a Laplace-Beltrami  operator $\Delta = \tau_2^2 (\p_{\tau_1}^2 + \p_{\tau_2}^2)$ acting on a modular graph function $\cC(\tau)$ so that Stokes's theorem may be used to evaluate its integral over $\cM_L$,
\bea
\label{intLap}
\int _{\cM_L} { d^2 \tau \over \tau_2^2} \, \Delta \cC = \int _0 ^1 d \tau _1 \p_{\tau_2} \cC (\tau) \Big |_{\tau_2=L}
\eea
 The expressions for $\cB_{(p,q)}$, simplified along the lines discussed above, are given by,
\bea
\label{Bupto6}
\cB_{(0,0)} & = & 1
\\
\cB_{(1,0)} & = &  E_2
\no \\
3 \, \cB_{(0,1)} & = & 5 E_3 + \zeta (3)
\no \\
\cB_{(2,0)}  & = & \Delta  C_{2,1,1} - 10  E_4 +  2 E_2^2 
\no \\
180 \, \cB_{(1,1)}  & =  & 70 \, \Delta C_{3,1,1} - 1612 E_5 + 580 E_2E_3 + 60 E_2 \zeta(3) +29 \zeta (5)
\eea
up to weight 5 for which there is a single kinematic contribution for each weight. At weight~6, two different kinematic arrangements contribute, 
\bea
\label{Bupto6a}
6 \, \cB_{(3,0)} & = & \Delta \Big ( - 9 C_{2,2,1,1}  + 6   C_{4,1,1} + 156  C_{3,2,1} +41  C_{2,2,2} +18E_3^2 + 9 E_2 E_4 \Big ) 
\no \\ && \quad 
+ 72 E_2 C_{2,1,1}   -12 E_3^2- 36 E_2 E_4 -2652 E_6
\no \\
27 \, \cB_{(0,2)} & = & \Delta \Big ( 9  C_{2,2,1,1} - 6   C_{4,1,1} +258   C_{3,2,1} +64   C_{2,2,2} - 18E_3^2 - 9 E_2 E_4  \Big ) 
\no \\ && \quad 
-36 E_2 C_{2,1,1}  + 483 E_3^2 + 30 \zeta (3) E(3) 
\no \\ && \quad
+ 18 E_2 E_4 +6E_2^3 - 3186 E_6 + 3 \zeta(3)^2 
\no \eea
We recall that the Eisenstein series $E_a$ may be defined by Kronecker-Eisenstein sums over the lattice $\Lambda_\tau ' = \Lambda_\tau  \setminus \{ 0 \}$ where $\Lambda_\tau = \ZZ \tau + \ZZ$,
\bea
E_a(\tau) = \sum _{p \in \Lambda_\tau  '} { \tau_2^a \over \pi^a |p|^{2a}}
\eea
as may the remaining modular graph functions of weight $w = a_1 + \cdots + a_r$, 
\bea
C_{a_1, \cdots, a_r} (\tau) = \sum _{p_1, \cdots, p_r \in \Lambda_\tau '} 
{ \tau_2^w \, \delta (\sum_r  p_r) \over \pi ^w |p_1|^{2a_1} \cdots |p_r|^{2a_r}}
\eea
The Eisenstein series is an eigenfunction of the Laplace-Beltrami operator, 
\bea
\label{LapE}
\Delta E_s(\tau) = s(s-1) E_s(\tau)
\eea
Various further identities satisfied by Eisenstein series and modular graph functions needed here may be found in Appendix \ref{sec:A}.

\subsection{The low-energy expansion of $\cA_L$ up to weight 6}

Putting all contributions together, neglecting terms exponential in $L$ or power-behaved in $L$ with non-vanishing exponent, we obtain,
\begin{align}
\cA_{(0,0)} & = { \pi \over 3} & \hskip 0.5in \cA_{(0,1)} & = { \pi \zeta (3) \over 9}
\no \\
\cA_{(1,0)} & = 0 & \hskip 0.5in \cA_{(1,1)} & = { 29 \pi \zeta(5)  \over 540} 
\end{align}
These contributions give the coefficients in the low-energy expansion respectively of the effective interactions $\cR^4$, $D^4 \cR^4$, and $D^6 \cR^4$ which all receive contributions from BPS states only, as well as of the effective interaction $D^{10}\cR^4$. The result $\cA_{(0,0)} = { \pi \over 3}$ is simply the volume of $\cM$, which is given by ${\rm Vol} (\cM)={\pi \over 3}$, and it will be natural to factor this number out of all contributions.  Having done so, we then see that all the above coefficients preserve transcendentality.

\sm

The remaining coefficients at weight four and six are given as follows,
\bea
\label{ALtot}
\cA_{(2,0)} & = &  
 { 4 \pi \zeta(3)  \over 45} \left [ 
\ln (2 L) + {\zeta '(4) \over \zeta (4) } - {\zeta '(3) \over \zeta (3) }  -{ 1 \over 4} \right ]
\\
\cA_{(3,0)} & = & { \pi \zeta (5) \over 315} \left [ 33 \ln (2L) - 2 { \zeta '(6) \over \zeta (6)}  -33  { \zeta '(5) \over \zeta (5)} 
+70  { \zeta '(4) \over \zeta (4)} - 35 { \zeta '(2) \over \zeta (2)} + {13 \over 2}  \right ]
\no \\
\cA_{(0,2)} & = &   { \pi \zeta(5) \over 405} \left [ 18 \ln (2L) + 23 { \zeta '(6) \over \zeta (6)}  - 18  { \zeta '(5) \over \zeta (5)} 
-10  { \zeta '(4) \over \zeta (4)} +5 { \zeta '(2) \over \zeta (2)} - {31 \over 12} \right ] +{ \pi \zeta (3)^2 \over 27}
\no
 \eea
Even upon factoring out ${\rm Vol}(\cM)$, it is more challenging to identify any patterns of transcendentality, and we shall postpone such discussion until we have also evaluated the contributions from $\cA_R$.

\newpage

\section{The exact non-analytic part $\cA_R$}
 \label{sec:3}
\setcounter{equation}{0}

In this section, we shall obtain the non-analytic part $\cA_R(L; s_{ij})$ of the genus-one amplitude arising from the integral of $\cB(s_{ij}|\tau)$ over $\cM_R$. To evaluate $\cA_R(L; s_{ij})$ we shall partition the integration over the four copies of the torus into six channels, and then perform the analytic continuation in $s_{ij}$ for each channel. A key result of this paper is that the low-energy expansion of $\cA_R$ may be obtained exactly as a power series to all orders in $s_{ij}$. 

\subsection{General structure of $\cA_R$}

The non-analytic part $\cA_R$ results from the contribution of the integral of $\cB(s_{ij}|\tau)$ over $\cM_{R}$, 
\bea
\cA_R(L; s_{ij}) = \int _{\cM_R} { d^2\tau \over \tau_2^2} \, \cB(s_{ij}|\tau)
= \int _0 ^1 d \tau_1 \int _L ^\infty { d \tau_2 \over \tau_2^2} \, \cB(s_{ij}|\tau)
\eea
where $\cB(s_{ij}|\tau)$ is given by (\ref{cB}). We shall parametrize the points $z_i$ on the torus by the real coordinates $\a_i, \b_i$ with $z_i = \a_i + \tau \b_i$ with $\a_i, \b _i \in \RR/\ZZ$ for $i=1,2,3,4$. We rearrange the Fourier decomposition of  the Green function $G(z|\tau)$ in (\ref{GFourier}) into  its constant Fourier mode $G_0(z|\tau) $ in the variable $\alpha$ plus the contribution of all the non-constant  Fourier modes $g(z|\tau)$,
\bea
G(z|\tau) = G_0(z|\tau) + g (z|\tau)
\eea
For the specific choice of region $0 \leq \b_i < 1$ the differences $\b= \b_i -\b_j$  lie in the interval $-1 < \b < 1$ where we have the following expressions for the two contributions to $G$,
\bea
G_0(z|\tau) & = & 2 \pi \tau_2 \left ( \b^2 - |\b| +{ 1 \over 6} \right )
\no \\
g (z|\tau) & = & \sum _{m\not = 0} { 1 \over |m|} \sum _k \exp \Big \{ 2 \pi i m (\alpha + \tau_1 \beta+ k \tau_1)
- 2 \pi \tau _2 |m(k+\beta)| \Big \}
\eea
We expose the independence of $G_0$ on the angles $\alpha_i$  by introducing $Q$ as follows, 
\bea
4 \pi \tau_2 Q (s_{ij} ; \beta _i) = \sum _{i<j} s_{ij} \, G_0 (z_{ij} |\tau)
\eea
In terms of $Q$ the function $\cA_R(L;s_{ij})$ may be recast in the following form,
\bea
\cA_R(L; s_{ij}) & = &  \int _L ^\infty { d \tau_2 \over \tau_2^2} \, 
\prod _{k=1}^4  \int _0 ^1 d \beta_k  \, e^{ 4 \pi \tau_2 Q(s_{ij}; \beta _i)}
\cF(s_{ij}; \beta _i , \tau_2) 
\no \\
\cF(s_{ij}; \beta _i , \tau_2)  & = & 
\int _0 ^1 d \tau_1  \prod _{k=1}^4 \int _0 ^1  d \alpha_k \, 
\exp \left \{ \sum _{i<j} s_{ij} g(z_i-z_j|\tau) \right \}
\eea
On the one hand, for non-zero $\beta$ (or more generally for $\beta \not \in \ZZ$), the non-constant Fourier part $g$ of the Green function is bounded by an exponential $\cO(e^{-2 \pi L |\beta|})$, while for $\beta \to 0$ it behaves as $- \ln (2 \pi \tau_2 |\b|)$. Therefore, any polynomial in $g(z_i-z_j |\tau)$ is integrable on  $\Sigma ^4 \times \cM_R$, and we may expand the integral over the exponential $\cF$ in powers of $s_{ij} g(z_i-z_j|\tau)$. 

\sm

On the other hand, the dependence on $G_0$ may not be so expanded since we have $G_0  \to \infty$ as $\tau_2 \to \infty$, and we shall treat its dependence exactly. The non-analytic dependence of $\cA_R$ on $s_{ij}$, in the form of branch cuts in $s_{ij}$ starting at $s_{ij}=0$,  arises precisely from the integral near the cusp of the exponential dependence on $Q(s_{ij}, \beta_i)$.

\subsection{Partitioning the integration into six channels}

In view of the presence of the absolute values  $|\beta_i-\beta_j|$ in the contributions of $G_0$ to $Q$, it is useful to partition the domain of integration $[0,1]^4$ over the variables $\beta_i$ into six regions, which physically correspond to the different scattering {\sl channels} \cite{DHoker:1994gnm}. Below we list these channels  together with the simplified form taken by the function $Q$ in each one of the channels,  
{\small \bea
\cD_{st} =  \Big \{ 0 \leq \b_1 \leq \b_2 \leq \b _3 \leq \b_4 =1 \Big \} 
& \hskip 0.2in & 
Q_{st} = s \b_1 (\b_3-\b_2) + t(\b_2-\b_1)(1-\b_3)
\no \\
\cD_{ts} = \Big \{ 0 \leq \b_3 \leq \b_2 \leq \b _1 \leq \b_4 =1\Big \}
& \hskip 0.2in &
Q_{ts} = t \b_3 (\b_1-\b_2) + s(\b_2-\b_3)(1-\b_1)
\no \\
\cD_{tu} =  \Big \{ 0 \leq \b_1 \leq \b_3 \leq \b _2 \leq \b_4 =1 \Big \} 
& \hskip 0.2in &
Q_{tu} = t \b_1 (\b_2-\b_3) + u(\b_3-\b_1)(1-\b_2)
\no \\
\cD_{ut} = \Big \{ 0 \leq \b_2 \leq \b_3 \leq \b _1 \leq \b_4 =1\Big \}
& \hskip 0.2in &
Q_{ut} = u \b_2 (\b_1-\b_3) + t(\b_3-\b_2)(1-\b_1)
\no \\
\cD_{us} =  \Big \{ 0 \leq \b_2 \leq \b_1 \leq \b _3 \leq \b_4 =1 \Big \} 
& \hskip 0.2in &
Q_{us} = u \b_2 (\b_3-\b_1) + s(\b_1-\b_2)(1-\b_3)
\no \\
\cD_{su} = \Big \{ 0 \leq \b_3 \leq \b_1 \leq \b _2 \leq \b_4 =1\Big \}
& \hskip 0.2in &
Q_{su} = s \b_3 (\b_2-\b_1) + u(\b_1-\b_3)(1-\b_2)
\qquad
\eea}
The non-analytic part of the amplitude $\cA_R$ is obtained by  summing the contributions from the six channels labelled $\cD_\cC$ above,
\bea
\label{Achannels}
\cA_R (L; s_{ij}) = \sum _{\cC} \cA_\cC (L; s_{ij})
\hskip 1in \cC =st, ts, tu, ut, us, su
\eea
where the non-analytic part in each channel $\cC$ is given by,
\bea
\cA_{\cC} (L; s_{ij} ) =  \int _L ^\infty { d \tau_2 \over \tau_2^2} \, 
\int _{\cD_\cC} d ^4 \beta  \, e^{ 4 \pi \tau_2 Q_\cC (s_{ij}; \beta _i)}
\cF(s_{ij}; \beta _i , \tau_2) 
\eea
Note that the function $\cF$ stays the same in all channels.

\sm

It will be convenient to use new coordinates $x_i$ for $i=1,2,3,4$ in each region in terms of which the domain is the same for all six regions, 
\bea
\cD = \Big \{ x_1, x_2, x_3, x_4 \geq 0, ~ x_1+x_2+x_3+x_4=1 \Big \}
\eea
and such that the dependence on $x_i$ of $Q_{st}$ and its permutations is the same for each domain, with the roles of $s,t,u$ permuted. Concretely, the changes of variables are as follows,
{\small \begin{align}
\cD_{st} &   &  x_1&=\beta _1, &  x_2&=\beta_2-\beta_1, &  x_3&= \beta _3 - \beta _2, & x_4 &= 1-\beta _3, & 
Q_{st} & = sx_1x_3+tx_2x_4
\no \\
\cD_{ts} &  &  x_1&=\beta _3, &  x_2&=\beta_2-\beta_3, &  x_3&= \beta _1 - \beta _2, & x_4 &= 1-\beta _1, &
Q_{ts} &= tx_1x_3+sx_2x_4
\no \\
\cD_{tu} &  &  x_1&=\beta _1, &  x_2&=\beta_3-\beta_1, &  x_3&= \beta _2 - \beta _3, & x_4 &= 1-\beta _2, &
Q_{tu} &=  t x_1x_3+ u x_2x_4
\no \\
\cD_{ut} &  &  x_1&=\beta _2, &  x_2&=\beta_3-\beta_2, &  x_3&= \beta _1 - \beta _3, & x_4 &= 1-\beta _1, &
Q_{ut} &= u x_1x_3+t x_2x_4
\no \\
\cD_{us} &  &  x_1&=\beta _2, &  x_2&=\beta_1-\beta_2, &  x_3&= \beta _3 - \beta _1, & x_4 &= 1-\beta _3, &
Q_{us} &= u x_1x_3+s x_2x_4
\no \\
\cD_{su} &  &  x_1&=\beta _3, &  x_2&=\beta_1-\beta_3, &  x_3&= \beta _2 - \beta _1, & x_4 &= 1-\beta _2, &
Q_{su} &= sx_1x_3+ux_2x_4
\end{align} }
In terms of the variables $x_i$, it is clear that the non-analytic part of the amplitude in each channel may be expressed in terms of a single function, whose arguments of $s,t,u$ get permuted for the other channels,
\bea
\label{channels}
\cA_{st}(L; s_{ij}) = \cA_\star(L; s,t) & \hskip 0.6in & \cA_{ts}(L; s_{ij}) = \cA_\star (L; t,s) 
\no \\
\cA_{tu}(L; s_{ij}) = \cA_\star (L; t,u) & \hskip 0.6in & \cA_{ts}(L; s_{ij}) = \cA_\star (L; u,t) 
\no \\
\cA_{us}(L; s_{ij}) = \cA_\star (L; u,s) & \hskip 0.6in & \cA_{ts}(L; s_{ij}) = \cA_\star (L; s,u) 
\eea
The function $\cA_\star$ may be expressed as follows,
\bea
\cA_\star(L;s,t) = \int _L ^\infty { d \tau_2 \over \tau_2^2} \int _{[0,1]^4} [d x] \, e^{4 \pi \tau _2 Q_{st}} \cF(s_{ij};\beta_i, \tau_2)
\eea
where the integration measure in $x_i$ is given by, 
\bea
[dx] = dx_1 \, dx_2\, dx_3 \, dx_4  \, \delta (1-x_1-x_2-x_3-x_4)
\eea
The function $\cF$ is expressed in terms of $s,t$ by eliminating $u$ in terms of $s,t$,  and the variables $\beta_i$ are given in terms of $x_i$ by the change of variables appropriate for channel $st$, namely $\beta_1=x_1$, $\beta_2=x_1+x_2$, $\beta_3=x_1+x_2+x_3$, and $\beta _4=x_1+x_2+x_3+x_4=1$.

\subsection{Expanding $\cF$ in powers of $s,t,u$}

Any power of the  non-constant Fourier mode  $g$ of the Green function is  integrable and the part dependent only on $g$ may be expanded in a power series in $s,t,u$. Using the abbreviation $g_{ij} = g(z_i-z_j|\tau)$, we obtain,  
\bea
\cF (s_{ij}; \b_i, \tau_2)= \int _0 ^1 d \tau_1 \prod _{\kappa=1}^4 \int _0  ^1 d \alpha _\kappa \, 
\prod_{i<j} \left ( \sum_{N_{ij}=0}^\infty { 1 \over N_{ij} !} \, s_{ij}^{N_{ij}} \, g_{ij}^{N_{ij}} \right )
\eea 
Since the integral in $\alpha_i$ vanishes when only a single factor $g$ depends on $\alpha_i$, no vertex can terminate a single $g$. For example, up to third order in $s,t,u$, we have, 
\bea
\cF & = & \int _0 ^1 d \tau_1 \prod _{k=1}^4 \int _0  ^1 d \alpha _k \, 
\bigg ( 1+ {s^2\over 2} (g_{12}^2+g_{34}^2 ) + {t^2\over 2} (g_{14}^2+g_{23}^2) + {u^2\over 2} (g_{13}^2 + g_{24}^2) 
 \\ && \hskip 1in
+ stu \Big ( g_{12} \, g_{23} \, g_{31} + g_{13} \, g_{34} \, g_{41} 
+ g_{12} \, g_{24} \, g_{41} + g_{23} \, g_{34} \, g_{42} \Big  )
 \bigg ) + \cO(s_{ij}^4)
\no
\eea 
To higher order, the expansion rapidly becomes unwieldy, and it is unknown how to compute the remaining integrals over $\beta_i$ and $\tau_2$ exactly. However, we are seeking here only the contributions that are of order $L^0$ and $\ln L$. The number of terms contributing to these orders is drastically reduced, as we shall show in the next subsection.

\subsection{Contributions involving factors of $g_{13}$ or $g_{24}$ are suppressed}

We begin by showing that, in the contribution to second order in $s,t,u$ given above, the contributions involving  $g_{13}$ and $g_{24}$ are suppressed by inverse powers of $L$ for large $L$. We shall then prove a Lemma stating that any contribution which involves a factor of $g_{13}$ and/or $g_{24}$ is suppressed by powers of $L$ for large $L$.
By interchanging $s$ and $t$, the cases for $g_{13}$ and $g_{24}$ are equivalent to one another, and we shall concentrate on $g_{13}$.

\sm

The  integral over $\alpha _i$ and $\tau_1$ evaluates  as follows,
\bea
\label{f21}
\cF_{13}^{(2)} =  \half \int _0^1 d \tau_1 \prod _{\kappa=1}^4 \int _0^1 d \alpha _\kappa \, g_{13} ^2 
= \sum _{m=1}^\infty { 1 \over m^2} \, \sum _{ k \in \ZZ} e^{-4 \pi \tau_2 m |k+x_2+x_3 |}
\eea
Since $x_2+x_3\geq 0$, the contributions from $k \not=0,-1$ will be uniformly exponentially suppressed and of order $\cO(e^{-4 \pi \tau_2})$. Within this approximation,  the sum may be restricted to its contributions from $k=0,-1$, and we obtain, 
\bea
\cF_{13}^{(2)} = \sum _{m=1}^\infty { 1 \over m^2} 
\Big ( e^{-4 \pi \tau_2 m (x_2+x_3)} + e^{-4 \pi \tau_2 m (x_1+x_4)}  \Big ) 
\eea
The contribution to $\cA_\star$ of the two terms are equal to one another upon swapping $(x_1,x_2)$ with $(x_3,x_4)$ and, for fixed value of $m \not=0$, is given by the following integral, 
\bea
\label{Lambda13}
\Lambda _{13} (m) = 2 \int _L ^\infty { d \tau _2 \over \tau_2^2} \int _{[0,1]^4}  [dx]  \, e^{4 \pi \tau _2 [ - m(x_2+x_3) +s x_1 x_3+t x_2 x_4] }   
\eea
Integrating over $x_4$ by using the $\delta$-functions sets $x_4=1-x_1-x_2-x_3$, and integrating next over $x_3$, we find,
\bea
\Lambda _{13} (m) & = &  
\int _L ^\infty  { d \tau _2 \over 2 \pi \tau_2^3} \int _0^1 dx_2 \int _0 ^{1-x_2}  dx_1    
{ e^{4 \pi \tau _2 [ - mx_2 +t x_2 (1-x_1-x_2)]}   \over  m- sx_1 + tx_2 }   
\no \\ &&
- \int _L ^\infty  { d \tau _2 \over 2 \pi \tau_2^3} \int _0^1 dx_1 \int _0 ^{x_1} dx_2    
{  e^{4 \pi \tau _2 [ - mx_1 +s (1-x_1) (x_1-x_2)]}  \over  m- s(1-x_1) + tx_2 } 
\eea
where we have used the change of variables $x_1 \to 1-x_1$ in the second line. Changing variables from $x_1, x_2$ to $x,y$ defined by $(x_1,x_2)=((1-x)y,x)$  in the first line, and $(x_1, x_2) = (x,xy)$ in the second line, we obtain,  
\bea
\Lambda _{13} (m) & = &  
\int _L ^\infty  { d \tau _2 \over 2 \pi \tau_2^3} \int _0^1 dx (1-x) \int _0 ^1  dy    \, 
{ e^{-4 \pi \tau _2 x [ m - t  (1-x)(1-y)]}   \over  m+t -(1-x)(sy +t)  }   
\no \\ &&
- \int _L ^\infty  { d \tau _2 \over 2 \pi \tau_2^3} \int _0^1 dx \, x \int _0 ^1  dy   \,  
{  e^{-4 \pi \tau _2 x [ m - s (1-x) (1-y)]}  \over  m-s +x(s+ty) } 
\eea
Since the Taylor series in powers of $s,t$ has domain of convergence $|s|,|t|,|u|<1$, we may  assume that $|s|, |t| < \ep < 1$ for some $\ep$. As a result,  the brackets in the exponentials as well as the denominator, are uniformly bounded from below by $m-\ep$ and from above by $m +\ep$. Hence the absolute values of both integrals are bounded from above  by, 
\bea
\int _L ^\infty  { d \tau _2 \over 2 \pi \tau_2^3} \int _0^1 dx     \, 
{ e^{-4 \pi \tau _2 x (m - \ep)}   \over  m - \ep  }  
= \int _L ^\infty  { d \tau _2 \over 8 \pi^2 \tau_2^4} { 1 - e^{-4 \pi \tau_2 (m-\ep)} \over (m-\ep)^2}
< { 1 \over 24 \pi^2 L^3(m-\ep)^2}
\eea 
and both integrals are suppressed by powers of $L$ of order $\cO(L^{-3})$, and may be omitted to all orders in $s,t,u$. The generalization of this result is given by the following Lemma.

\sm

{\lem
\label{lemma1}
{\sl Any contribution to $\cA_\star$ from the series expansion of $\cF$ in powers of $s,t,u$ which involves a factor of $g_{13}$ or a factor of $g_{24}$ is suppressed by at least $L^{-3}$ for large $L$.}}

\sm

The proof will be given in subsection \ref{proofs} and uses the suppression provided by a single factor of $g_{13}$ or $g_{24}$ to bound the entire integral over $x_i$ and $\tau_2$.

\subsection{Contributions involving $g_{ij}g_{ik}$ are suppressed for $j\not= k$}

Clearly, when the product $g_{ij}g_{ik}$ contains a factor of $g_{13}$ or $g_{24}$, the contribution is suppressed by Lemma \ref{lemma1}. Thus, there are no contributions to third order in $s,t,u$. To fourth order in $s,t,u$ the factor $g_{ij} g_{jk}$ occurs in two types of contributions, namely the square graphs generated by $g_{12}g_{23} g_{34} g_{41}$ and the touching bubbles graph generated by $g_{12}^2g_{14}^2$ and its permutations. 
We shall discuss the case of the square graph here before extending the validity of the results to a general Lemma.

\sm

For the square graph we have, 
\bea
\label{f1234}
\cF_{1234}^{(4)} =  
\int _0^1 d \tau_1 \prod _{\kappa=1}^4 \int _0^1 d \alpha _\kappa \, g_{12} \, g_{23} \, g_{34} \, g_{41} 
= \sum _{m=1}^\infty { 2 \over m^4} \, \sum _{ k_i \in \ZZ} \delta (\sum _i k_i) e^{-2 \pi \tau_2 m S_{1234}}
\eea
where the argument of the exponent is given by,
\bea
S_{1234} = |k_1+\beta_1-\beta_2| + |k_2+\beta _2-\beta _3| + |k_3+\beta _3-\beta_4| + |k_4 + \beta _4-\beta_1|
\eea
and the $\delta$-function arises from the integration over $\tau_1$. Since the differences of $\beta$ variables in the first three terms are negative we see that we must have $k_1,k_2,k_3 = 0,1$ lest the contribution be exponentially suppressed. In the last term the difference of $\beta$-variables is positive, so we must have $k_4=0,-1$  and thus $k_4=-k_1-k_2-k_3$ where at most one of the integers $k_1, k_2, k_3$ can take the value $+1$. The final result is as follows,
\bea
\label{f1234a}
\cF_{1234}^{(4)} =  \sum _{m=1}^\infty { 2 \over m^4} \sum_{i=1}^4 e^{-4 \pi \tau_2 (x_1+x_2+x_3+x_4-x_i)} 
\eea
Each term in this sum is bounded from above by $\cF_{13}^{(2)}$ and hence its integral over $x_i$ and $\tau_2$ is suppressed by at least three powers of $L$, by  the same arguments we used to show that $\cF^{(2)}_{13}$ led to such suppression. The generalization of this result is given by the following Lemma. 

\sm

{\lem
\label{lemma2}
{\sl Any contribution to $\cA_\star(L;s,t)$ from the Taylor series expansion of $\cF$ in powers of $s,t,u$ which involves at least one factor of $g_{ij} g_{jk}$ with $j \not=k$ is suppressed by at least three powers of $L$ for large $L$.}}

\begin{figure}[h]
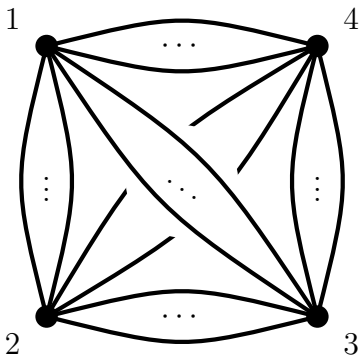

\begin{center}
\tikzpicture[scale=0.9]
\scope[xshift=-5cm,yshift=0cm]
\draw  [ultra thick] (2,0) node{$\cdots$};
\draw  [ultra thick] (0,0) .. controls (2, 0.5) .. (4,0);
\draw  [ultra thick] (0,0) .. controls (2, -0.5) .. (4,0);
\draw  [ultra thick] (0,2) node{$\vdots$};
\draw  [ultra thick] (0,0) .. controls (0.5, 2) .. (0,4);
\draw  [ultra thick] (0,0) .. controls (-0.5, 2) .. (0,4);
\draw  [ultra thick] (4,2) node{$\vdots$};
\draw  [ultra thick] (4,0) .. controls (4.5, 2) .. (4,4);
\draw  [ultra thick] (4,0) .. controls (3.5, 2) .. (4,4);
\draw  [ultra thick] (2,4) node{$\cdots$};
\draw  [ultra thick] (0,4) .. controls (2, 4.5) .. (4,4);
\draw  [ultra thick] (0,4) .. controls (2, 3.5) .. (4,4);
\draw  [ultra thick] (0,0) .. controls (1.5, 2.5) .. (4,4);
\draw  [ultra thick] (0,0) .. controls (2.5, 1.5) .. (4,4);
\draw  [ultra thick, fill, color=white] (2,2) circle (0.8);
\draw  [ultra thick] (0,4) .. controls (1.6, 1.6) .. (4,0);
\draw  [ultra thick] (0,4) .. controls (2.4, 2.4) .. (4,0);
\draw  [ultra thick] (2,2) node{$\ddots$};
\draw [thick, fill=black] (0,0) circle(0.15);
\draw [thick, fill=black] (4,0) circle(0.15);
\draw [thick, fill=black] (0,4) circle(0.15);
\draw [thick, fill=black] (4,4) circle(0.15);
\draw (-0.5,4.4) node{$1$};
\draw (-0.5,-0.4) node{$2$};
\draw (4.5,-0.4) node{$3$};
\draw (4.5,4.4) node{$4$};
\endscope
\endtikzpicture
\caption{Decomposition of a general modular graph function contributing to the channel $\cA_*(L;s,t)$, where the constant Fourier part $G_0$ of the Green function is treated exactly, and each edge represents the non-constant Fourier part $g$ of the Green function, with $N_{ij}$ edges connecting vertices $i$ and $j$.  \label{fig:1}}
\end{center}
\end{figure}

\subsection{Proofs of the Lemmas}
\label{proofs}

The proof of both lemmas proceeds by bounding the corresponding integrals by simpler integrals which may be bounded from above by inverse powers of $L$. A general term contributing to $\cA_\star$ takes the following form, for integer exponents $N_{ij}\geq 0$,
\bea
\label{CC}
\cC= \int _L ^\infty { d \tau_2 \over \tau_2^2} \int _{[0,1]^4} [d x] \, e^{4 \pi \tau _2 Q_{st}} 
\int _0 ^1 d \tau_1 \prod _{\kappa=1}^4 \int _0  ^1 d \alpha _\kappa \prod _{1\leq i<j\leq 4} g(z_i-z_j|\tau)^{N_{ij}}
\eea
We shall represent each factor $g$ by an independent infinite sum over variables $m$ and $k$, which we label as follows. For each pair $i<j$, we introduce $N_{ij}$ integers  $a_{ij}$ which label each Green function with $1 \leq a_{ij} \leq N_{ij}$. The corresponding summation variables $m$ and $k$ will be labelled by $m_{ij} ^{a_{ij}}$ and $k_{ij}^{a_{ij}}$. Carrying out the integrals over $\a_\kappa$ and $\tau_1$ gives,
\bea
\cC & =& \sum_{m_{ij} ^{a_{ij}} \not=0} \, \sum_{ k_{ij}^{a_{ij}}} \delta (K) \delta (M)  \prod _{i<j} 
\prod _{a_{ij}=1}^{N_{ij}} { 1 \over |m_{ij}^{a_{ij}}|}
\int _L ^\infty { d \tau_2 \over \tau_2^2} \int _{[0,1]^4} [d x] \, e^{4 \pi \tau _2 Q_{st}} 
\no \\ && \qquad \times 
\exp \Big \{ - 2 \pi \tau_2 \sum_{i<j} \sum _{a_{ij}=1}^{N_{ij}} |m_{ij}^{a_{ij}}| \cdot |k_{ij} ^{a_{ij}} + \beta _i-\beta _j| \Big \}
\eea
where the Kronecker $\delta(M)$ and $\delta(K)$ are given by,
\bea
\delta (K) & = & \delta \bigg ( \sum _{i<j} \sum_{a_{ij} =1}^{N_{ij}} m^{a_{ij}}_{ij} k_{ij} ^{a_{ij}} \bigg )
\no \\
\delta (M) & = & \delta(M_{12}+M_{13}+M_{14}) \, \delta (M_{12} - M_{23}-M_{24} )\, \delta(M_{13}+M_{23}-M_{34})
\no \\
M_{ij} & = & \sum _{a_{ij}=1}^{N_{ij}} m_{ij} ^{a_{ij}}
\eea
For $s,t$ real and negative, the integrand is real and  positive and the integral is convergent.

\subsubsection{Proof of Lemma \ref{lemma1}}

To prove Lemma \ref{lemma1}, we assume that $N_{13}\geq 1$ (or equivalently that $N_{24}\geq 1$). Since $-1 \leq \beta _i - \beta _j\leq 0$, only the values $k_{ij} ^{a_{ij}}=0,1$ can contribute and we restrict the sum over $k$ accordingly. Next, we bound the exponential on the second line of (\ref{CC}) by setting $m_{ij} ^{a_{ij}}=0$ for all $i<j$ such that $(i,j) \not= (1,3)$, so that we have,
\bea
\cC & \leq & \sum_{m_{ij} ^{a_{ij}} \not=0} \, \sum_{ k_{ij}^{a_{ij}}} \delta (K) \delta (M)  \prod _{i<j} 
\prod _{a_{ij}=1}^{N_{ij}} { 1 \over |m_{ij}^{a_{ij}}|}
\int _L ^\infty { d \tau_2 \over \tau_2^2} \int _{[0,1]^4} [d x] \, e^{4 \pi \tau _2 Q_{st}} 
\no \\ && \qquad \times 
\exp \bigg \{ - 2 \pi \tau_2 \sum_{a_{13}=1}^{N_{13}} |k_{13} ^{a_{13}} -x_2-x_3| \bigg \}
\eea
When $k_{13}^{a_{13}}=0$ or $1$ respectively, the argument of the exponential is bounded from below by $-2 \pi \tau_2 (x_2+x_3)$ or $-2 \pi \tau_2 (x_1+x_4)$. In either case the resulting integral of (\ref{Lambda13}) is suppressed by inverse powers of $L$. The remaining sums over $k_{ij}^{a_{ij}}$ and $m_{ij}^{a_{ij}}$ may be bounded from above by convergent sums, and are thus convergent. 

\subsubsection{Proof of Lemma \ref{lemma2}}

To prove Lemma \ref{lemma2} we assume that $N_{13}=N_{24}=0$ since contributions which do not satisfy these constraints are suppressed by inverse powers of $L$ in view of Lemma \ref{lemma2}.  We shall investigate (\ref{CC}) for  $N_{12}, N_{14} \not=0$, the other three cases being analogous. As with the proof of Lemma \ref{lemma1}, we take $s,t$ to be real and negative, in which case the integrand of (\ref{CC})  is positive and the integrals convergent. Only the values $k_{ij}^{a_{ij}}=0,1$ can contribute and we restrict the sum over $k$ accordingly. We bound the integrand from above by setting $m_{23}^{a_{23}}=m_{34}^{a_{34}}=0$ and $m_{12}^{a_{12}}=m_{14}^{a_{14}}=1$, so that we have, 
\bea
\cC & \leq & \sum_{m_{ij} ^{a_{ij}} \not=0} \, \sum_{ k_{ij}^{a_{ij}}} \delta (K) \delta (M)  \prod _{i<j} 
\prod _{a_{ij}=1}^{N_{ij}} { 1 \over |m_{ij}^{a_{ij}}|}
\int _L ^\infty { d \tau_2 \over \tau_2^2} \int _{[0,1]^4} [d x] \, e^{4 \pi \tau _2 Q_{st}} 
\no \\ && \qquad \times 
\exp \bigg \{ - 2 \pi \tau_2
 \sum_{a_{12}=1}^{N_{12}} |k_{12} ^{a_{12}} -x_2| 
-2 \pi \tau_2 \sum_{a_{14}=1}^{N_{14}} |k_{14} ^{a_{14}} -1+x_1|\bigg \} 
\eea
The argument of the exponential in the second line above, which we shall denote by $\cL$, may now be bounded from below as follows for the different allowed values of $k_{12}^{a_{12}}$ and $k_{14}^{a_{14}}$, 
\bea
(k_{12}^{a_{12}}, k_{14}^{a_{14}}) = (0,0) && \cL \geq - 2 \pi \tau_2 ( x_2+x_3)
\no \\
(k_{12}^{a_{12}}, k_{14}^{a_{14}}) = (1,0) && \cL \geq - 2 \pi \tau_2 
\no \\
(k_{12}^{a_{12}}, k_{14}^{a_{14}}) = (0,1) && \cL \geq - 2 \pi \tau_2 ( x_1+x_2)
\no \\
(k_{12}^{a_{12}}, k_{14}^{a_{14}}) = (1,1) && \cL \geq - 2 \pi \tau_2 ( x_1+x_4)
\eea
For the case on the second line, we see immediately that we have exponential suppression in $L$, while for the remaining three cases, the integrals are of the form (\ref{Lambda13}) and thus are suppressed by inverse powers of $L$. In each case, the remaining coefficient sum over $m$ and $k$ variables is independent of $L$ and convergent.
Thus, any contribution with a factor of $g_{ij}g_{ik}$ and $j \not=k$ is suppressed at least by inverse powers of $L$, which proves Lemma \ref{lemma2}.

\subsection{Summary of the non-analytic part to order $L^0$ and $\ln L$}

\begin{figure}[h]
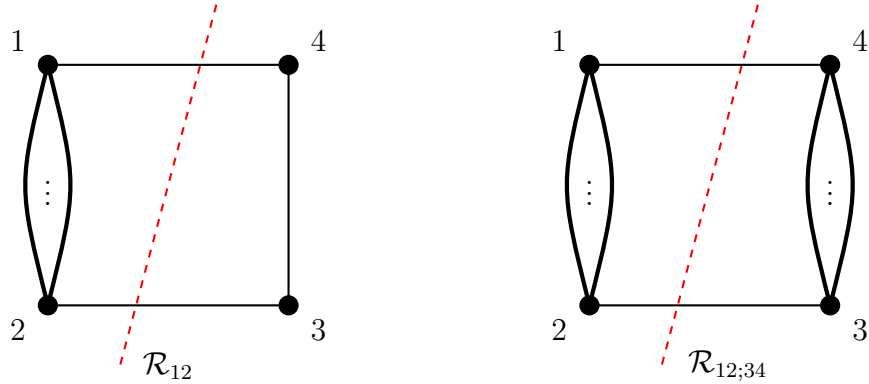

\begin{center}
\tikzpicture[scale=0.8]
\scope[xshift=-5cm,yshift=0cm]
\draw  [thick] (0,0) -- (4,0);
\draw  [ultra thick] (0,0) .. controls (0.5, 2) .. (0,4);
\draw  [ultra thick] (0,0) .. controls (-0.5, 2) .. (0,4);
\draw  [ultra thick] (0,2) node{$\vdots$};
\draw  [thick] (4,0) -- (4,4);
\draw  [thick] (0,4) -- (4,4);
\draw [thick, fill=black] (0,0) circle(0.15);
\draw [thick, fill=black] (4,0) circle(0.15);
\draw [thick, fill=black] (0,4) circle(0.15);
\draw [thick, fill=black] (4,4) circle(0.15);
\draw (-0.5,4.4) node{$1$};
\draw (-0.5,-0.4) node{$2$};
\draw (4.5,-0.4) node{$3$};
\draw (4.5,4.4) node{$4$};
\draw (2,-1) node{$\cR_{12}$};
\draw [thick, color=red, dashed] (2.8, 5) -- (1.2,-1);
\endscope
\scope[xshift=4cm,yshift=0cm]
\draw  [ thick] (0,0) -- (4,0);
\draw  [ultra thick] (0,0) .. controls (0.5, 2) .. (0,4);
\draw  [ultra thick] (0,0) .. controls (-0.5, 2) .. (0,4);
\draw  [ultra thick] (4,2) node{$\vdots$};
\draw  [ultra thick] (4,0) .. controls (4.5, 2) .. (4,4);
\draw  [ultra thick] (4,0) .. controls (3.5, 2) .. (4,4);
\draw  [ultra thick] (0,2) node{$\vdots$};
\draw  [thick] (0,4) -- (4,4);
\draw [thick, fill=black] (0,0) circle(0.15);
\draw [thick, fill=black] (4,0) circle(0.15);
\draw [thick, fill=black] (0,4) circle(0.15);
\draw [thick, fill=black] (4,4) circle(0.15);
\draw (-0.5,4.4) node{$1$};
\draw (-0.5,-0.4) node{$2$};
\draw (4.5,-0.4) node{$3$};
\draw (4.5,4.4) node{$4$};
\draw (2.3,-1) node{$\cR_{12;34}$};
\draw [thick, color=red, dashed] (2.8, 5) -- (1.2,-1);
\endscope
\endtikzpicture
\caption{Contributions $\cR_{12}$ and $\cR_{12;34}$  which remain from the decomposition of a general modular graph function of the channel $\cA_*(L;s,t)$ of Figure \ref{fig:1} after implementation of Lemmas \ref{lemma1} and \ref{lemma2}. The thick black edges indicate the non-constant part $G_0$ of the Green function, and  represent the exchanges of massive states. The thin black edges schematically indicate the effect of the exactly treated constant Fourier part $g$ of the Green function, and represent the exchanges of massless states. Both graphs contribute to producing a massless cut in the $s$-channel, represented by the dashed lines. \label{fig:2}  }
\end{center}
\end{figure}

The non-analytic part $\cA_R(L,s_{ij})$ is given by a sum of contributions from six different channels in (\ref{Achannels}) each one of which has been expressed in terms of a single function $\cA_*(L,s,t)$ in (\ref{channels}). Lemmas \ref{lemma1} and \ref{lemma2} reduce the number of possible terms in $\cA_*(L,s,t)$ by showing that any contribution which contains a factor of $g_{13}$, $g_{24}$ or a corner $g_{ij}g_{ik}$ with $j \not=k$ is suppressed by inverse powers of $L$ for large $L$, and thus may be omitted. The remaining contributions are  represented schematically in Figure \ref{fig:2}. This simplification allows us to evaluate $\cA_*(L,s,t)$ exactly to this order, a result summarized by Theorem \ref{theorem1} below.

We introduce the following notation for the integrals over angles $\tau_1, \alpha _i$,
\bea
\cF_{ij} ^{(N)} (\beta_i |\tau_2) & = &  \int _0^1 d\tau_1 \prod _{\kappa=1}^4 \int _0 ^1 d \alpha _\kappa \, g_{ij}^N
\no \\
\cF_{ij;kl} ^{(M,N)} (\beta _i |\tau_2) & = &  \int _0^1 d\tau_1 \prod _{\kappa=1}^4 \int _0 ^1 d \alpha _\kappa \, 
g_{ij}^M g_{kl}^N
\eea
and their further integrals over $\beta_i, \tau$, 
\bea
\cR_{12} ^{(N)} (L; s,t) & = & 
\int_L^\infty { d \tau_2 \over \tau_2^2} \int _{[0,1]^4} [dx] e^{4 \pi \tau_2 Q_{st}} \cF_{12}^{(N)} (\beta_i |\tau_2)
\no \\ 
\cR_{12;34} ^{(M,N)} (L; s,t) & = & 
 \int_L^\infty { d \tau_2 \over \tau_2^2} \int _{[0,1]^4} [dx] e^{4 \pi \tau_2 Q_{st}}  \cF_{12;34}^{(M,N)} (\beta_i |\tau_2)
\eea
The integrals are defined for $M,N \geq1$, but vanish whenever either $N=1$ and/or $M=1$. 
{\thm
\label{theorem1}
{\sl The function $\cA_*(L; s,t)$, to all orders in $s$ and $t$, and to order $L^0$ and $\ln L$ for large  $L$, is given by the following Taylor series in $s,t$,
\bea
\label{Astar}
\cA_\star(L; s,t) = \sum _{N=2}^\infty 2 { s^N \over N!} \cR_{12}^{(N)}(L; s,t) 
+ \sum _{M,N=2} ^\infty { s^{M+N} \over M! \, N!} \cR_{12;34}^{(M,N)} (L; s,t) + (s \leftrightarrow t) 
\eea
where the function $\cR^{(N)}_{12} (s,t)$ is given by,
\bea
\label{R12a}
\cR_{12}^{(N)}  & = & -  2 \pi  \sum _{k=0}^\infty \sum _{\ell=0}^k C_{k,\ell}  S(N,k+1) (-2s)^{k-\ell+2} (2t)^\ell  
\no \\ && \hskip 0.6in
\times \Big ( \ln(- 4 \pi Ls) + \Psi (k-\ell+3)  - 2 \Psi (k+7)  \Big )
\eea
and the function $\cR_{12;34}^{(M,N)} (s,t)$ takes the form, 
\bea
\label{R1234a}
\cR_{12;34}^{(M,N)}  & = &  
2 \pi   \sum _{k=0}^\infty  
\sum _{\ell_1=0}^ \infty \sum _{\ell_2=0}^ \infty D_k (\ell_1, \ell_2) S(M,k+\ell_1+1) S(N,k+\ell_2+1) 
  (-2s)^{k+\ell_1+\ell_2+3}  (-2t)^k 
\no \\ && \qquad \times
\Big (  \ln (- 4 \pi Ls)  + \Psi (k+\ell_1+\ell_2+4)  - 2 \Psi(2k+\ell_1+\ell_2 +8) \Big ) 
\eea
The rational-valued coefficients $C,D$ are given as follows,
\bea
\label{CD}
C_{k,\ell} & = & { k!  ~ (k-\ell+2)!  \over (k+6)! ~   (k-\ell)!} 
\no \\
D_k(\ell_1, \ell_2) & = & {  (k+\ell_1)! ~  (k+\ell_2) ! ~ (k+\ell_1+\ell_2+3) !  
\over k! ~  \ell_1 ! ~  \ell_2 ! ~   (2k+\ell_1+\ell_2+7) ! }
\eea
and the functions $S(N,k)$ are multiple infinite sums defined by, 
\bea
\label{SNk}
S(N,k) = \sum _{{m_r \not=0 \atop r=1, \cdots, , N}}
{  \delta (\sum_r m_r)  \over | m_1 \cdots m_N| \left ( |m_1| + \cdots |m_N| \right )^k }
\eea
The theorem will be proven in Appendix \ref{sec:B}.}}

\sm

A few remarks are in order. First, the $L$-dependence allowed by the theorem may be restated equivalently that all contributions which are exponential in $L$ are omitted, as are all terms which are power-behaved in $L$ with non-vanishing either positive or negative exponents. 

\sm

Second, an exact formula for $S(m,n)$ in terms of \multi zeta-values was given in Appendix A.3 of \cite{Green:2008uj} along with a proof by Zagier, 
\bea
\label{Szeta}
S(m,n) = m! \sum _{{a_1, \cdots, a_r \in \{ 1,2\} \atop a_1+\cdots + a_r =m-2}}
2^{2(r+1) -m-n} \zeta(n+2,a_1, \cdots, a_r)
\eea
Note that various powers of 2 appear in the expressions of (\ref{R12a}) and (\ref{R1234a})  in view of the factor of $\half$ in the definition of $m$ and $n$ in (\ref{Fm}) and (\ref{mn}) in Appendix \ref{sec:B}. 

\sm

Third, we have used the identities $\cF_{34}^{(N)}  = \cF_{12}^{(N)}$ and $\cF_{23}^{(N)}  = \cF_{14}^{(N)} $ 
to simplify the expression on the first line, and to produce the factor of 2 multiplying $\cR_{12}^{(N)}$.

\subsection{Contributions to $\cA_R(L;s_{ij})$ up to order $\cO(s_{ij}^6)$}

In this subsection, we shall evaluate $\cA_R(L;s_{ij})$ up to and including order $\cO(s_{ij}^6)$, which is the same order to which the analytic part is currently available in explicit form. The contribution arising from $\cR_{12;34}^{(M,N)}(L;s,t)$ in Theorem \ref{theorem1}, which contributes for $M,N\geq 2$, is multiplied by a factor $s^4$ while $\cR_{12;34}^{(M,N)}(L;s,t)$ itself has an explicit factor of $s^3$. Thus, this part of $\cA_*$ starts at $\cO(s^7)$ and does not contribute to order $\cO(s_{ij}^6)$. 

\sm

Since $\cR_{12}^{(N)}(L;s,t)$ in Theorem \ref{theorem1} has a  prefactor of $s^2$, all contributions to order $\cO(s_{ij}^6)$ arise from $2 \leq N \leq 4$ and, for any such $N$, from $k\leq 4-N$. They are given by,
\bea
\label{R12ex}
\cR_{12}^{(N)} (L;s,t) & = &   8 \pi s^2 \sum _{k=0}^{4-N} \sum _{\ell=0}^k C_{k,\ell}  S(N,k+1) (-2s)^{k-\ell} (2t)^\ell  
\no \\ && \qquad
\times \Big ( - \ln(- 4 \pi Ls) - \Psi (k-\ell+3)  + 2 \Psi (k+7)  \Big )
\eea
For the small values of $N$ needed here  the function $S(N,k+1)$, which was defined in (\ref{SNk}) and expressed in terms of MZVs in (\ref{Szeta}),  evaluates as follows,
\begin{align}
S(2,1) & =  \zeta(3) & \hskip 0.4in S(2,2) & =  \half \zeta (4) & \hskip 0.4in S(3,2) & = 6 \zeta (5) - 3 \zeta (2) \zeta (3)
\no \\
S(2,3) & = { 1 \over 4} \zeta (5) &  S(3,1) & = { 3 \over 2} \zeta (4)  &  S(4,1) & =  30 \, \zeta (5) - 12 \, \zeta (2) \, \zeta (3)
\end{align}
The contribution of order $\cO(s_{ij}^4)$ arises solely from $N=2$ and $k=0$ and is given by, 
\bea
s^2 \cR_{12}^{(2)} \Big |_{k=0} &= & { \pi \zeta (3) s^4 \over 45 } \left ( - \ln (- 4 \pi L s) - \gamma +{17 \over 5} \right )
\eea
The contributions to order $\cO(s_{ij}^5)$ arise from $N=2$ with $k=1$, and from $N=3$ with $k=0$. Adding the corresponding terms gives,  
\bea
s^2 \cR_{12}^{(2)} \Big |_{k=1} + {s^3 \over 3} \cR^{(3)}_{12} \Big |_{k=0}
=  { \pi \zeta (4)  s^4(s+2t) \over 630} \left ( - \ln (-4 \pi L s) - \gamma +{129 \over 35} \right )
 \eea
Upon adding the  contributions from the different channels, one is led to adding to the above contribution the one in which $t$ and $u$ are swapped. The sum is proportional to $(s+2t)+(s+2u)=0$, and hence the contribution to $\cA_R$ of order $\cO(s_{ij}^5)$ cancels. 

\sm

The contributions to order $\cO(s_{ij}^6)$ arise from $N=2$ with $k=2$, from $N=3$ with $k=1$, and from $N=4$ with $k=0$, which are individually given by, 
\bea
s^2 \cR_{12}^{(2)} \Big |_{k=2} & = &  
{ \pi \zeta(5) \over 1260} \, s^4 (6s^2-3st+t^2)  \Big ( -\ln (-4 \pi L s) - \gamma \Big ) 
\no \\ && \qquad 
+ { \pi \zeta (5)  \over 176400} \, s^4 (2816s^2-1513st+551t^2)
\no \\
{s^3 \over 3} \cR_{12}^{(3)} \Big |_{k=1} & = & 
{ \pi (-4 \zeta (5) + 2 \zeta (2) \zeta (3)) \over 315} \, s^5(3s-t)  \Big ( -\ln (-4 \pi L s) - \gamma \Big ) 
\no \\ && \qquad
+{\pi (-4 \zeta (5) + 2 \zeta(2) \zeta (3)) s^5 (352 s -129 t) \over 11025}
\no \\
{s^4 \over 12}  \cR_{12}^{(4)} \Big |_{k=0} & = &
{ \pi (5 \zeta (5) -2 \zeta(2) \zeta(3))  \over 90}  \, s^6 \Big ( -\ln (-4 \pi L s) - \gamma + { 17 \over 5} \Big ) 
\eea
Using the following relations between symmetric polynomials,
\bea
\sigma _4 = \half \sigma _2^2 \hskip 1in \sigma _6 = {1 \over 4} \sigma _2^3 + {1 \over 3 } \sigma _3^2
\eea  
 all terms involving  $\zeta(2)\zeta(3)$ cancel upon adding the contributions  from the six channels.    

\sm

Adding all contributions to order $\cO(s_{ij}^6)$ we find, 
\bea
\label{ARtot}
\cA_R(L;s_{ij}) & =  & { 4 \pi \zeta (3) s^4 \over 45 } \Big ( - \ln (- 4 \pi L s) - \gamma  \Big ) 
+ \hbox{ 2 cyclic perms of } s,t,u
\no \\ &&
+   { (84   s^6 + 2 s^4 \sigma _2)\pi   \over 1260} \zeta(5) \Big ( - \ln (-4 \pi Ls) - \gamma  \Big ) 
+ \hbox{ 2 cyclic perms of } s,t,u
\no \\ &&
+ { 34 \pi \zeta(3) \sigma _2^2 \over 75} + { 73 \pi \zeta(5)  \over 1225} \sigma _2^3  + {1423 \pi \zeta(5)  \over 18900} \sigma _3^2
\eea
where the instruction for cyclic permutation applies only to the first two lines. Note that the combination $84 s^6 + 2 s^4 \sigma _2$ may be expressed in terms of the combination found in \cite{Green:2008uj}, using the following relation 
$84  s^6 + 2 s^4 \sigma _2 = 87 s^6 +s^4 (t-u)^2$. These results will be used to study the full amplitude $\cA_L+\cA_R$  in Section \ref{sec:5}.

\section{Absence of irreducible \multi zeta-values in $\cA_R$}
 \label{sec:4}
\setcounter{equation}{0}

In this section, we shall use the results of Theorem \ref{theorem1}, which gives the expression for $\cA_*(L;s,t)$ to orders $L^0$ and $\ln L$ for large $L$ and to all orders in $s,t,u$, to prove  that the coefficients of the Taylor expansion in $s,t,u$ of the corresponding expression for $\cA_R(L;s_{ij})$  are all free of irreducible MZVs and are polynomials in odd zeta-values only. In fact we shall obtain relatively simple expressions for these functions in terms of integrals with integrands that involve the Virasoro tree-level amplitude (\ref{A0}).

\sm

The absence of irreducible MZVs in the  coefficient of the discontinuity of the amplitude is a direct consequence of unitarity  for the non-analytic part of $\cA_R(L;s_{ij})$ which is proportional to $\ln (-s)$ in the $s$-channel. However, there appears to be no physical argument that irreducible MZVs should also be absent from the analytic part of $\cA_R(L;s_{ij})$, but Theorem \ref{theorem2} below shows that this is nonetheless so.  The proof will proceed along the lines of a parallel result for the Laurent polynomial in $\tau_2$ of the modular graph functions $D_N$ for arbitrary $N$, which was given recently in \cite{DHoker:2019xef}.

\sm

It will be convenient to rearrange the decomposition of the non-analytic part $\cA_R(L;s_{ij})$ into a sum of terms that are individually free of irreducible MZVs. While the decomposition of (\ref{Achannels}) along with the symmetry under $s\leftrightarrow t$ of $\cA_*(L;s,t)$ gave,
\bea
\cA_R(L;s_{ij}) = 2 \cA_*(L;s,t) + 2 \cA_*(L;t,u) + 2 \cA_*(L;u,s)
\eea
we shall instead use the decomposition, 
\bea
\label{ARstar}
\cA_R(L;s_{ij}) = 2 \, \sum_{i=1}^2 \Big ( \mA_* ^{(i)} (L;s,t,u) +  \mA_* ^{(i)} (L;t,u,s) +  \mA_* ^{(i)} (L;u,s,t) \Big ) 
\eea
where the individual components are given by,
\bea
\label{mAi}
\mA_* ^{(1)} (L;s,t,u) & = &  
\sum _{N=2}^\infty 2 { s^N \over N!} \left ( \cR_{12}^{(N)}(L; s,t) + \cR_{12}^{(N)}(L; s,u) \right )  
\no \\
\mA_* ^{(2)} (L;s,t,u) & = &  
\sum _{M,N=2} ^\infty { s^{M+N} \over M! \, N!} \left ( \cR_{12;34}^{(M,N)} (L; s,t) +  \cR_{12;34}^{(M,N)} (L; s,u) \right ) 
\eea
The results are summarized by  Theorem \ref{theorem2} below, through which these functions will be expressed as integrals over a function $W$, 
\bea
\label{VS}
W(s,t) = {1 \over stu} \left ( {\Gamma (1-s) \Gamma (1-t) \Gamma (1-u) \over \Gamma (1+s) \Gamma (1+t) \Gamma (1+u)} -1 \right ) \hskip 1in u=-s-t
\eea
which is closely related to the Virasoro amplitude (\ref{A0}).

{\thm 
\label{theorem2}
{\sl The coefficients of the Taylor series in $s,t,u$ of the functions $\mA_* ^{(i)}(L;s,t,u)$ are free of irreducible MZVs and are polynomials in odd zeta-values only, with rational coefficients. These results may be seen explicitly from the following expressions for $\mA_* ^{(i)} (L;s,t,u)$.
\bea
\label{thm2a}
\mA_* ^{(1)} (L;s,t,u)  & = &   
- 4 \pi s^2 \left ( \mA (s,t,u;0) \, \ln(- 4 \pi Ls)  +{\p \over \p \ep}  \mA (s,t,u;\ep) \Big |_{\ep=0} \right )
\no \\
\mA_*^{(2)}(L;s,t,u) & = & - 4 \pi s^3 \left ( \mB (s,t,u;0) \, \ln (- 4 \pi Ls) + {\p \over \p \ep}  \mB(s,t,u;\ep) \Big |_{\ep=0} \right )
\eea
where $\mA (s,t,u;\ep)$ and $\mB (s,t,u;\ep)$ are independent of $L$ and given in terms of $W$ of (\ref{VS})  by,\footnote{We are grateful to Oliver Schlotterer for pointing out a missing factor of 4 in the expression for $\mB$ in (4.6), which originates from a missing factors of 2 in equations (4.28-31).}
\bea
\label{thm2b}
\mA (s,t,u;\ep)  & = &  
{ 2 s^2 \over \Gamma (3+\ep)} \int _0 ^1 \!\! dx \int _0 ^1 \!\! dy \,  
 x^{5+2\ep} y^{2+\ep} (1-y)^{2+\ep}  W(s,-sxy+t(1-x))
 \no \\
\mB(s,t,u;\ep) &  =  &
s^4 \sum_{k=0}^\infty  s^{k} t^k 
\int _0 ^1 dx \, { x^{k+3+\ep}(1-x)^{k+3+\ep} \over 2 \, \Gamma (k+4 +\ep) \, k! } 
 \left ( { \p ^k \over \p \mu ^k} W(s,\mu - sx) \Big |_{\mu=0} \right )^2
\eea
 }}

In the remainder of this section we shall prove Theorem \ref{theorem2}.  In appendix~\ref{sec:C} we will check that the discontinuity of $\mA (s,t,u;0) \, \ln(- 4 \pi Ls) $   across the $s$-channel branch cut matches the discontinuity  required by unitarity following the procedure in \cite{Green:2008uj}.  We have not verified that the discontinuity of $\mB (s,t,u;0) \, \ln (- 4 \pi Ls)$ matches that required by unitarity, although this is undoubtedly the case.  We stress that although unitarity directly implies that the discontinuity of the low-energy expansion of the amplitude has no irreducible MZVs, it does  not  imply the absence of MZVs in the remaining, analytic, contributions to $\mA_* ^{(i)}(L;s,t,u)$. {Therefore,  the result obtained in Theorem \ref{theorem2} is significantly stronger than what is required by unitarity alone, since the theorem implies that  the complete expressions   $\mA_* ^{(1)} (L;s,t,u)$ and $\mA_*^{(2)}(L;s,t,u)$  are free of irreducible MZVs.

\subsection{Calculation of  $\mA_* ^{(1)}(L;s,t,u)$}

The starting point for the calculation of the Taylor series of $\cR_{12}$ of Theorem \ref{theorem1} was given in (\ref{R12N}) of Appendix \ref{sec:B}, and may be used to construct  $\mA_* ^{(1)}(L;s,t,u)$ with the help of (\ref{mAi}) as well as (\ref{Lam12}) and (\ref{Cee}).  It will be convenient to collect these contributions and express them in the form of the first line in (\ref{thm2a}), with the function $\mA (s,t,u;\ep)$ given by, 
\bea
\label{mAep}
\mA (s,t,u;\ep) = \sum _{N=2}^\infty  { s^N \over N!} \sum _{{m_r \not=0 \atop r=1, \cdots, , N}}
{ \delta (\sum_r m_r)  \over | m_1 \cdots m_N|} \,  \mC (s,t,u;m;\ep)
\eea
where $2m = |m_1| + \cdots + |m_N|$. The  function $\mC$ is given by,
 \bea
\mC (s,t,u;m;\ep) = \int _0 ^1 dx \int _0 ^1 dy \, \left ( { x^{5+2\ep} y^{2+\ep} (1-y)^{2+\ep} \over m +s xy - t(1-x)}
+ { x^{5+2\ep} y^{2+\ep} (1-y)^{2+\ep} \over m +s xy - u(1-x)} \right )
\eea
To perform the sums over $m_r$, we introduce an angular integration over the variable $\alpha$ to enforce the vanishing of the sum of the $m_r$ variables and another integration over the variable $2 \beta$ to exponentiate the denominators in $\mC(s,t,u;m;\ep)$ using the formula,
\bea
{ 1 \over  m+sxy-t(1-x)} = \int _0 ^\infty 2 d \beta \, e^{ - \beta ( |m_1|+\cdots + |m_N| + 2 sxy - 2 t(1-x) )}
\eea
The integrations over $\alpha, \beta$ decouple the sums over $m_r$ from one another, and these sums may be carried out using the following elementary formula, 
\bea
\label{logsum}
\sum _{m_r \not=0} { 1 \over |m_r|} \, e^{ 2 \pi i m_r \alpha - \beta |m_r|} = 
- \ln \left  | 1 - e^{-\beta +2 \pi i \alpha} \right |^2
\eea
Carrying out also the sum over $N$ gives the following expression, 
\bea
\mA(s,t,u;\ep)  & = &   \int _0 ^1 dx \int _0 ^1 dy \,  x^{5+2\ep} y^{2+\ep} (1-y)^{2+\ep} 
\int _0 ^1 \!\! d \alpha \int _0^\infty  \!\! 2 d \beta 
\no \\ && \quad \times 
\Big (  |1-e^{-\beta + 2 \pi i \alpha} |^{-2s} -1 \Big )  \Big ( e^{-2 \beta (sxy-t(1-x))} + e^{-2 \beta (sxy-u(1-x))} \Big )
\eea
where the subtraction of 1 in the large parentheses accounts for the absence of the $N=0$ term in the sum over $N$ (\ref{mAep}), and we have also used the observation that the above integral produces a vanishing $N=1$ term.
Changing variables from $\alpha, \beta$ to $z= e^{-\beta + 2 \pi i \alpha}$,  we obtain,  
\bea
\mA (s,t,u;\ep) & = & { 1 \over \pi }
\int _0 ^1 dx \int _0 ^1 dy \,  x^{5+2\ep} y^{2+\ep} (1-y)^{2+\ep} \int _{|z|\leq 1}   {d^2 z  \over |z|^2} \,
\no \\ &&
\times  \Big (  |1-z |^{-2s} -1 \Big ) 
\Big (   |z|^{ 2sxy-2t(1-x)} +  |z|^{ 2sxy-2u(1-x)} \Big) 
\eea
When recast in terms of $s$ and $t$, the exponent of the second term of the second set of large parentheses takes the following form, 
\bea
2sxy-2u(1-x) =  2s -2sx(1-y) +2t(1-x)
\eea
Upon letting $y \to 1-y$, a transformation under which the  integration over $y$ is invariant, we recover the integrand given by the first term in the parentheses, but for $z \to z^{-1}$. Putting both contributions together produces an integral over the entire complex plane, 
\bea
\mA (s,t,u;\ep)  & = &  { 1 \over \pi }
\int _0 ^1 \!\! dx \int _0 ^1 \!\! dy \,  x^{5+2\ep} y^{2+\ep} (1-y)^{2+\ep} 
\bigg  [ \int _{\CC}   {d^2 z  \over |z|^2} \,  |1-z |^{-2s}  \,  |z|^{ -2t_1} 
\no \\ &&
\hskip 1.2in 
- \int _{|z|\leq 1}  d^2 z \, \Big ( |z|^{-2-2t_1} + |z|^{-2-2u_1} \Big ) \bigg ] 
\eea
where we  set, 
\bea
t_1=-sxy+t(1-x) \hskip 1in u_1 = -s -t_1
\eea
The first integral over $z$ is essentially Shapiro's representation of the Virasoro amplitude \cite{Shapiro:1970gy}, 
\bea
\int _\CC d^2 z \, |z|^{-2-2t_1} |1-z|^{-2s} 
=
{ \pi s \over t_1 u_1} \, { \Gamma (1-s) \Gamma (1-t_1) \Gamma (1-u_1) \over \Gamma (1+s) \Gamma (1+t_1) \Gamma (1+u_1) } 
\eea
while the subtraction terms given by the integrals over the unit disc are elementary and given by $\pi s /(t_1u_1)$. 
Assembling all contributions, and expressing the result in terms of the function $W$ of (\ref{VS}) proves the first part of Theorem \ref{theorem2} for the representation of $\mA (s,t,u;\ep)$ and thus $\mA_* (L;s,t,u)$.
Expressing the function $W$ in terms of odd zeta-values, as was done for the tree-level amplitude in (\ref{A0a}), then readily allows us to complete the proof that the coefficients of the Taylor expansion in $s,t,u$ of $\mA_* ^{(1)} (L;s,t,u) $, including non-analytic and analytic parts, are free of irreducible MZVs, and are polynomials in odd zeta-values only, with rational coefficients.

\subsection{Calculation of  $\mA_* ^{(2)}(L;s,t,u)$}

We shall take as a starting point the expression for $\mA^{(2)}(L;s,t,u)$ given in (\ref{mAi}) and recast $\cR_{12;34}^{(M,N)}$ in terms of the functions $\Lambda _{12;34}$ evaluated in Appendix \ref{sec:B}, 
\bea
\mA^{(2)}(L;s,t,u) = \sum_{M,N=2}^\infty { s^{M+N} \over M! \, N!} 
\sum_{{m_r \not=0 \atop 1\leq r \leq  M}} \sum _{{n_s\not=0 \atop 1\leq s \leq N}} 
{ \delta (\sum_r m_r) \delta (\sum_s n_s) \Lambda (L;s,t,u;m,n) 
\over |m_1 \cdots m_M \, n_1 \cdots n_N | }
\eea
where $\Lambda (L;s,t,u;m,n)= \Lambda_{12;34}(L;s,t;m,n)+ \Lambda_{12;34}(L;s,u;m,n)$, each terms of which was calculated  in (\ref{B23}), so that we find the following expression for $\Lambda$,
\bea
\label{Lam23}
\Lambda(L;s,t,u;m,n) & = &
\int _L ^\infty  { d \tau_2 \over 4 \pi \tau_2^3} \int _0 ^1  dx_1 \int _0 ^{1-x_1} \!\!\! dx_2 \, 
{ e^{4 \pi \tau_2[ - m x_2+sx_1(1-x_1-x_2)]}  \over n +sx_1-tx_2} 
\no \\ &&
+ \int _L ^\infty  { d \tau_2 \over 4 \pi \tau_2^3} \int _0 ^1  dx_1 \int _0 ^{1-x_1} \!\!\! dx_2 \, 
{ e^{4 \pi \tau_2[ - m x_2+sx_1(1-x_1-x_2)]}  \over n +sx_1-ux_2} 
\eea
and where $2m=|m_1|+\cdots + |m_M|$ and $2n=|n_1|+\cdots + |n_N|$. Changing variables from $x_1$ to $x=x_1$ in the first integral and from $x_1$ to $x=x_1+x_2$ in the second integral and eliminating $u=-s-t$ in favor of $s$ and $t$, we obtain, 
\bea
\Lambda(L;s,t,u;m,n)  & = & \int _L ^\infty  { d \tau_2 \over 4 \pi \tau_2^3} \int _0 ^1  dx \int _0 ^{1-x} dx_2 \, 
{ e^{4 \pi \tau_2[ - m x_2+sx(1-x-x_2)]}   \over n +sx-tx_2}
\no \\ && 
+ \int _L ^\infty  { d \tau_2 \over 4 \pi \tau_2^3} \int _0 ^1  dx \int _0 ^x dx_2 \, 
{ e^{4 \pi \tau_2[ - m x_2+s(x-x_2)(1-x)]}   \over n +sx + tx_2}
\eea
Both contributions are analytic in $t$ and may be expanded in a Taylor series in $t$.   The integrations in $x_2$ may now be extended to $\infty$, up to exponentially suppressed terms which we neglect, and we obtain, 
\bea
\Lambda(L;s,t,u;m,n) & = &   \sum _{k=0}^\infty k! \, t^k 
\int _L ^\infty  { 16 \pi^2 \, d \tau_2 \over (4 \pi  \tau_2)^{k+4} } \int _0 ^1  dx \, 
{  e^{4 \pi \tau_2sx(1-x)}  \over  (m+sx)^{k+1} (n+sx)^{k+1}}
 \\ &&
+ \sum _{k=0}^\infty k! \, (-t)^k 
\int _L ^\infty  { 16 \pi^2 \, d \tau_2 \over (4 \pi  \tau_2)^{k+4}} \int _0 ^1  dx \,
{  e^{4 \pi \tau_2sx(1-x)]}  \over  (m+s(1-x))^{k+1} (n+sx)^{k+1}}
\no \eea
Integrating over $\tau_2$ produces an incomplete $\Gamma$-function, of which we shall retain only those terms which are of order $L^0$ and $\ln (L)$, 
\bea
\int _L ^\infty  { 16 \pi^2 \, d \tau_2 \over (4 \pi  \tau_2)^{k+4} } \,  e^{4 \pi \tau_2sx(1-x)} 
= 4 \pi { (sx(1-x))^{k+3} \over (k+3)!} \Big (  ( \Psi (k+4) - \ln [ - 4 \pi Ls x(1-x) ] \Big )
\eea
Using this result for the contributions from the incomplete $\Gamma$-function, we express $\mA^{(2)}(L;s,t,u) $ in the convenient form of  (\ref{thm2a}), analogous to the expression given there for $\mA^{(1)}(L;s,t,u)$, and the function  
$\mB(s,t,u;\ep)$ is given by,
\bea
\mB(s,t,u;\ep) = \sum _{M,N=2}^\infty { s^{M+N} \over M! \, N!} 
\sum_{{m_r \not=0 \atop 1\leq r \leq  M}} \sum _{{n_s\not=0 \atop 1\leq s \leq N}} 
{ \delta (\sum_r m_r) \, \delta (\sum_s n_s) 
\over |m_1 \cdots m_N \, n_1 \cdots n_N | }  
\mD(s,t,u;m,n;\ep)
\eea
with $\mD$ given by,
\bea
\label{mD}
\mD(s,t,u,;m,n;\ep) & = & 
   \sum _{k=0}^\infty { k! \, s^{k} t^k \over \Gamma (k+4+\ep)} 
 \int _0 ^1  dx \, {  x^{k+3+\ep} (1-x)^{k+3+\ep}   \over  (m+sx)^{k+1} (n+sx)^{k+1}}
\\ && 
+ \sum _{k=0}^\infty { k! \, s^{k}  (-t)^k \over \Gamma (k+4+\ep)} 
\int _0 ^1  dx \, {  x^{k+3+\ep} (1-x)^{k+3+\ep}  \over  (m+s(1-x))^{k+1} (n+sx)^{k+1}}
\no \eea
The sums over the variables $m_r$ and $n_r$ are analogous to one another, and may be carried out in terms of a single family of functions defined by,
\bea
\mI_k (s,x) =  \sum _{N=2}^\infty {s^N \over N!} 
\sum_{{n_r \not=0 \atop 1\leq r \leq  N}}{  k! \, \delta (\sum_r n_r)  \over |n_1 \cdots n_N  | \,   (n+sx)^{k+1}}
\eea
In terms of the functions $\mI_k$, the first integral in (\ref{mD}) is over $\mI_k(s,x)^2$ while the second integral is over $(-)^k \mI_k(s,1-x) \mI_k(s,x)$. Symmetrizing in $x \to 1-x$, we obtain, 
\bea
\label{mBmI}
\mB(s,t,u;\ep)   =  
\sum_{k=0}^\infty  s^{k} t^k 
\int _0 ^1 dx \, { x^{k+3+\ep}(1-x)^{k+3+\ep} \over 2 \, \Gamma (k+4 +\ep) \, k! } 
 \Big ( \mI_k(s,x) + (-)^k \mI_k(s,1-x)  \Big )^2
\eea

\subsubsection{Calculation of the functions $\mI_k(s;x)$}

To carry out the summations over $n_1, \cdots, n_N$ for given $N$ we again introduce an angular integration over a variable $\alpha$ to enforce the vanishing sum of the $n_r$, and a further integration over a variable $2 \beta$ to exponentiate the denominator so that we obtain the following alternative integral representation for $\mI_k(s,x)$, 
\bea
\mI_k(s,x) =  \sum _{N=2}^\infty {s^N \over N!}  \int _0 ^1 d \alpha \int _0 ^\infty 2 d \beta \, (2 \beta )^k 
\sum_{{n_r \not=0 \atop 1\leq r \leq  N}}
{e^{2 \pi i \alpha  (n_1 + \cdots + n_N)} e^{-2\beta (n+xs)}  \over |n_1 \cdots n_N  | } 
\eea
The sums over $n_r$ have now been decoupled, and may be carried out using (\ref{logsum}). 
Carrying out the sum over $N$ as well gives, 
\bea
\mI_k(s,x)=   \int _0 ^1 d \alpha \int _0 ^\infty 2 d \beta (2 \beta )^k \,  e^{- 2 \beta sx}
\Big ( |1-e^{-\beta+ 2 \pi i \alpha}|^{-2s} -1 \Big ) 
\eea
Changing variables from $\alpha,\beta$ to $z= e^{-\beta  +2 \pi i \alpha}$ gives, 
\bea
\mI_k(s,x) =  \int _{|z|\leq 1}  {d ^2 z \over  \pi |z|^2}  (-\ln |z|^2 )^k   |z|^{ 2  sx}
\Big ( |1-z|^{-2s} -1 \Big ) 
\eea
Evaluating now the special combination that occurs in (\ref{mBmI}), we find, 
\bea
\mI_k(s,x) + (-)^k \mI_k(s,1-x) 
& = &
\int _{|z|\leq 1}  {d ^2 z \over  \pi |z|^2}  (-\ln |z|^2 )^k \,   |z|^{ 2  sx}
\Big ( |1-z|^{-2s} -1 \Big ) 
\no \\ &&
+ \int _{|z|\leq 1}  {d ^2 z \over  \pi |z|^2}  (\ln |z|^2 )^k \,  |z|^{ 2  s(1-x)}
\Big ( |1-z|^{-2s} -1 \Big ) 
\qquad
\eea
we observe that the integrands involving $ |1-z|^{-2s}$ are mapped into one another under $z \to 1/z$ so that this part of both integrals may be combined into a single integral over the entire complex plane, 
\bea
\mI_k(s,x) + (-)^k \mI_k(s,1-x) 
& = &
\int _\CC  {d ^2 z \over  \pi |z|^2}  (-\ln |z|^2 )^k \,   |z|^{ 2  sx}  |1-z|^{-2s}
 \\ &&
- \int _{|z|\leq 1}  {d ^2 z \over  \pi |z|^2}  (-\ln |z|^2 )^k \,  \left (  |z|^{ 2  sx} + (-)^k |z|^{2s(1-x)} \right )
\no
\eea
The sum of these integrals is proportional to a $k$-order derivative of the function $W$ of (\ref{VS}), 
\bea
\mI_k(s,x) + (-)^k \mI_k(s,1-x)  =  s^2 { \p^k \over \p \mu^k} W(s,\mu-sx) \Big |_{\mu=0}
\eea
from which we readily obtain the second line of (\ref{thm2b}), thereby completing the proof of Theorem \ref{theorem2}.

\section{Transcendentality Properties}
 \label{sec:5}
\setcounter{equation}{0}

In this subsection, we shall begin by adding together the results for  $\cA_L$ and $\cA_R$ to produce  the full genus-one four-graviton amplitude $\cA^{(1)}(s_{ij})$, up to order $\cO(s_{ij}^6)$.\footnote{This result includes corrections and extensions of the expressions given in \cite{Green:2008uj}.} We shall then use these explicit expressions to investigate transcendentality and spell out the assumptions required to have uniform transcendentality to order $\cO(s_{ij}^6)$ at genus-one, consistently with uniform transcendentality at tree-level. The result will be the {\sl transcendentality assignments} 1-4 described in the introduction.

\sm

Although we have determined the exact expression for $\cA_R$ to all orders in~$s_{ij}$, we  do not  have explicit formulas for $\cA_L$ beyond order $\cO(s_{ij}^6)$.   In the following  we will derive a number of partial and indirect results, on the basis of which  we will formulate a conjecture on the structure of $\cA_L$ to higher orders in $s_{ij}$. Lemma~\ref{lemma5} states that the term linear in $\tau_2$ in the Laurent polynomial of $\cB(s_{ij}|\tau)$ may be deduced from $\cA_R$ exactly  to all orders in $s_{ij}$, and is free of multiple zeta-values. Theorem~\ref{thm6} gives the contribution to $\cA_L$ from any two-loop modular graph function, a result obtained in \cite{DHoker:2019mib}. This lemma and theorem will be used to motivate a conjecture on the transcendentality properties of $\cA_L$ to all orders in $s_{ij}$.

\subsection{The full genus-one amplitude to order $\cO(s_{ij}^6)$}
\label{genone}

Assembling the contributions to $\cA_L$ in (\ref{ALtot}) and to $\cA_R$ in (\ref{ARtot}) gives $\cA^{(1)}(s_{ij})$ with the help of (\ref{A1ALAR}) for the total amplitude  to order $\cO(s_{ij}^6)$. We see explicitly that all $L$-dependence cancels as required for the consistency of the calculation. It is instructive to rearrange  the total amplitude as the sum 
of ``analytic" and ``non-analytic" pieces,  
\bea
\label{suman}
\cA^{(1)}(s_{ij}) = 2\pi \Big ( \cA_L(L;s_{ij}) + \cA_R(L;s_{ij}) \Big ) = 
\cA_{{\rm an}}(s_{ij}) +\cA_{{\rm non-an}}(s_{ij}) 
\eea
The reason for the quotation marks on  analytic and non-analytic is that the non-analytic piece actually contains also analytic contributions, so that the nomenclature is natural and suggestive but not entirely precise.
 The ``analytic'' piece $ \cA_{{\rm an}} (s_{ij})$   is given by, 
\bea
\label{Aan}
 \cA_{{\rm an}} (s_{ij}) = { 2 \pi^2 \over 3} \left ( 1 + {\zeta (3) \sigma _3 \over 3} + 
 {29 \zeta (5) \sigma _2 \sigma _3 \over 180} + { \sigma _3^2 \over 18} \, \zeta (3)^2 + \cO(\sigma_2^2\sigma_3)\right ) 
 \eea
 and the  ``non-analytic'' piece $ \cA_{{\rm non-an}} (s_{ij})$ has the form, 
 \bea
\label{Anonan}
 \cA_{{\rm non-an}} (s_{ij}) = {2\pi^2 \over 3}  \left ( \hat \cA_{{\rm sugra}} +\hat \cA_4 +\hat \cA_6  +  \cO(s_{ij}^7) \right )
 \eea
 The lowest order term $\hat \cA_{{\rm sugra}}$ is a regularized version of the ten-dimensional one-loop supergravity amplitude. 
  The contribution of order $\cO(s_{ij}^4)$ is given by, 
  \bea
\label{Ahatfour}
\hat \cA_4 & = & { 4 s^4 \over 15} \, \zeta (3)  \left ( - \ln (- 2 \pi s) + Z_4 +{63 \over 20} \right ) +
\hbox{ 2 cyclic perms of } s,t,u
\eea
while the contribution of order $\cO(s_{ij}^6)$ takes the form,
\bea
\hat \cA_6 & = & - { 84 s^6 + 2 s^4 \sigma _2 \over 420} \, \zeta (5) \,( \ln (-2\pi s)-Z_6) + \hbox{ 2 cyclic perms of } s,t,u
\no \\ &&
+ {\sigma _2^3 \over 18} \, \zeta (5) \, \left ( -Z_6-Z_5+Z_4+Z_3+{8339 \over 2450} \right )
\no \\ &&
+ { \sigma _3^2 \over 54} \, \zeta (5) \, \left ( Z_6 + Z_5 - Z_4 - Z_3 +{ 51011 \over 420} \right )
\label{Ahatsix}
\eea
The instruction to add permutations of $s,t,u$ applies only to the first line of (\ref{Ahatsix}).
The combination $Z_n$ was given in the introduction, and is repeated here for convenience,
\bea
Z_n = { \zeta '(n) \over \zeta(n)} - { \zeta '(n-1) \over \zeta(n-1)} - \gamma
\eea
where $\gamma$ is the Euler constant. The discontinuity of $ \cA_{{\rm non-an}} (s_{ij}) $, namely the  coefficients of the $\ln(-2\pi s)$ terms in (\ref{Ahatfour}) and  (\ref{Ahatsix}), reproduce those obtained in \cite{Green:2008uj}. They  have also recently been reproduced from ${\cal N}=4$ supersymmetric Yang--Mills theory by considering a flat-space limit of $AdS_5\times S^5$  \cite{Alday:2018pdi}.

\subsection{Transcendentality assignments up to order $\cO(s_{ij}^6)$}
\label{sec:52}

Recall that the tree-level amplitude exhibits a consistent grading by transcendental weight by assigning weight $n$ to the Riemann zeta-value $\zeta(n)$, weight $+1$ to $\pi$, weight $-1$ to the kinematic variables $s,t,u$, and thus weight $-k$ to the symmetric polynomial $\sigma _k =s^k+t^k+u^k$.  Consistency between tree-level and genus-one amplitudes resulting from unitarity requirea that we maintain these assignments at genus-one, as expressed in the first and second transcendentality assignments in the introduction.

\sm

Inspection of the ``analytic" part $\cA_{{\rm an}} (s_{ij})$ in  (\ref{Aan})  shows that all terms inside the parenthesis  have  zero  total transcendental weight.

\sm

Inspection of the ``non-analytic" part $\cA_{{\rm non-an}} (s_{ij})$ in  (\ref{Anonan}) shows that the supergravity contribution has weight zero, but $\hat \cA_4$ and $\hat \cA_6$, given in (\ref{Ahatfour}) and (\ref{Ahatsix}), involve $\ln (-2 \pi s)$ as well as the combinations $Z_n$.  The argument of $\ln(-2 \pi s)$ has weight zero, as is required for the argument of a transcendental function. Furthermore, the terms $\hat \cA_4$ and $\hat \cA_6$ do not depend separately on $\ln (-2 \pi s)$ and $Z_n$, but may be written as a sum of $\ln(-2 \pi s) - Z_n$ and differences $Z_m-Z_n$. Thus, a minimal requirement for $\hat \cA_4$ and $\hat \cA_6$ to have weight zero is given by assumption 3 of the introduction, that the combinations $\ln (-2 \pi s)- Z_n$ have weight 1 for all $n \geq 4$.
The validity of assigning weight one to differences $Z_m-Z_n$ for $m > n \geq 4$ may be argued using the following identity,
\bea
Z_m - Z_n = {d \over d \ep} \ln \left ( { \zeta (m+\ep) \zeta (n-1+\ep) \over \zeta (m-1+\ep) \zeta (n+\ep)} \right ) \bigg |_{\ep =0} 
\eea
The total weight of the argument of the logarithm vanishes for all values of $\ep$, and assigning weight 1 to the logarithm  for any value of $\ep$ then shows that $Z_m-Z_n$ has weight 1. With the help of these assignments all terms in  $\hat \cA_4$ and $\hat \cA_6$, with the exception of the rational additions, have weight zero, consistently with the transcendental weight of (\ref{Aan}). 

\sm

Finally, we address the weight assignments of the rational additive terms. Inspection of their origin in the calculation of $\cA_L$ in Appendix \ref{sec:A} reveals that they arise from the integration of  modular graph functions whose term in $\tau_2$ in the Laurent polynomial is non-vanishing. More precisely they arise in (\ref{Easq})  as differences of the $\Psi$-function at different integers. Inspection of the calculation of the non-analytic part $\cA_R$ in Theorem \ref{theorem1} and Appendix \ref{sec:B} reveals that the rational additions also all arise from differences between $\Psi$-functions at integers. Such differences equal differences of harmonic sums, 
\bea
\label{PsiH}
\Psi(m+1)-\Psi(n) = H_m - H_n 
\hskip 1in 
H_m = \sum _{k=1}^m { 1 \over k}
\eea
It has been argued in the literature on transcendentality in quantum field theory amplitudes (see for example \cite{Kotikov:2002ab,Beccaria:2009vt} for early papers), and in particular amplitudes of the $\cN=4$  supersymmetric  Yang-Mills theory in four dimensions, that one should assign transcendentality one to the harmonic sum functions $H_m$. With this final assumption, namely transcendentality assignment 4. of the introduction, the full genus-one amplitude $\cA^{(1)}(s_{ij})$ has uniform transcendentality up to order $\cO(s_{ij}^6)$.

\subsection{Further noteworthy features of the amplitude to order $\cO(s_{ij}^6)$}

The following additional features of the expressions (\ref{suman}) -- (\ref{Ahatsix}) for the genus-one amplitude to order $\cO(s_{ij}^6)$ are worth noting.
\begin{itemize}
\item
In (\ref{Aan}) we have factored out the volume of the fundamental domain $\cM$ in the Poincar\'e metric, and collected together all contributions to a given order $w$ in $s_{ij}$. The coefficient of a term of order $w$ is a polynomial in odd zeta-values with total weight~$w$. Assuming this is a general property accounts for the absence of  terms proportional to $\sigma_2$ and  $\sigma_2^2$ since there are no weight-two  or weight-four combinations of odd zeta-values.
\item
Similarly, the coefficient of a term in (\ref{Anonan}) of order $(s_{ij})^w$ or $(s_{ij})^w  \log (-2 \pi s)$  is proportional to an odd zeta-value of weight $w-1$.  This is consistent with the absence of a term proportional to $s^5 \log (-2\pi s)$, which would have a coefficient  that  would be a weight-four combination of zeta-values.
\item
There is a $\zeta(3)^2\,\sigma_3^2$ term in (\ref{Aan}) but the potential term proportional to $\zeta(3)^2\,\sigma_2^3$ is absent. There is no obvious reason for the vanishing of this coefficient.  However, it is notable that the series of terms that occurs in (\ref{Aan}) is the same series of terms  that arises in the low-energy expansion of the tree-level Virasoro amplitude (\ref{A0b}) multiplied by $\sigma_3$,  although the rational coefficients are different.  
\end{itemize}}

\subsection{Transcendentality of $\cA_R$ to all orders in $s_{ij}$}

The function $\cA_R (L;s_{ij})$ is obtained in (\ref{ARstar}) as a symmetrization in $s,t,u$ 
of the functions $\mA^{(1)} _*(L;s,t,u)$ and $\mA^{(2)} _*(L;s,t,u)$, which in turn are obtained in (\ref{thm2a}) and (\ref{thm2b}) from the functions $\mA(s,t,u;\ep)$ and $\mB(s,t,u;\ep)$ and their first derivative with respect to $\ep$, both evaluated at $\ep =0$. Inspection of these formulas and their expansion in powers of $s,t,u$ may then be used to prove the following theorem.

{\thm{
\label{lemma4}
{\sl Using only the transcendentality assignments required at tree-level, namely assumptions 1. and 2. of the introduction, the functions $  \mA(s,t,u;0)$ and $\mB(s,t,u;0)$ have weight  one and two, respectively. Using in addition assumption 4., namely that harmonic sum functions have weight one, the following functions have weight two and three, respectively,
\bea
&& {\p \over \p \ep}  \mA(s,t,u;\ep) \Big |_{\ep=0}  - \gamma \, \mA(s,t,u;0) 
\no \\
&&  {\p \over \p \ep} \mB(s,t,u;\ep) \Big |_{\ep=0}  - \gamma \, \mB(s,t,u;0) 
\eea
where $\gamma$ is Euler's constant. }}}

\sm

The lemma below will relate the term linear in $\tau_2$ of the particular combinations of modular graph functions that enter into the $\cA_L$ part of the genus-one amplitude   to the behavior obtained exactly for the $\cA_R$ part of the amplitude. Combining the Taylor expansion of $\cB(s_{ij}|\tau)$ in powers of $s_{ij}$ in (\ref{Bpq}) and the Laurent expansion of each coefficient in $\tau_2$ given in (\ref{bpq}), we define the term of order $\tau_2$ as follows,
\bea
\cB(s_{ij}|\tau) \Big |_{\tau_2}  = \mb^{(1)}(s_{ij}) \tau_2 
\hskip 1in  \mb^{(1)}(s_{ij}) = \sum _{p,q=0}^\infty \mb^{(1)} _{p,q} \, { \sigma _2 ^p \, \sigma _3^q \over p! \, q!}
\eea

{\lem
\label{lemma5}
\sl {The coefficient $\mb^{(1)}(s_{ij}) $ of  $\tau_2$ in the Laurent polynomial of $\cB(s_{ij}|\tau)$ is given by,
\bea
\mb^{(1)}(s_{ij})  = 8 \pi s^2  \Big ( \mA(s,t,u;0) + s \mB(s,t,u;0) \Big ) + \hbox{2 cyclic perms of } s,t,u
\eea
Explicit expressions  for $\mA$ and $\mB$ are given in (\ref{thm2b}). The coefficient $\mb^{(1)}(s_{ij})$  has vanishing transcendental weight, and is free of irreducible multiple zeta-values. }}

\sm

The proof of this proposition proceeds by inspection of (\ref{ARstar}), (\ref{VS}), (\ref{thm2a}), and (\ref{thm2b}). The coefficient $\mb^{(1)}(s_{ij})$  produces the entire logarithmic dependence of $\cA_L(L;s_{ij})$, and it is given by $\mb^{(1)}(s_{ij}) \ln (L)$. This contribution must be exactly the opposite of the $\ln(L)$ contribution in $\cA_R$, which is readily deduced from (\ref{ARstar}), (\ref{VS}), (\ref{thm2a}), and (\ref{thm2b}), and given by the expression in \ref{lemma5}. By  inspection of (\ref{VS}) we see that the weight of $W$ is 3 so that the weight of the coefficient of $\tau_2$ vanishes. Finally, the explicit expressions for $\mA(s,t,u;0)$ and $\mB(s,t,u;0)$ readily show that the coefficients in their Taylor expansion in $s,t,u$ are free of irreducible MZVs, thereby completing the proof of the proposition. 

\sm

Since the coefficients of the Laurent polynomial of a modular graph function of sufficiently high weight include irreducible multiple zeta-values it is likely, though so far unproven,  that the coefficient of its term linear in $\tau_2$ will also involve irreducible multiple zeta-values. Therefore, Lemma \ref{lemma5} provides a non-trivial constraint on the structure of the particular combinations of modular graph functions which enter into string amplitudes.

\subsection{Integrating general modular functions over $\cM_L$} 

Modular graph functions, defined through Kronecker-Eisenstein series, have polynomial growth at the cusp. Specifically,  the Laurent polynomial of an arbitrary modular graph function $\cC(\tau)$ of weight $w$ (such as $\cB_{(p,q)} (\tau)$ with $w=2p+3q$ in (\ref{bpq})), is given by, 
\bea
\cC(\tau) = \sum _{k=1-w}^w \mc_k \tau_2^k + \cO(e^{-2 \pi \tau_2}) 
\eea
where $\mc_k$ has weight $w$.  The integral of $\cC(\tau)$ over $\cM_L$ has polynomial growth in $L$ which may be extracted to leave a convergent integral over the fundamental domain $\cM$. Following~\cite{DHoker:2019mib}, we associate to $\mC(\tau)$ a modular function $\hat \cC(\tau)$ of the same weight $w$, defined by, 
\bea
\hat \cC(\tau) = \cC(\tau)  - { 45 \, \mc_1 \over \pi \zeta(3)} \left ( C_{2,1,1} -{ 2 \over 3} E_4  \right ) 
 - \sum _{k=2}^w { \mc_k \, \pi^k \over 2 \zeta (2k)} E_k(\tau) 
 \eea
By construction $\hat \cC(\tau)$ tends to a constant  at the cusp, and is therefore integrable on $\cM$ with the Poincar\'e volume element.  

{\lem
\label{lemma6}
{\sl The contributions of order $L^0$ and $\ln L$ to the  integral over $\cM_L$ of $\cC(\tau)$ are obtained as follows,
\bea
\int _{\cM_L} { d^2 \tau \over \tau_2^2} \, \cC(\tau) = \int _\cM { d^2 \tau \over \tau_2^2} \, \hat \cC(\tau)  
+  \mc_1 \Big (  \ln (2L) + \gamma + Z_4 \Big )
\eea
where contributions whose dependence on $L$ is exponential or power-behaved with non-vanishing exponents have been omitted. }}

The  integral of $\hat \cC(\tau)$ over $\cM$ is finite, and may be evaluated using the Poincar\'e series for $\hat \cC(\tau)$ and the standard unfolding trick. For two-loop modular graph functions, this procedure was carried out completely  \cite{DHoker:2019mib}, and the results will be summarized in the next subsection. For the general case, we shall resort to a conjecture in subsection \ref{sec:57}.

\subsection{Weight of integrals of two-loop modular graph functions}

It was shown in \cite{DHoker:2019txf} that an arbitrary two-loop modular graph function of weight $w$ may be decomposed into a linear combination with integer coefficients of  a subset of all two-loop modular graph functions of weight $w$ of the form, 
\bea
\cC_{u,v;w} (\tau) 
=\sum _{p_r \in \Lambda '} { \tau_2 ^w \, \delta (p_1+p_2+p_3) \over \pi^w \, p_1^u \, \bar p_2 ^v \, p_3^{w-u} \, \bar p_3 ^{w-v}}
\eea
where $\Lambda ' = \Lambda \setminus \{ 0 \}$ and $\Lambda = \tau \ZZ + \ZZ$. The coefficient $\mc_1$ of $ \tau_2$  in the Laurent polynomial of $\cC_{u,v;w}(\tau)$ for $w$ even is given by,
\bea
\mc_1 = 8 (-)^{{w \over 2}} { \zeta (w) \zeta (w-1) \over (2 \pi)^{w-1}}  \binom{u+v-2}{u-1}  \binom{w-2}{w-u-v} 
\eea
Its weight is $w$. The corresponding term vanishes for all two-loop modular graph functions whose weight $w$ is odd.
We shall denote by $\cS_{u,v;w} $ the integral of $\cC_{u,v;w}$ over $\cM_L$,
\bea
\cS_{u,v;w}= \int _{\cM_L} { d^2 \tau \over \tau_2^2} \, \cC_{u,v;w}(\tau)
\eea

{\thm
\label{thm6}
{\sl Up to contributions which behave as a power of $L$ with non-zero positive or negative exponent, the value of $\cS_{u,v;w}$  is given respectively for even and odd weight $w$ by,
\begin{align}
\label{thm6f}
\hbox{even } & &
\cS_{u,v;w} & = 
\mc_1 \Big ( \ln (2L) + \gamma + Z_w \Big )
+    { \zeta(w)\zeta(w-1)  \over (2 \pi)^{w-1}} \binom{u+v-2}{u-1}  \cG_{u,v;w}  
\no \\
\hbox{odd } &  &
\cS_{u,v;w} & =  4 \pi  {\zeta (w)  \zeta (w-1) \over (2 \pi)^{w-1}}
 (-)^{{w+1 \over 2}} \binom{w-2}{w-u-v} \binom{u+v-2}{u-1} 
\end{align}
where $\cG_{u,v;w} $ is a linear combination, with integer coefficients, of harmonic sums of weight one. Since the weight of $\mc_1$ is $w$ the weight of $\cS_{u,v;w}$ is $w+1$.}}

\sm

The proof of formulas (\ref{thm6f}) was given in \cite{DHoker:2019mib}. Using our transcendental weight assignment, the harmonic sum functions $\cG_{u,v;w}$ have weight one; the term proportional to $\ln(2L)+ \gamma+Z_w$ will combine with the logarithmic contributions from $\cA_R$ to produce a term proportional to $\ln (-2 \pi s) - Z_w$ which has weight one by our assumption 3. Since $\mc_1$ has weight $w$ all contributions to the full genus-one amplitude 
arising from the restricted class of two-loop modular graph functions will have total weight $w+1$.

\subsection{Conjecture on transcendentality  to all orders in $s_{ij}$}
\label{sec:57}

While the integral representation for the integrand $\cB(s_{ij}|\tau)$ of the genus-one amplitude is explicitly known, 
we do not have an algorithm to extract the terms of order $L^0$ in  its integral over $\cM_L$ in any explicit manner, beyond the case of two-loop modular graph functions. Therefore we have no general  formulas for $\cA_L$ beyond $\cO(s_{ij}^6)$ on which we can test the properties of transcendental weight assignments explicitly.  However, experience with the integration of modular graph functions of weight up to six and two-loop modular graph functions of general weight suggest the following  conjecture. 

 {\conj
\label{intweight}
{\sl The genus-one amplitude is given by the sum in (\ref{suman}) of an ``analytic" piece $\cA_{{\rm an}}(s_{ij})$ of the form,
\bea
\label{con-an}
\cA_{{\rm an}} (s_{ij})  =  {2  \pi ^2 \over 3} \sum_{p,q=0}^\infty \mC_{(p,q)} { \sigma _2^p \, \sigma _3^q \over p! \, q!}
\eea
where $\mC_{(p,q)}$ is a sum of multiple zeta-values of weight $w=2p+3q$, 
and a ``non-analytic" piece $\cA_{{\rm non-an}}(s_{ij})$  of the form, 
\bea
\label{con-non}
\cA_{{\rm non-an}} (s_{ij}) & = & 2 \pi  \bigg ( 8 \pi s^2 \Big ( \mA(s,t,u;0) + s \mB(s,t,u;0 \Big ) 
\Big ( \ln (-2 \pi s) - Z_4 \Big ) 
\no \\ && \hskip 0.2in  + \mM (s,t,u) + \mN(s,t,u) \bigg )   + \hbox{2 cyclic perms of } s,t,u
\eea   
The terms $\mM$ and $\mN$ are characterized by the following properties. The function $\mM(s,t,u)$ is a linear combination, with rational coefficients, of terms each of which is a product of total weight zero of a multiple zeta-value, kinematic variables $\sigma _2, \sigma _3$, and a finite harmonic sum of weight one. The function $\mN(s,t,u)$ is a linear combination, with rational coefficients, of terms each of which is a product of total weight zero of a multiple zeta-value times the first derivative of the logarithm of a ratio of multiple zeta-values whose weight vanishes.}}

\sm

The motivations for the various parts of the conjecture are as follows. 
\begin{itemize}
\item 
The structure of the ``analytic" piece $\cA_{{\rm an}}(s_{ij})$ is motivated on the one hand by the results obtained up till order $\cO(s_{ij}^6)$ in (\ref{Aan}), and on the other hand by the fact that the terms of order $\tau_2^0$ in $\cB(s_{ij}|\tau)$ are all linear combinations of products of multiple zeta-values times powers of $s_{ij}$ of total weight zero. Therefore, we know that such terms must arise at every order of the Taylor expansion in $s_{ij}$.
\item 
Part of the structure of the  ``non-analytic" piece, namely the first line of (\ref{con-non}), has actually been proven since it is dictated by unitarity and has been computed exactly in section \ref{sec:4}. Thus, the heart of the conjecture for the ``non-analytic" part is  the structure of the functions $\mM$ and $\mN$. 
\item 
The  function $\mN$ emerges from the integration of the Poincar\'e seed of the modular graph functions over the semi-infinite strip $-\thalf \leq \tau_1 \leq \thalf$ and $\tau_2 >0$. Proceeding by successive partial fraction decomposition of the seed function, and summing over the $n$-variables in each loop momentum $p = m \tau +n$, one will end up with poles of first order and poles of higher orders, multiplied by exponentials. It is the integrations of the simple pole terms that produce single logarithms of $m$ and $n$ under the summation over all the lattice momenta $m,n$, and thus give rise to single derivatives of multiple zeta-values. The fact that the terms linear in $\tau_2 $ have been eliminated in $\hat \cC$ guarantees that the derivatives may be organized as a single derivative of the logarithm of a ratio of multiple zeta-values whose total weight vanishes. These general remarks are borne out in the special cases of low weight $w \leq 6$ and two-loop modular graph functions, in which cases we had theorems.
\item
 The function $\mM$ also emerges from the integration of the linear term in $\tau_2$, and specifically from the analytic continuation and asymptotics of the incomplete $\Gamma$-function at small argument.  The conjectured form is confirmed by the cases of low weight $w \leq 6$ and two-loop modular graph functions, but a general derivation of the result is so far out of reach. As discussed in the introduction, we stress that the finite harmonic sum functions always arise as coefficients of  infinite series, to which a definite weight may be assigned.   This is manifestly the case  for the contributions from $\cA_R$ since  they are given by infinite power series' in the kinematic variables $s,t,u$. However, the $\cA_L$ contributions have so far  only been evaluated up to order $\cO(s_{ij}^6)$, and we do not have knowledge of the complete Taylor series to all orders in $s,t,u$. For this reason,  although the structure of the contributions from $\cA_L$ at low orders is suggestive, the general structure  remains  conjectural.
\end{itemize}

\newpage

\section{Discussion and outlook}
 \label{sec:6}
\setcounter{equation}{0}

The results of  this  paper have pointed to some interesting systematics of the coefficients in the low-energy expansion of genus-one four-graviton scattering amplitudes in  Type II superstring  theory.  An important part of the determination of these coefficients involves disentangling the non-analytic threshold dependence from the analytic part.  This involved combining the integral over the cut-off fundamental domain, $\cM_L$, with the integral over its complement, $\cM_R$, which is a small neighborhood of the cusp.

\begin{itemize}

\item
We have established the precise expressions for the coefficients of  terms of order $s^4$ in the low-energy expansion (correcting numerical errors in \cite{Green:1999pv, Green:2008uj}) as well as the coefficients of order $s^5$ and $s^6$.  These results, which are summarized in (\ref{Aan}) and (\ref{Anonan}),  possess interesting transcendentality properties.      
A  striking feature of the explicit coefficients in the genus-one low-energy expansion is the vanishing of the coefficient of $ \zeta(3)^2\,\sigma_2^3 $.   Only the term proportional to $\zeta(3)^2\,\sigma_3^2 $ contributes to the term of order $s^6$.
As a consequence, to this low order, the pattern of terms contributing to the analytic part of the genus-one low-energy expansion in (\ref{Aan}) is the same as the pattern in the tree-level  expansion (\ref{A0b}) if the latter is multiplied by $\sigma_3$.  

\item
It is a simple consequence of unitarity that the coefficients of the logarithmic terms in the low-energy expansion of the four-graviton  amplitude do  not contain irreducible MZVs.  This follows simply from the fact that the discontinuity of the genus-one amplitude across the massless two-particle threshold is proportional to the square of the tree-level amplitude, but the coefficients of the tree-level amplitude low-energy expansion only involve polynomials in ordinary odd Riemann zeta-values.  However, we also showed the stronger result  that irreducible MZVs are  absent from the analytic term that arises from the integral over $\cM_R$, which is by no means obvious.

\item 
 Irreducible MZVs might  be expected to arise in the  analytic terms in the low energy expansion of the  ten-dimensional genus-one four-graviton amplitude that arise from the integral over ${\cal M}_L$. We know from \cite{Zerbini}  that the coefficients of terms in the  Laurent expansion around $\tau_2\to \infty$ of certain modular graph functions have coefficients  that are irreducible MZVs, which suggests that the integral  over ${\cal M}_L$ will  give rise to irreducible MZV's  in the coefficients at high enough order in the  low-energy expansion, as well as in the compactified theory.  However, if the correspondence between the structure of the tree-level expansion and the analytic part of the genus-one expansion noted above were to persist to all orders it would imply that the irreducible MZVs cancel in the sum of all contributions at a given order in the low-energy expansion.

\item
 Although this paper  has not been concerned with the genus-two contribution to the amplitude  some consequences of two-loop unitarity follow in a straightforward manner.  In particular,  figure~\ref{fig:threedisc} illustrates a discontinuity in the channel with three massless particles, which is proportional to the product of two on-shell five-particle tree amplitudes integrated over phase space of the  intermediate three-particle state.  It is known that the coefficients in the low-energy expansion of the  five-particle closed-string amplitude contain irreducible single-valued  MZVs, starting at weight-11  \cite{SCHLotterer:2012ny} and these therefore enter into the expression for the discontinuity across the  logarithmic branch cut.  In other words, whereas irreducible MZV's do not arise in the coefficients of the logarithmic terms in the low energy expansion of the genus-one four-graviton amplitude, they do arise in the expansion of the genus-two four-graviton amplitude.  
 
\begin{figure}[h]
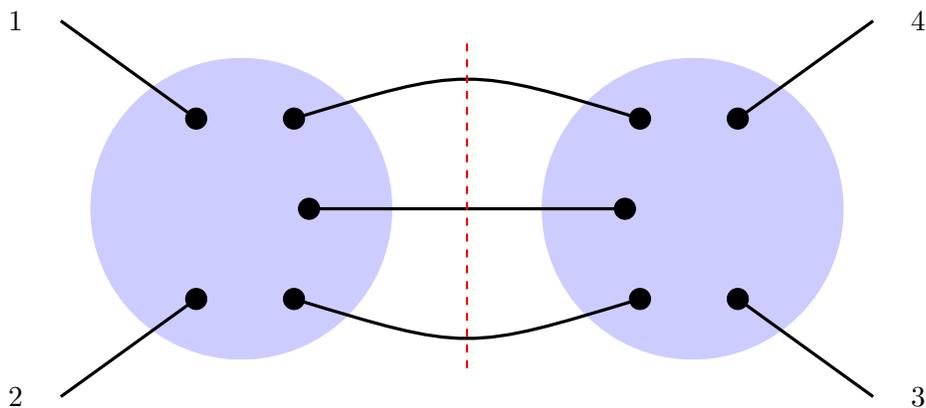

\begin{center}
\tikzpicture[scale=2]
\scope[xshift=-0cm,yshift=-0.cm]
%

\filldraw [blue!20]  (-1.5,0) ellipse (1.0 and 1.0);
\filldraw [blue!20]  (1.5,0) ellipse (1.0 and 1.0);

\draw [very thick] (-1.15,0.6)  .. controls (0, 0.95) ..  (1.15,0.6) ;
\draw [very thick] (-1.05,-0.0) -- (1.05, -0.0) ;
\draw[very thick]  (-1.15,-0.6)  .. controls (0, -0.95) ..  (1.15, -0.6) ;
\draw[very thick]  (2.7,1.25)-- (1.8,0.60) ;
\draw[very thick]  (-2.7,-1.25) -- (-1.8,-0.60) ;
\draw [very thick] (2.7,-1.25)-- (1.8,-0.60) ;
\draw [very thick] (-2.7,1.25)-- (-1.8,0.60) ;
\draw [thick,dashed,red]   (0.0,1.1) -- (0.0,-1.1);
\draw (-3.0, 1.25) node{\small 1};
\draw (-3.0, -1.25) node{\small  2};
\draw (3.0, -1.25) node{\small  3};
\draw (3.0, 1.25) node{\small 4};
\draw [fill=black]  (1.8,0.60) circle [radius=.07];
\draw [fill=black](1.8, -0.6)  circle [radius=.07];
\draw [fill=black](-1.8, 0.6)  circle [radius=.07];
\draw [fill=black](-1.8, -0.6)  circle [radius=.07];
\draw [fill=black]  (-1.15,0.6)   circle [radius=.07];
\draw [fill=black] (-1.05,0.0)  circle [radius=.07];
\draw [fill=black](-1.15,-0.6)   circle [radius=.07];
\draw [fill=black] (1.15,0.6) circle [radius=.07];
\draw [fill=black] (1.05,-0.0) circle [radius=.07];
\draw [fill=black] (1.15,-0.6)  circle [radius=.07];

\endscope
\endtikzpicture
\end{center} 
\caption{The three-particle discontinuity of a genus-two amplitude is proportional to the square of a tree-level five-particle amplitude integrated over phase space for the intermediate three particle states. }
\label{fig:threedisc}
\end{figure}

\item
The analysis in this paper was for Type II superstring amplitudes, but it would be interesting to find out whether and how transcendentality extends to the case of Heterotic string amplitudes, which have only half the space-time supersymmetry of the Type II strings. Modular  graph forms, introduced in \cite{DHoker:2016mwo,DHoker:2016quv}, are required from the outset in the Heterotic string, and a number of results have already been obtained for their role in its genus-one amplitudes \cite{Schlotterer:2016cxa,Basu:2017nhs,Basu:2017zvt, Gerken:2018jrq}.

\item 
We conclude with some comments on transcendentality. In quantum field theory, the concept of transcendentality is intimately linked with the interpretation of Feynman integrals as iterated integrals in the sense of Chen \cite{Chen} (see also \cite{Goncharov:1998kja}) and  ``periods" in the sense of Kontsevich and Zagier \cite{KonZag}. The period integrals that show up in quantum field theory include zeta-values, multiple zeta-values, polylogarithms and multiple polylogarithms, as well as periods on various algebraic varieties  (for a useful set of lecture notes, see  \cite{Duhr:2014woa}). The associated motivic iterated integrals and multiple zeta-values may be endowed with a powerful Hopf algebra structure on which the weight can de defined unambiguously \cite{Stieberger:2013wea,Broedel:2015hia}. Transcendentality restricts the structure of the amplitudes in terms of periods. In superstring theory, iterated integral representations in terms of elliptic functions have been developed for open superstring amplitudes \cite{Broedel:2013tta,Broedel:2014vla}, and the role of the single-valued projection to closed superstrings has been studied recently further in  \cite{Schlotterer:2018zce,brown7,Vanhove:2018elu}. The results obtained in this paper on transcendentality suggest that a powerful structure may underlie loop amplitudes for closed superstrings.

\end{itemize}

\newpage

\appendix

\section{Calculation of $\cA_{(p,q)}(L)$ up to weight 6}
 \label{sec:A}
\setcounter{equation}{0}

In this appendix we shall evaluate the coefficients $\cA_{(p,q)}(L)$ of the low-energy expansion of the analytic part $\cA_L(s_{ij};L)$ os the genus-one four-graviton amplitude. The coefficients $\cA_{(p,q)}(L)$ are obtained as integrals over $\cM_L$ of the modular graph functions $\cB_{(p,q)}(\tau)$, whose derivation up to weight 6 we first review from \cite{Green:2008uj}.

\subsection{Calculation of the modular graph functions $\cB_{(p,q)}(\tau)$}

To obtain the modular graph functions $\cB_{(p,q)}(\tau)$, we expand the exponential integrand which defines $\cB(s_{ij}|\tau)$ in formula (\ref{cB})  in powers of $s_{ij}$ to order $w=2p+3q$, and we have,\footnote{The variables used in \cite{Green:2008uj} are related to the ones used here by $\cB_{(p,q)} = p! \, q! \, j^{(p,q)}$.  }
\bea
\label{Bpqw}
\sum _{{p, q \geq 0 \atop
2p+3q=w}}  \cB _{(p,q)} (\tau) \,  \frac{\sigma _2^p \sigma _3^q}{p!\, q!}
=
{ 1 \over w!} \prod _{\kappa=1}^4 \int _{\Sigma } { d^2 z_\kappa\over \tau_2} \,
\Big (  s g_{12} + s g_{34} + t g_{14} + t g_{23} + u g_{13} + u g_{24} \Big )^w
\eea
Since both sides of (\ref{Bpqw})  are homogeneous of degree $w$ in $s,t,u$, we may parametrize $s,t,u$ by a single variable $x$, setting for example $s=1+x$, $t=-1+x$ and $u=-2x$, and expressing the invariants $\sigma _2= 2+6x^2$ and $\sigma _3 = 6x(1-x^2)$ in terms of $x$. All contributions may then be decomposed in terms of the following family of integrals, 
\bea
\cK_{m,n} = \prod _{\kappa=1}^4 \int _{\Sigma } { d^2 z_\kappa \over \tau_2}
\Big (  g_{12} + g_{34} - g_{14} - g_{23}  \Big )^m
\Big ( g_{12}+g_{34} + g_{14}+g_{23} - 2 g_{13} - 2 g_{24} \Big )^n
\eea 
One trivially has $\cB_{(0,0)}=1$, while for low values of $q$ one has the following expressions,
\bea
\cB_{(p,0)} & = & 
{p! \, \cK_{2p,0} \over 2^p \cdot (2p)!} 
\no \\
\cB_{(p,1)}  & = & { p! \,  \cK_{2p+2,1} \over 6 \cdot 2^p \cdot (2p+2)!} 
\no \\
\cB_{(p,2)}  & = & { p!  \, \cK_{2p+4,2}  \over 36 \cdot 2^p \cdot (2p+4)!} 
- {p! \, \cK_{2p+6,0} \over 12 \cdot 2^p \cdot (2p+5)!} 
\eea
Expanding the integrands, one may identify each term in the expansion with a modular graph function given by a single graph. Recall that any integral  in which a single Green function ends on a vertex must vanish since the integral of the Green function on the torus vanishes. 
Expressing these integrals in terms of modular graph functions we obtain, 
\bea
\label{cB1}
\cB_{(1,0)} & = & E_2
\no \\
3\, \cB_{(0,1)} & = &  D_3 + 4 E_3 
\no \\
12 \, \cB_{(2,0)}  & = &   D_4 + 9 E_2^2 + 6 E_4   
\no \\
72 \, \cB_{(1,1)} & = &  D_5 - 12 D_{2,2,1}+ 16 D_{3,1,1}+ 12 C_{3,1,1} -24 C_{2,2,1}  
 + 14 E_2 D_3  + 48 E_2 E_3 
\no \\
240 \, \cB_{(3,0)} & = &
D_6 + 45 E_2 D_4 - 10 D_3^2 + 120 C_{3,1,1,1} + 90 E_2^3 - 90 D_{2,2,1,1}
\no \\
540 \, \cB_{(0,2)} & = & D_6 + 60 C_{2,1,1,1,1} + 120 D_{1,1,1,1;1,1} -720 D_{2,1,1,1;1}
+ 60 D_{2,2,2} 
\no \\ &&
+ 270 D_{2,2,1,1} - 120 C_{3,1,1,1} +360 C_{2,2,1,1} - 120 D_{3,2,1}
\no \\ &&
 -15 E_2 D_4 +50 D_3^2 + 240 E_3 D_3 - 30 E_2^3
\eea
We have used the following correspondences between the $C$ and $D$ functions and identities,
\bea
C_{2,1,1,1,1} & = & D_{4,1,1}
\no \\
C_{2,2,1,1} & = & D_{1,1,1,1;2}
\no \\
D_{2,1,2,1} & = & D_{2,2,1,1,}
\no \\
C_{3,1,1,1} & = & D_{3,1,1,1}
\eea
The graphical representations of $D_{2,2,2}$ and $D_{2,1,1,1;1}$ are given by, 
\bea
\label{fig11}
\tikzpicture[scale=1.1]
\scope[xshift=-5cm,yshift=-0.4cm]
\draw (-1.3,0) node{$D_{2,2,2} \, =$};
\draw[thick]   (0,0.5) node{$\bullet$} ..controls (1,0.75) .. (2,0.5) node{$\bullet$} ;
\draw[thick]   (0,0.5) node{$\bullet$} ..controls (1,0.25) .. (2,0.5) node{$\bullet$} ;
\draw[thick]   (0,0.5) node{$\bullet$} ..controls (0.8,0) .. (1,-0.8) node{$\bullet$} ;
\draw[thick]   (0,0.5) node{$\bullet$} ..controls (0.5,-0.5) .. (1,-0.8) node{$\bullet$} ;
\draw[thick]   (2,0.5) node{$\bullet$} ..controls (1.2,0) .. (1,-0.8) node{$\bullet$} ;
\draw[thick]   (2,0.5) node{$\bullet$} ..controls (1.5,-0.5) .. (1,-0.8) node{$\bullet$} ;
\draw (5.6,0) node{$D_{2,1,1,1;1} \, =$};
\draw[thick]   (7,-0.7) node{$\bullet$} -- (8.5,-0.7) node{$\bullet$} ;
\draw[thick]   (7,-0.7) node{$\bullet$} -- (7,0.7) node{$\bullet$} ;
\draw[thick]   (7,0.7) node{$\bullet$} -- (8.5,0.7) node{$\bullet$} ;
\draw[thick]   (8.5,0.7) node{$\bullet$} ..controls (8.2,0) .. (8.5,-0.7) node{$\bullet$} ;
\draw[thick]   (8.5,0.7) node{$\bullet$} ..controls (8.8,0) .. (8.5,-0.7) node{$\bullet$} ;
\draw[thick]   (7,-0.7) node{$\bullet$} -- (8.5,0.7) node{$\bullet$} ;
\endscope
\endtikzpicture
\eea
The graph $D_{2,2,1,1}$ may be represented in two different graphical ways,
\bea
\label{fig12}
\tikzpicture[scale=1.1]
\scope[xshift=-5cm,yshift=-0.4cm]
\draw (5.6,0) node{$D_{2,2,1,1} \, =$};
\draw[thick]   (7,-0.7) node{$\bullet$} -- (8.5,-0.7) node{$\bullet$} ;
\draw[thick]   (7,-0.7) node{$\bullet$} ..controls (6.7,0) ..  (7,0.7) node{$\bullet$} ;
\draw[thick]   (7,-0.7) node{$\bullet$} ..controls (7.3,0) ..  (7,0.7) node{$\bullet$} ;
\draw[thick]   (7,0.7) node{$\bullet$} -- (8.5,0.7) node{$\bullet$} ;
\draw[thick]   (8.5,0.7) node{$\bullet$} ..controls (8.2,0) .. (8.5,-0.7) node{$\bullet$} ;
\draw[thick]   (8.5,0.7) node{$\bullet$} ..controls (8.8,0) .. (8.5,-0.7) node{$\bullet$} ;
\draw  (9.4,0) node{$=$} ;
\draw[thick]   (10,0.7) node{$\bullet$} -- (12,0.7) node{$\bullet$} ;
\draw[thick]   (11,-0.7) node{$\bullet$} ..controls (10.1,0) ..  (10,0.7) node{$\bullet$} ;
\draw[thick]   (11,-0.7) node{$\bullet$} ..controls (10.9,0) ..  (10,0.7) node{$\bullet$} ;
\draw[thick]   (11,-0.7) node{$\bullet$} ..controls (11.1,0) ..  (12,0.7) node{$\bullet$} ;
\draw[thick]   (11,-0.7) node{$\bullet$} ..controls (11.9,0) ..  (12,0.7) node{$\bullet$} ;
\draw  (11,0.7) node{$\bullet$} ;
\endscope
\endtikzpicture
\eea
and the graph $D_{1,1,1,1;1,1}$ is a tetrahedron that can be represented by 
\bea
\label{fig11}
\tikzpicture[scale=1.1]
\scope[xshift=-5cm,yshift=-0.4cm]
\draw (-1.3,0) node{$D_{1,1,1,1;1,1}\, =$};
\draw[thick]   (0,-0.7) node{$\bullet$} -- (1.5,-0.7) node{$\bullet$} ;
\draw[thick]   (0,-0.7) node{$\bullet$} -- (0,0.7) node{$\bullet$} ;
\draw[thick]   (0,0.7) node{$\bullet$} -- (1.5,0.7) node{$\bullet$} ;
\draw[thick]   (1.5,-0.7) node{$\bullet$} -- (1.5,0.7) node{$\bullet$} ;
\draw[thick]   (1.5,-0.7) node{$\bullet$} -- (0,0.7) node{$\bullet$} ;
\draw[thick]   (1.5,-0.7) node{$\bullet$} -- (0,0.7) node{$\bullet$} ;
\filldraw [white!20]  (0.75,0) ellipse (.1 and .1);
\draw[thick]   (0,-0.7) node{$\bullet$} -- (1.5,0.7) node{$\bullet$} ;
\endscope
\endtikzpicture
\eea

\subsection{Identities between modular graph functions}

To integrate these modular graph functions over $\cM_L$ we simplify the expressions for the coefficients $\cB_{(p,q)}$ obtained in (\ref{cB1}) using the identities between modular graph functions  derived systematically  in \cite{DHoker:2016quv} up to weight 6 included. Earlier derivations of some of these identities include \cite{DHoker:2015gmr} for  two-loop modular graph functions, \cite{DHoker:2015sve} for $D_4$, \cite{DHoker:2016mwo} for all modular graph functions of weight four and five, \cite{Basu:2016kli} for the use of slightly different methods, \cite{Basu:2015ayg,DHoker:2016quv, Kleinschmidt:2017ege} for tetrahedral graphs, and \cite{Broedel:2018izr} for the differential identity for $C_{2,2,1,1}$. 
A more formal context for the identities between modular graph functions has been developed in \cite{Brown:2017qwo,Brown2}.

\sm

As much as possible, we shall express modular graph functions as Laplace-Beltrami operators acting on  modular functions, as such integrals may be carried out using (\ref{intLap}). 

\sm

For weights $w \leq 5$, we have the following algebraic identities, 
\bea
D_3 & = & E_3 + \zeta (3)
\no \\
D_4  & = &  24 C_{2,1,1}+ 3E_2^2 - 18 E_4
\no \\
D_5 & = & 60 C_{3,1,1} + 10 E_2 C_{1,1,1} -48 E_5 + 16 \zeta (5)
\no \\
40 D_{3,1,1} & = & 300 C_{3,1,1} + 120 E_2 E_3 - 276 E_5 + 7 \zeta (5)
\no \\
10 D_{2,2,1} & = & 20 C_{3,1,1} - 4 E_5 + 3 \zeta (5)
\no \\
30 C_{2,2,1} & = & 12 E_5 + \zeta (5)
\eea
and  differential identities, expressed as follows for later convenience,
\bea
\label{diff211}
2 C_{2,1,1} & = &  \Delta C_{2,1,1}   + E_2^2 - 9 E_4
\no \\
60 C_{3,1,1} & = & 10 \Delta C_{3,1,1}   + 40 E_2 E_3  - 172 E_5 - \zeta (5)
\eea
For weight $w=6$ the algebraic identities  involving only dihedral graphs are, 
{\small \begin{align} 
\label{6di}
D_6 & =  15 E_2 D_4 - 30 E_2^3 + 10 C_{1,1,1}^2 + 60 D_{4,1,1} - 720 C_{2,2,1,1} - 240 E_3 C_{1,1,1} 
\no \\ & \quad
+ 720 E_2 E_4 + 1440 E_3^2  + 5280 C_{3,2,1} 
- 360 E_2 C_{2,1,1} + 1280 C_{2,2,2} - 3380 E_6 
\no \\ 
2 C_{3,1,1,1} & =  - 3 C_{2,2,1,1} + 9 E_{2} E_{4} + 6 E_{3}^2 + 18 C_{4,1,1}
+ 24 C_{3,2,1}+ 2 C_{2,2,2} -32 E_{6}  
\no\\
3 D_{4,1,1}  & =  
 109 C_{2,2,2} + 408 C_{3,2,1} + 36 C_{4,1,1} +18 E_2 C_{2,1,1} 
 + 12 E_3 C_{1,1,1} - 211 E_6 
\end{align}}
The algebraic  identities involving trihedral and tetrahedral graphs are given by, 
{\small \begin{align}
\label{6tri}
3 D_{2,2,2}
& =    18 C_{2,2,1,1}  + 58 C_{2,2,2} + 192 C_{3,2,1}   
+ 3 E_2^3 - 24 E_3^2    - 18 E_2 E_4  - 46 E_6
\no\\
2 D_{3,2,1} & = 
 - 18 C_{2,2,1,1}  + 36 C_{4,1,1} + 69 C_{2,2,2} + 288 C_{3,2,1}
 + 6 E_2 C_{2,1,1} + 18 E_2E_4 + 36 E_3^2 - 183 E_6
\no\\
3 D_{2,2,1,1} & =  
 - 6 C_{2,2,1,1} + 10 C_{2,2,2} + 48 C_{3,2,1} + 12 C_{4,1,1} + 6 E_2 E_4 + 12 E_3^2   - 40 E_6 
\no \\
18 D_{2,1,1,1;1} & =  
  9 C_{2,2,1,1}  + 20 C_{2,2,2}  + 60 C_{3,2,1}  - 9 E_2 E_4 - 18 E_3^2  + 10 E_6
  \no \\
  3 \cD_{1,1,1;1,1,1} & = C_{2,2,2} + 12 C_{3,2,1} - 4 E_6
\end{align}}
Finally, there are a number of differential identities, 
\bea
24 \, C_{3,2,1} & = & \Delta (6 C_{3,2,1} + C_{2,2,2} ) - 84 E_6 + 12 E_3^2
\no \\
3 \, C_{2,2,2} & = & \Delta (3 C_{3,2,1} + C_{2,2,2} ) - 36 E_6
\no \\
24 C_{4,1,1} & = & \Delta (2 C_{4,1,1}  -4 C_{3,2,1} - C_{2,2,2} ) + 8 E_6 - 4 E_3^2 + 12 E_2 E_4
\no \\
18 C_{2,2,1,1} & = & 9 \Delta \tilde  C_{2,2,1,1} 
 -14  C_{2,2,2} +48 C_{3,2,1} - 36 C_{4,1,1} 
\no \\ &&
- 36 C_{2,1,1} E_2 +6 E_2^3 + 72 E_2 E_4 + 180 E_3^2 +284 E_6
\eea
where we have defined,
\bea
\tilde C_{2,2,1,1}= C_{2,2,1,1} -2E_3^2-E_2E_4
\eea 
Using the above identities, we eliminate the non-Eisenstein functions (except for $C_{2,1,1}$ in its product with $E_2$)   in terms of the Laplacian-Beltrami operator acting on certain modular graph functions. The results are given in (\ref{Bupto6}).  The  integrals of $\cB_{(p,q)}$ to weight~6 have thus been reduced to integrals 
with an exposed Laplace-Beltrami operator which may be performed using (\ref{intLap}), plus terms which are linear bilinear and trilinear in Eisenstein series, and the term $E_2 C_{2,1,1}$. We now proceed to evaluating the remaining integrals.

\subsection{Integrals with an exposed Laplace-Beltrami operator}

Several contributions to $\cB_{(p,q)}$ involve an exposed Laplace-Beltrami operator acting on a modular graph function which we shall generically denote by  $\cC(\tau)$. The modular graph function in question is of weight $w=2p+3q$ and thus has a Laurent polynomial of the form, 
\bea
\cC(\tau) = \sum ^w _{k=1-w} \mc _k \tau _2 ^k + \cO(e^{-2 \pi \tau_2})
\eea
For example, in terms of the variable $y=\pi \tau_2$, we have, 
\bea
\label{C211}
C_{2,1,1}  (\tau) =  { 2 y^4 \over 14175} + { \zeta(3) y \over 45} + +{ 5 \zeta (5) \over 12 y} -{ \zeta (3)^2 \over 4 y^2} +{9 \zeta (7) \over 16 y^3} + \cO(e^{-2 \pi \tau_2})
\eea
The integral of  $\Delta \cC(\tau)$ is given by (\ref{intLap}) and we find,
\bea
\int _{\cM_L} { d^2 \tau \over \tau_2^2} \, \Delta \cC(\tau) = \int _0^1 { d \tau_1} \, \p_{\tau_2} \cC(\tau) \Big |_{\tau_2=L} = \mc_1 + \cO(L^{\pm 1})
\eea
Using the expressions for the Laurent polynomials of the modular graph functions $C_{2,1,1}$, $C_{3,1,1}$, $C_{4,1,1}$, $C_{3,2,1}$, $C_{2,2,2}$ and $C_{2,2,1,1}$, which were  given respectively in formulas (3.41), (3.42) and (3.43) of \cite{DHoker:2016quv}, we find the following integrals, 
\begin{align}
 \int _{\cM_{L} } { d^2 \tau \over \tau_2^2} \, \Delta C_{2,1,1} & =  {\pi \zeta (3) \over 45} 
 & \hskip 0.5in 
 \int _{\cM_{L} } { d^2 \tau \over \tau_2^2} \, \Delta C_{3,2,1} & =  { \pi \zeta (5) \over 630} 
\no \\
\int _{\cM_{L} } { d^2 \tau \over \tau_2^2} \, \Delta C_{3,1,1} & =  0
& \hskip 0.5in 
\int _{\cM_{L} } { d^2 \tau \over \tau_2^2} \, \Delta C_{2,2,2} & =  0
\no \\
\int _{\cM_{L} } { d^2 \tau \over \tau_2^2} \, \Delta C_{4,1,1} & =  - { \pi \zeta(5) \over 1890} 
& \hskip 0.5in
\int _{\cM_{L} } { d^2 \tau \over \tau_2^2} \, \Delta \tilde C_{2,2,1,1} & =  { \pi \zeta (5) \over 63} 
\end{align}
where we have omitted to write $+  \cO(L^{\pm 1})$ which applies in each case.

\subsection{Integrals of Eisenstein series and their products}

Integrals of Eisenstein series and their pairwise and triple products in  $\cM_L$ were evaluated by Zagier in \cite{Zagier} by generalizing the standard Rankin-Selberg method for cuspidal functions to functions of polynomial growth at the cusp. Here, we shall give an alternative derivation based on the use of the differential equation satisfied by the Eisenstein series, given in (\ref{LapE}),  and their expansion at the cusp,
\bea
E_a(\tau) = { 2 \zeta(2a) \over \pi^a} \tau_2^a +{2 \Gamma ( a-\half) \zeta(2a-1) \over \Gamma (a) \pi^{a-\half} \tau_2^{a-1}} + \cO(e^{-2 \pi \tau_2})
\eea
To make contact with Zagier's notations, we define the functions $\zeta^*(s)$ and $E^*(\tau,s)$ by, 
\bea
\label{Estar}
\zeta ^*(a) = { \Gamma (a/2) \over \pi^{a/2}} \, \zeta (a) 
\hskip 1in 
E^*(\tau,a) = { \Gamma (a) \over 2 } E_a (\tau) 
\eea
which satisfy the simple functional relations $\zeta ^*(1-a) = \zeta ^*(a)$ and $E^*(\tau, 1-a) = E^*(\tau, a)$.
The Laurent polynomial of $E^*(\tau, a)$ near the cusp reflects the symmetry $a\leftrightarrow 1-a$, 
\bea
E^*(\tau, a) = \zeta ^*(2a) \tau_2^a + \zeta^* (2a-1) \tau_2^{1-a} + \cO(e^{- 2 \pi \tau_2})
\eea
and the  Eisenstein series $E^*(\tau,a)$ clearly  satisfies the differential equation of (\ref{LapE}). 

\sm

The integral of a single Eisenstein series is obtained by  integrating of $a(a-1)E^*(\tau,a) = \Delta E^* (\tau,a)$ using (\ref{intLap}) and the asymptotic expression for $E^*(\tau,a)$ near the cusp, 
\bea
a(a-1) \int _{\cM_L} { d^2 \tau \over \tau_2^2} E^*(\tau,a) 
= a \zeta ^*(2a)L^{a-1} + (1-a) \zeta^*(2a-1) L^{-a} 
\eea
The cases of interest here are for integer $a\geq 2$ so that the first term is a strictly positive power of $L$ and the second is a strictly negative power of $L$, both of which will cancel against opposite $L$-dependence of $\cA_R$.  Thus the integral effectively vanishes.

\sm

The integral of a bilinear $E_aE_b$  in Eisenstein series is computed analogously by integrating the total derivative term $E_a \Delta E_b - E_b \Delta E_a$ and we find,  
\bea
\label{EaEb}
\int _{\cM_L} { d^2 \tau \over \tau_2^2} E^*(\tau, a) E^*(\tau,b)  =  \sum _{x=a,1-a}\sum_{y=b,1-b} 
\zeta ^*(2x) \zeta ^*(2y) \, { L^{x+y-1} \over x+y -1} + \cO(e^{-2 \pi L})
\eea
Since the cases of interest here are for integer $a,b \geq 2$, all terms are either strictly positive or strictly negative powers of $L$ and may be omitted whenever $a \not=b$. For $b=a$, we take a limit of (\ref{EaEb}) as $b \to a$, and there now arise terms of orders $L^0$ and $\ln L$, given as follows,
\bea
\int _{\cM_L} { d^2 \tau \over \tau_2^2} E^*(\tau, a)^2 = 
2 \zeta ^*(2a) \zeta ^*(2a-1) \left [ \ln L + { \zeta^{*\prime} (2a) \over \zeta ^*(2a)} 
- { \zeta ^{* \prime} (2a-1) \over \zeta ^*(2a-1)} \right ]
\eea
up to strictly positive or strictly negative powers of $L$ and exponential contributions, which we omit in view of earlier arguments. Deriving the following relation from (\ref{Estar}),
\bea
\label{zstar}
{ \zeta ^{*\prime}(a) \over \zeta ^*(a)} = {\zeta '(a) \over \zeta (a)} + \half \Psi (a/2) - \half \ln \pi
\eea
we obtain an expression for the integral of $E_a (\tau)^2$. Neglecting all exponentially suppressed terms  and all power-behaved terms in $L$ with non-zero exponent, and the duplication formula for the $\Psi$-functions   $\Psi (a)- \Psi (a-\half) = 2\Psi (a) - 2\Psi (2a-1) + 2\ln 2$, we obtain, 
\bea
\label{Easq}
\int _{\cM_{L} } { d^2 \tau \over \tau_2^2} E_a(\tau)^2 
& = & 
 { 16 \Gamma (2a-1) \zeta (2a) \zeta (2a-1) \over \Gamma (a)^2 (2 \pi)^{2a-1}} \bigg  [ 
\ln (2L) + {\zeta '(2a) \over \zeta (2a) } - {\zeta '(2a-1) \over \zeta (2a-1) } 
\no \\ && \hskip 2in
 - \Psi (2a-1) + \Psi (a) \bigg  ]
\eea
which for the cases needed here reduces to, 
\bea
\int _{\cM_{L} } { d^2 \tau \over \tau_2^2} E_2(\tau)^2 
& = & 
 { 2 \pi \zeta(3)  \over 45} \left [ 
\ln (2L) + {\zeta '(4) \over \zeta (4) } - {\zeta '(3) \over \zeta (3) }  -\half \right ]
\no \\
\int _{\cM_{L} } { d^2 \tau \over \tau_2^2} E_3(\tau)^2 
& = & 
 {  \pi \zeta(5)  \over 315} \left [ 
\ln (2L) + {\zeta '(6) \over \zeta (6) } - {\zeta '(5) \over \zeta (5) }  - { 7 \over 12} \right ]
\eea
up to non-zero power terms in $L$ and exponentially suppressed terms.

\sm

Finally, we shall quote from \cite{Zagier}, without giving the derivation, the result for the integral of a trilinear in the Eisenstein series that we shall need here, 
\bea
\label{EEE}
\int _{\cM_L} { d^2 \tau \over \tau_2^2} E^*(\tau, a) E^*(\tau,b) E^*(\tau,c)
 =  
\zeta^*(w-1) \zeta^*(w-2a) \zeta^*(w-2b) \zeta ^*(w-2c)
\eea
where $w=a+b+c$ for integer $a,b,c \geq 2$ so that all other terms have non-zero powers of $L$ or are exponentially suppressed in $L$. For the special case of interest here we have $a=b=c=2$ and converting back to the integral of $E_2^3$, we find, 
\bea
\int _{\cM_{L} } { d^2 \tau \over \tau_2^2} E_2(\tau)^3 
 =  { \pi \zeta (5) \over 36}
\eea
up to non-zero power terms in $L$ and exponentially suppressed terms.

\subsection{Integral of  $ E_2 \, C_{2,1,1}$}

We need the following integral evaluated at $a=2$,
\bea
K(a) = \int _{\cM_L} { d^2 \tau \over \tau_2^2} \, E_a(\tau) C_{2,1,1}(\tau) 
= { 2 \over \Gamma(a)} \int _{\cM_L} { d^2 \tau \over \tau_2^2} \, E^*(\tau,s) C_{2,1,1}(\tau)
\eea
Multiplying both sides by $a(a-1)\Gamma(a)/2$, using the eigenfunction equation for $E_a$, and integrating by parts to pick up the boundary term at $\tau_2=L$, we find,
\bea
\half a(a-1) \Gamma (a) K(a) & = &
 \int _0 ^1 d \tau _1 \Big [ \p_{\tau_2} E^*(\tau, a) C_{2,1,1}(\tau) - E^*(\tau, a) \p_{\tau_2}  C_{2,1,1}(\tau) \Big ] _{\tau _2 = L}
\no \\ &&
+ \int _{\cM_L} { d^2 \tau \over \tau_2^2} \,E^*(\tau,a) \Delta C_{2,1,1}(\tau)
\eea
Using the differential equation (\ref{diff211}) and rearranging the term arising from $C_{2,1,1}$, we find, 
\bea
\half (a-2) (a+1) \Gamma (a) K(a) & = &
 \int _0 ^1 d \tau _1 \Big [ \p_{\tau_2} E^*(\tau, a) C_{2,1,1}(\tau) - E^*(\tau, a) \p_{\tau_2}  C_{2,1,1}(\tau) \Big ] _{\tau _2 = L}
\no \\ &&
+ \int _{\cM_L} { d^2 \tau \over \tau_2^2} \,E^*(\tau,a) \Big (3 E^*(\tau, 4)  - 4E^*(\tau, 2)^2 \Big ) 
\eea
The contribution of the $E^*(\tau,4)$ term on the last line produces power-behaved $L$-dependence with non-zero exponents and may be omitted. The other part may be evaluated using  (\ref{EEE}) within the same approximation, 
\bea
 \int _{\cM_L} { d^2 \tau \over \tau_2^2} \,E^*(\tau,a) \Big (3 E^*(\tau, 4)  - 4E^*(\tau, 2)^2 \Big ) 
 = - 4 \zeta ^*(a+3)  \zeta ^*(4-a) \zeta ^*(a)^2 
 \eea
To evaluate the contribution on the first line, we use the Laurent expansion of $C_{2,1,1}$ given in (\ref{C211}),
and keep only power-behaved terms whose exponent vanishes as $a\to 2$, we find, 
\bea
 \int _0 ^1 d \tau _1 \Big [ \p_{\tau_2} E^*(\tau, a) C_{2,1,1}(\tau) - E^*(\tau, a) \p_{\tau_2}  C_{2,1,1}(\tau) \Big ] _{\tau _2 = L}
= { 5 \zeta (5) \over 12 \pi } (a+1) \zeta ^*(2a) L^{a-2}
\eea
Putting all together, and dividing by a factor of $(a+1)$, we find, 
\bea
 (a-2)  \Gamma (a) K(a) = { \pi \zeta (5) \over 36 } \left ( { \zeta ^*(2a) \over 3 \zeta ^*(4)} L^{a-2}
-  {\zeta ^*(a+3)  \zeta ^*(4-a) \zeta ^*(a)^2 \over (a+1) \zeta ^*(5) \zeta ^*(2)^3} \right )
\eea
Both sides vanish for $a=2$, and the derivative in $a$ at $a=2$ evaluates as follows, 
\bea
\int _{\cM_L} { d^2 \tau \over \tau_2^2} \, E_2(\tau) C_{2,1,1}(\tau)  = 
{ \pi \zeta (5) \over 108 } \left ( \ln (2L)  + 2 {\zeta '(4) \over \zeta (4)} 
- {\zeta '(5) \over \zeta (5)}  -  {\zeta '(2) \over \zeta (2)}   \right )
\eea
up to non-zero power terms in $L$ and exponentially suppressed terms.

\newpage

\section{Calculation of $\cA_R$ and proof of Theorem \ref{theorem1}}
 \label{sec:B}
\setcounter{equation}{0}

In this Appendix, we shall prove Theorem \ref{theorem1} and carry out the evaluation of the integrals $\cR_{12}^{(N)}$ and $\cR_{12;34}^{(N)}$ in terms of which the non-analytic part $\cA_*(L;s,t)$ was expressed to all orders in $s$ and $t$ and order $L^0$ and $\ln L$ in $L$.

\subsection{Reducing $\cR_{12}^{(N)}$}

To evaluate the integrand factor $\cF_{12}^{(N)}$ we express each factor of $g_{12}$ as an infinite sum over $k,m$-variables, and carry out the integrations over the variables $\a, \tau_1$, 
\bea
\cF_{12}^{(N)}   =   \sum _{{m_r \not=0, k_r  \atop r=1, \cdots, , N}} 
{ \delta (\sum_r m_r) \delta (\sum_r  k_r m_r)  \over | m_1 \cdots m_N|} \, \exp \left ( 
- 2 \pi \tau_2 \sum _{r=1}^N |m_r | \, |k_r + x_2| \right )
\eea
Any  contribution which contains two unequal values of $k_r$ is exponentially suppressed in $\tau_2$, so that all $k_r$ must be equal to one another, and the factor $\delta (\sum_r k_r m_r)$ becomes redundant. Since we have $0\leq x_2 \leq 1$, only the values $k_r=0$ for all $r$ and $k_r=-1$ for all $r$ survive,
\bea
\label{Fm}
\cF_{12}^{(N)}  =  \sum _{{m_r \not=0 \atop r=1, \cdots, , N}}
{ \delta (\sum_r m_r)  \over | m_1 \cdots m_N|} \left ( e^{-4 \pi \tau_2 m x_2} + e^{-4 \pi \tau_2 m (1-x_2)} \right )
\hskip 0.6in 
m = \half \sum _{r=1}^N |m_r|
\eea
In terms of the remaining integral, 
\bea
\Lambda _{12} (L;s,t;m) = \int_L^\infty { d \tau_2 \over \tau_2^2} \int _{[0,1]^4} [dx] e^{4 \pi \tau_2 Q_{st}}
\Big ( e^{-4 \pi \tau_2 m x_2} + e^{-4 \pi \tau_2 m (1-x_2)} \Big )
\eea
the function $\cR_{12}^{(N)}(L;s,t) $ is given as follows,
\bea
\label{R12N}
\cR^{(N)} _{12} (L;s,t) = \sum _{{m_r \not=0 \atop r=1, \cdots, , N}}
{ \delta (\sum_r m_r) \, \Lambda _{12}(L;s,t; m) \over | m_1 \cdots m_N|}
\eea
Since $1-x_2=x_1+x_3+x_4\geq x_1+x_4$ we see immediately that the contribution from the second term in parentheses in $\Lambda _{12}$  is suppressed by inverse powers of $L$, and may be neglected. Henceforth, we shall use $\Lambda_{12}$ built from the first term in the parentheses.

\subsection{Evaluating $\Lambda _{12}$}

Carrying out the integration over $x_3$ by using the $\delta$-function in $[dx]$ sets $x_3=1-x_1-x_2-x_4$, and subsequently carrying out the integral over $x_2$, we find, 
\bea
\Lambda _{12} (L;s,t;m) =  \int _L ^\infty \!\! { d \tau _2 \over 4 \pi \tau_2^3} \int _0^1 dx_4 \int ^{1-x_4} _0 \!\!\! dx_1  
{ e^{4 \pi \tau_2 sx_1(1-x_1-x_4)} - e^{-4 \pi \tau_2 (1-x_1-x_4)  (m - t x_4 )} \over m + s x_1 -t x_4}
\eea
For  $|s|,|t| < \ep < 1$ we place an upper bound on the contribution from the second exponential above,  using the procedure used earlier to place an upper bound on $\Lambda_{13}$, given by, 
\bea
\int _L ^\infty { d \tau _2 \over 4 \pi \tau_2^3} \int _0^1 dx_4 \int ^{1-x_4} _0 dx_1  
{  e^{-4 \pi \tau_2 (1-x_1-x_4)  (m -\ep )} \over m -\ep} < { 1 \over 48 \pi^2 L^3 (m-\ep)^2}
\eea
Hence the integral is suppressed by inverse powers of $L$. Therefore, up to this order, $\Lambda_{12}$ is given by the contribution from the first exponential only.  Setting $x_4 =1-x$ and then setting $x_1=xy$, the integral becomes, 
\bea
\Lambda _{12}(L;s,t;m) =   \int _L ^\infty { d \tau _2 \over 4 \pi \tau_2^3} \int _0^1 dx \, x \int ^1 _0 dy  \, 
{ e^{4 \pi \tau_2 sx^2y(1-y)}  \over m + s xy -t (1-x)}
\eea
Carrying out the integral over $\tau_2$  with the help of the incomplete $\Gamma$-function, defined by,
\bea
\Gamma (\a, x) = \int _x ^\infty  dz  \, z^{\a-1} \, e^{-z}
\eea
we find the following integral expression, 
\bea
\Lambda _{12}(L;s,t;m) =  4 \pi s^2   \int _0^1 dx \int ^1 _0 dy  \, 
{ x^5 y^2(1-y)^2 \Gamma \Big (-2, - 4 \pi Ls x^2y(1-y) \Big ) \over m + s xy -t (1-x)}
\eea
The power series expansion  in $sL$  may be obtained from the asymptotics of $\Gamma(-2,x)$ for small $x$, which is  
derived from the asymptotic expansion of $\Gamma (0,x)=-\gamma - \ln x + \cO(x)$ (see  \cite{bateman}),
and the recursion relation for $\Gamma (\alpha, x)$ on $\alpha$, and is given by, 
\bea
 \Gamma (-2,x) =  - \half  \ln x - \half \gamma +{3 \over 4}  
- \sum _{{k=0 \atop k \not = 2}} ^\infty { (-)^k x^{k-2} \over k! \, (k-2)}
\eea
The contributions for $k=0,1$ are suppressed by negative powers of $L$, which we neglect, while those for $k \geq 3$ have strictly positive powers of $L$ which will be cancelled by the contributions from the analytic part of the expansion.
Therefore, retaining only the contributions of orders $L^0$ and $\ln L$, we obtain, 
\bea
\label{Lam12}
\Lambda_{12}(L;s,t;m) = 
- 2 \pi s^2 \left ( C(m,0) \left \{ \ln(- 4 \pi Ls) + \gamma - {3 \over 2} \right \} + \p_\ep C(m,\ep)\Big |_{\ep=0} \right )
\eea
where the function $C$ is given as follows,
\bea
\label{Cee}
C(m,\ep) =    \int _0^1 dx \int ^1 _0 dy  \, 
{ x^{5+2 \ep}  y^{2+\ep} (1-y)^{2+\ep}   \over m + s xy -t (1-x)} 
\eea
To evaluate $C$, we carry out a double expansion of the denominator in powers of $s$ and $t$, 
\bea
C(m,\ep) = \sum _{k=0}^\infty \sum _{\ell=0}^k \binom{k}{\ell} { (-s)^{k-\ell} t^\ell \over m^{k+1}} 
\int _0^1 dx \int _0 ^1 dy \, x^{k-\ell+5+2\ep} (1-x)^\ell y^{k-\ell+2 + \ep} (1-y)^{2 +\ep}
\eea
The radius of convergence  is $|m|$ in both $s,t$.
Carrying out the integrations, we find, 
\bea
C(m,\ep) = \sum _{k=0}^\infty \sum _{\ell=0}^k { \Gamma (k+1) \Gamma (k-\ell+3+\ep) \Gamma (3+\ep) 
\over \Gamma (k+7+2\ep) \Gamma (k-\ell+1)} { (-s)^{k-\ell} t^\ell \over m^{k+1}} 
\eea
One readily deduces, 
\bea
\Lambda_{12} = 
- 4 \pi s^2 \sum _{k=0}^\infty \sum _{\ell=0}^k C_{k,\ell}  { (-s)^{k-\ell} t^\ell \over m^{k+1}} 
\Big ( \ln(- 4 \pi Ls) + \Psi (k-\ell+3)  - 2 \Psi (k+7)  \Big )
\eea
where $\Psi (x) = \Gamma '(x) /\Gamma (x)$ and  the rational coefficients $C_{k,\ell}$ are given in (\ref{CD}).
Combining the definition of $\cR_{12}^{(N)}$ in (\ref{R12N}) and the evaluation of $\Lambda_{12}(s,t;m)$, we may now evaluate $\cR_{12}^{(N)} (s,t)$ by summing over the variables $m_r$, in terms of the multiple infinite sums $S(N,k)$ given in (\ref{SNk}), which gives the explicit expression for $\cR_{12}^{(N)} (L;s,t)$ in Theorem \ref{theorem1}.

\subsection{Reducing $\cR_{12;34}^{(M,N)}$}

We begin by carrying out the integrals in $\alpha_\kappa$ and $\tau_1$ in $\cF_{12;34}^{(M,N)}$, 
\bea
\cF_{12;34}^{(M,N)} 
& = &
\sum_{{m_r \not=0 \atop 1\leq r \leq  M}} \sum _{{n_s\not=0 \atop 1\leq s \leq N}} \sum _{k_r,\ell_s}
{ \delta (\sum_r m_r) \delta (\sum_s n_s) \delta (\sum_r k_r m_r+\sum_s \ell_s n_s)
\over |m_1 \cdots m_N \, n_1 \cdots n_N | }
\no \\ && \qquad \times
\exp \left ( - 2 \pi \tau_2    \sum _{r=1}^M |m_r| \, |k_r + x_2| 
-2 \pi \tau_2  \sum _{s=1}^N |n_s| \, |\ell_s + x_4|  \right )
\eea
Any contribution for which $k_r \not = k_{r'}$ with $r' \not = r$ is exponentially suppressed, and similarly for the $\ell_s$ variables. Thus, all $k_r$ must take equal value $k$ and, independently, all $\ell_s$ must take equal value $\ell$. Since $0 \leq x_2, x_4 \leq 1$, $k$ and $\ell$ can take the values $0,-1$ independently.  As a result, the Kronecker $\delta (\sum_r k_r m_r+\sum_s \ell_s n_s)$ becomes redundant, and the sums over $k$ and $\ell$ take the following simplified form,
\bea
\Big ( e^{-4\pi \tau_2 m x_2} + e^{-4\pi \tau_2 m (1-x_2)} \Big ) 
\Big ( e^{-4\pi \tau_2 n x_4} + e^{-4\pi \tau_2 n (1-x_4)} \Big ) 
\eea
where $m,n$ have been defined by,
\bea
\label{mn}
m = \half \sum _{r=1}^M |m_r| 
\hskip 1in 
n = \half \sum_{s=1}^N |n_s|
\eea
The contributions from $e^{-4\pi \tau_2 m (1-x_2)}$ and $e^{-4\pi \tau_2 m (1-x_4)}$ give terms in $\cA_*$ which are suppressed by inverse powers of $L$, and may be omitted. The expression for $\cF_{12;34}^{(M,N)}$ reduces to,
\bea
\cF_{12;34}^{(M,N)} =
\sum_{{m_r \not=0 \atop 1\leq r \leq  M}} \sum _{{n_s\not=0 \atop 1\leq s \leq N}} 
{ \delta (\sum_r m_r) \, \delta (\sum_s n_s) 
\over |m_1 \cdots m_N \, n_1 \cdots n_N | } \, 
e^{- 4 \pi \tau_2 [mx_2+ nx_4]}
\eea

\subsection{Calculating $\Lambda_{12;34}$}

The remaining integral over $\tau_2$ and $x_i$ is given by,
\bea
\Lambda_{12;34}(L;s,t;m,n) = 
\int _L ^\infty { d \tau_2 \over \tau_2^2} 
\int _{[0,1]^4} [dx] \, e^{4 \pi \tau_2 [ -mx_2-nx_4 + s x_1x_3+tx_2x_4]} 
\eea
 Integrating over $x_3$ using the $\delta$-function sets $x_3=1-x_1-x_2-x_4$ and then over $x_4$ gives, 
\bea
\int _L ^\infty  { d \tau_2 \over 4 \pi \tau_2^3} \int _0 ^1  dx_1 \int _0 ^{1-x_1} dx_2 \, 
{ e^{4 \pi \tau_2[ - m x_2+sx_1(1-x_1-x_2)]} -  e^{4 \pi \tau_2 [ -mx_2-(1-x_1-x_2) (n-tx_2)]}  \over n +sx_1-tx_2}
\eea
Assuming $|s|, |t| < \ep < 1$, the integral arising from the second exponential is bounded by, 
\bea
 \int _L ^\infty  { d \tau_2 \over 4 \pi \tau_2^3} \int _0 ^1 dx_1 \int _0 ^{1-x_1}  dx_2 \, 
{  e^{- 4 \pi \tau_2 (1-x_1-x_2) (n-\ep)} \over n-\ep} 
\leq   { 1 \over 48 \pi^2 L^3 (n-\ep)^2}
\eea
and therefore is suppressed by negative powers of $L$ and may be omitted.  The remaining contribution to $\Lambda _{12;34}$  is given by,
\bea
\label{B23}
\Lambda_{12;34}(L;s,t;m,n) = 
\int _L ^\infty  { d \tau_2 \over 4 \pi \tau_2^3} \int _0 ^1  dx_1 \int _0 ^{1-x_1} dx_2 \, 
{ e^{4 \pi \tau_2[ - m x_2+sx_1(1-x_1-x_2)]}  \over n +sx_1-tx_2}
\eea
It is analytic in $t$ and may be expanded as follows,  
\bea
 \Lambda _{12;34} (L;s,t;m,n) =  \sum _{k=0}^\infty t^k 
\int _L ^\infty  { d \tau_2 \over 4 \pi  \tau_2^3} \int _0 ^1  dx_1 \int _0 ^{1-x_1} \!\!\! dx_2 \, x_2^k \,
{  e^{4 \pi \tau_2 [-m x_2+ sx_1(1-x_1-x_2)]}  \over  (n+sx_1)^{k+1}}
\eea
To carry out the integral in $x_2$ we notice that the integrand is exponentially damped for large $x_2$ by the $m x_2$ term in the argument of the exponential. Therefore, up to exponential contributions which we neglect, we may extend the integration region for $x_2$ to $\infty$. The integral is then readily carried out, and we find, 
\bea
\label{Lpp}
 \Lambda _{12;34}=  16 \pi^2  \sum _{k=0}^\infty k ! \, t^k 
\int _L ^\infty \!\!\! { d \tau_2 \over (4 \pi  \tau_2)^{k+4}} \int _0 ^1  dx 
{  e^{4 \pi \tau_2 s x(1-x)}   \over  (m+sx)^{k+1} (n+sx)^{k+1}}
\eea
Next, we carry out the integral over $\tau_2$ in terms of the incomplete $\Gamma$-function, 
\bea
\label{Lppp}
 \Lambda _{12;34} =  4 \pi   \sum _{k=0}^\infty k ! \, t^k (-s)^{k+3} 
\int _0 ^1  dx 
{   x^{k+3} (1-x)^{k+3} \, \Gamma (-k-3, - 4 \pi Ls x(1-x))  \over  (m+sx)^{k+1} (n+sx)^{k+1}} 
\eea
All the $L$-dependence is now concentrated in the incomplete $\Gamma$-function. Since in this calculation we are omitting all positive and negative powers of $L$, we need to extract from $\Gamma(-\nu,y)$, for $\nu \in \NN$, only the terms of order $y^0$ and $\ln(y)$, which we shall denote by $\hat \Gamma (-\nu,y)$. They may be calculated from the general recursion relation, 
\bea
\Gamma (-\nu+1,y)= - \nu \Gamma (-\nu,y) +{ e^{-y} \over y^\nu}
\eea
and we find, 
\bea
(-)^\nu \nu! \, \hat \Gamma (-\nu,y) = - \ln (y)  + \Psi (\nu+1) \hskip 0.5in \nu \in \NN
\eea
Hence the terms of order $L^0$ and $\ln L$ we need from the incomplete $\Gamma$-function are,
\bea
\Gamma (-k-3, - 4 \pi Ls x(1-x)) = 
{ (-)^{k+3} \over (k+3)!} \Big (  \Psi (k+4) - \ln [ - 4 \pi Ls x(1-x) ]  \Big )
\eea
Putting all together, we find, 
\bea
\label{Lppp}
 \Lambda _{12;34} =  4 \pi   \sum _{k=0}^\infty { k ! \, t^k s^{k+3} \over (k+3)!}
\int _0 ^1  dx \, 
{   x^{k+3} (1-x)^{k+3}  \Big (  \Psi (k+4)  - \ln [ - 4 \pi Ls x(1-x) ] \Big )  \over  (m+sx)^{k+1} (n+sx)^{k+1}} 
\eea
To evaluate the integral we express it in terms of the following family of integrals, 
\bea
K_k (s, \ep) = {  k! \over (k+3)!} 
\int _0 ^1  dx \, { x^{k+3+\ep } (1-x)^{k+3+\ep }   \over  (m+sx)^{k+1} (n+sx)^{k+1}} 
\eea
and we have,
\bea
\Lambda _{12;34} = 4 \pi   \sum _{k=0}^\infty  t^k s^{k+3} 
\left [ K_k(s,0) \Big ( - \ln (-4 \pi Ls) + \Psi (k+4) \Big ) - \p_\ep K_k(s,\ep) \big |_{\ep=0} \right ] 
\eea
To calculate $K_k(s,\ep)$ we expand both denominators using the formula, 
\bea
{ 1 \over (1+s/M)^{k+1} } = \sum _{\ell=0}^ \infty {\Gamma (k+1+\ell) \over \Gamma (k+1) \ell !} \left (-{s \over M} \right )^\ell
\eea
and then perform the integrals in $x$, so that $K_k(s,\ep)$ is given by,
\bea
K_k (s, \ep) & = &  
\sum _{\ell_1=0}^ \infty \sum _{\ell_2=0}^ \infty 
 {  \Gamma (k+4+\ep) \Gamma (k+\ell_1+\ell_2+4 + \ep)  \Gamma (2k+\ell_1+\ell_2+8) \over \Gamma (k+4)  \Gamma (k+\ell_1+\ell_2+4) \Gamma (2k+\ell_1+\ell_2+8 + 2 \ep)}
\no \\ && \qquad \times D_k (\ell_1, \ell_2) 
{ (-s)^{\ell_1+\ell_2} \over m^{k+1+\ell_1}  n ^{k+1+\ell_2}} 
\eea
where the coefficients $D_k(\ell_1, \ell_2)$ are given in (\ref{CD}). The parts we need are, 
\bea
K_k (s, 0) =  
\sum _{\ell_1=0}^ \infty \sum _{\ell_2=0}^ \infty D_k (\ell_1, \ell_2)  {   (-s)^{\ell_1+\ell_2}  \over  m^{k+1+\ell_1}  n ^{k+1+\ell_2}}
\eea
and,
\bea
\p_\ep K_k(s,0)  
& = &
\sum _{\ell_1=0}^ \infty \sum _{\ell_2=0}^ \infty D_k (\ell_1, \ell_2)  {   (-s)^{\ell_1+\ell_2}  \over  m^{k+1+\ell_1}  n ^{k+1+\ell_2}}
\Big ( \Psi (k+\ell_1+\ell_2+4)
\no \\ && \hskip 1.5in
 + \Psi (k+4) - 2 \Psi(2k+\ell_1+\ell_2 +8) \Big )
\eea
The resulting expression for $\Lambda _{12;34}$ is then given by, 
\bea
\Lambda _{12;34} & = & - 4 \pi   \sum _{k=0}^\infty  
\sum _{\ell_1=0}^ \infty \sum _{\ell_2=0}^ \infty D_k (\ell_1, \ell_2)  { (-s)^{\ell_1+\ell_2}  s^{k+3}  t^k \over m^{k+1+\ell_1}  n ^{k+1+\ell_2}}
\no \\ && \quad \times
\Big (  \ln (- 4 \pi Ls)  + \Psi (k+\ell_1+\ell_2+4)  - 2 \Psi(2k+\ell_1+\ell_2 +8) \Big )
\eea
Finally, the functions $\cR_{12;34}^{(M,N)}(L;s,t)$ are given by,
\bea
\cR_{12;34}^{(M,N)} = 
\sum_{{m_r \not=0 \atop 1\leq r \leq  M}} \sum _{{n_s\not=0 \atop 1\leq s \leq N}} 
{ \delta (\sum_r m_r) \delta (\sum_s n_s) \Lambda _{12;34} (L;s,t;m,n) 
\over |m_1 \cdots m_M \, n_1 \cdots n_N | }
\eea
Substituting the value for $\Lambda _{12;34}(m,n)$ derived above, we see that for each $k, \ell_1, \ell_2$, the sums over $m_r$ and $n_s$ factorize, and we may express their values in terms of the $S$-functions of (\ref{SNk}), resulting in the expression for $\cR_{12;34} ^{(M,N)} (L;s,t)$ of (\ref{R1234a}).

\newpage

\section{The discontinuity calculated from unitarity}
\setcounter{equation}{0}
\label{sec:C}

\begin{figure}[h]
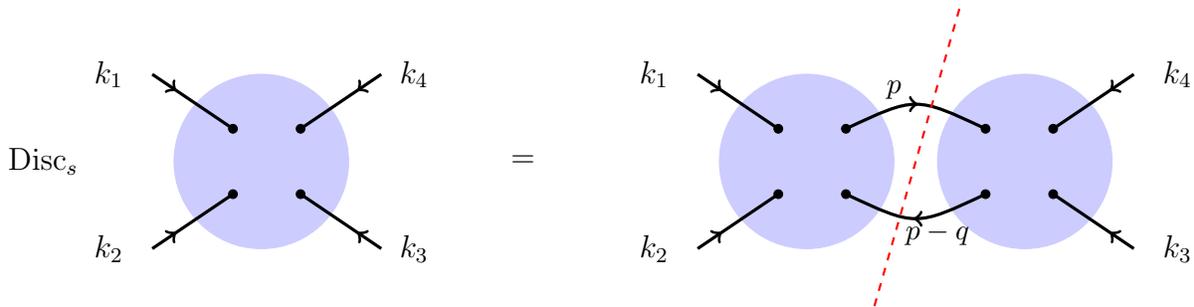

\begin{center}
\tikzpicture[scale=2.9]
\scope[xshift=-1.5cm,yshift=-0.4cm]
%
\filldraw [blue!20]  (-0.5,0) ellipse (.40 and .40);

\draw [very  thick] (-1,-0.4) -- (-.63,-0.15) ;
\draw [very  thick] (-1,0.4) -- (-.63,0.15) ;
\draw [very  thick] (.05,0.4) -- (-.32,0.15) ;
\draw [very  thick] (.05,-0.4) -- (-.32,-0.15) ;
\draw [very  thick, ->] (-0.91,0.339) -- (-0.89,0.324);
\draw [very  thick, ->] (-0.91,-0.339) -- (-0.89,-0.324);
\draw [very  thick, ->] (-0.04,0.339) -- (-0.06,0.324);
\draw [very  thick, ->] (-0.04,-0.339) -- (-0.06,-0.324);
\draw (0.7, 0.0) node{$=$};
\draw (-1.5, 0.0) node{${\rm Disc}_s$};
\draw (-1.2, 0.4) node{$k_1$};
\draw (-1.2, -0.4) node{$k_2$};
\draw (0.2, -0.4) node{$k_3$};
\draw (0.2, 0.4) node{$k_4$};
\draw [fill=black] (-0.32,0.15) circle [radius=.02];
\draw [fill=black] (-0.32,-0.15) circle [radius=.02];
\draw [fill=black] (-.63,0.15) circle [radius=.02];
\draw [fill=black] (-.63,-0.15) circle [radius=.02];
\endscope
\scope[xshift=1cm,yshift=-0.4cm]
%
\filldraw [blue!20]  (-0.5,0) ellipse (.40 and .40);
\filldraw [blue!20]  (0.5,0) ellipse (.40 and .40);
\draw [very thick] (-0.32,0.15)  .. controls (0, 0.3) ..  (0.32,0.15) ;
\draw [very  thick] (-0.32,-0.15)  .. controls (0, -0.3) ..  (0.32,-0.15) ;
\draw [very  thick, ->] (-0.01,0.263) -- (0.01,0.263);
\draw [very  thick, ->] (0.01,-0.263) -- (-0.01,-0.263);
\draw [very  thick] (1,0.4) -- (.63,0.15) ;
\draw [very  thick] (-1,-0.4) -- (-.63,-0.15) ;
\draw [very  thick] (-1,0.4) -- (-.63,0.15) ;
\draw [very  thick] (1,-0.4) -- (.63,-0.15) ;
\draw [very  thick, ->] (-0.91,0.339) -- (-0.89,0.324);
\draw [very  thick, ->] (-0.91,-0.339) -- (-0.89,-0.324);
\draw [very  thick, ->] (0.91,0.339) -- (0.89,0.324);
\draw [very  thick, ->] (0.91,-0.339) -- (0.89,-0.324);
\draw [thick,dashed,color=red]   (0.2,0.7) -- (-0.2,-0.7);
\draw (-1.2, 0.4) node{$k_1$};
\draw (-1.2, -0.4) node{$k_2$};
\draw (1.2, -0.4) node{$k_3$};
\draw (1.2, 0.4) node{$k_4$};
\draw (-0.1, 0.33) node{\small $p$};
\draw (0.1, -0.33) node{\small $p-q$};

\draw [fill=black] (-0.32,0.15) circle [radius=.02];
\draw [fill=black] (-0.32,-0.15) circle [radius=.02];
\draw [fill=black] (0.32,0.15)circle [radius=.02];
\draw [fill=black] (0.32,-0.15) circle [radius=.02];
\draw [fill=black] (-.63,0.15) circle [radius=.02];
\draw [fill=black] (-.63,-0.15) circle [radius=.02];
\draw [fill=black] (.63,0.15)circle [radius=.02];
\draw [fill=black] (.63,-0.15) circle [radius=.02];
\endscope
\endtikzpicture
\end{center} 
\caption{The discontinuity of the genus-one amplitude. Each blue blob represents the tree-level contribution to the four-particle amplitude, momentum flows in the direction of the corresponding arrow, and the dashed line represents the $s$-channel cut.} \label{fig:3}
\end{figure}

In this appendix we will describe the manner in which unitarity  of the genus-one superstring amplitude relates the discontinuity of the branch cut across the real positive $s$-axis to a bilinear in the on-shell tree-level four-string amplitude.  This procedure was used to determine the coefficients of low order terms in the  low-energy expansion in \cite{Green:2008uj} and has been extended to order $s^{11}$ in \cite{Alday:2018pdi} .  Here we are interested in the all orders  structure of the discontinuity.  A graphic representation of the unitarity  relation is presented in Figure \ref{fig:3}.


\subsection{Kinematics of the two-particle discontinuity}

Quite generally, the full four-particle amplitude with  massless external states satisfies the non-linear two-particle discontinuity relation in ten space-time dimensions
\be\begin{split}
i \, {\rm Disc_s} &\, \bA (k_1,k_2,k_3,k_4,,\ep_1,\ep_2,\ep_3,\ep_4)  =   \frac{\pi \kappa_{10}^2}{2 \alpha'} 
\int {d^{10} p\over (2\pi)^{10}} \,\delta^{(+)}(p^2)\,\delta^{(+)}((q-p)^2)\\
& \times 
\sum_{\{\ep_r,\ep_{s}\}}  \bA^{\huge\dagger} (k_1,k_2,-p,p-q,,\ep_1,\ep_2,\ep_r,\ep_s)\,
 \bA (k_3,k_4,p,q-p ,\ep_3,\ep_4,\ep_r,\ep_s) 
\label{unitarity}
\end{split}\ee
 where the sum in $\ep_r,\ep_s$ is over all the two-particle massless maximal supergravity states and the momenta in this expression have been rescaled by $\sqrt{\alpha'}$ so they are dimensionless. The  factor   $\delta^{(+)}(p^2)\equiv \delta^{(D)}(p^2)\theta(p^0)$ imposes
the mass-shell condition, $p^2 =0$, and   $q=k_1+k_2$.
Equation (\ref{unitarity}) relates the discontinuity of the full amplitude across the threshold for two intermediate massless particles to the integral of the square of the amplitude.
This   relation takes a very special form in maximal supergravity (as it does in maximal
Yang--Mills), because of the self-replicating relation derived in
\cite{Bern:1998ug},
\be\begin{split}
 \left(\frac{\alpha'}{4} \right)^4  \sum_{\{\ep_r,  \ep_s\}}  R^4(k_1,k_2,p-q,-p, \ep_1,\ep_2,\ep_r,\ep_s)\,&
  R^4(p, q-p, k_3, k_4,,\ep_3,\ep_4,\ep_r,\ep_s)  \\
& = s^4\,  R^4 (k_1,k_2,k_3,k_4,\ep_1,\ep_2,\ep_3,\ep_4) 
\label{replica}
\end{split}\ee

\sm

The unitarity equation (\ref{unitarity}) can be expanded to any order in string perturbation theory, but we are here only interested in the lowest order, which  determines the discontinuity of the genus-one amplitude in terms of the tree-level amplitude.  
Substituting the tree-level   contribution to $\bA$ in both  factors on the right-hand side of (\ref{unitarity}), using (\ref{ampdef}),  gives the equation for the $s$-channel discontinuity of the one-loop amplitude,
\bea
i \,  {\rm Disc}_s \, \cA^{(1)} & = &  {\pi s^4 \over 2}  \int  \frac{d^{10} p}{(2\pi)^{10}} \,  \delta ^+ (p^2) \delta ^+  ( (q-p)^2  ) \cA^{(0)}(k_1, k_2, -p, -q+p) 
\no \\ && \hskip 1.5in \times \cA^{(0)}(k_3,k_4,p,q-p)
\label{stripped}
\eea
where $q=k_1+k_2$ and $ \cA^{(0)}$ is the tree-level on-shell amplitude (\ref{A0a}). The external momenta are all massless so that $k_i^2=0$, and the internal momentum $p$ satisfies the on-shell conditions $p^2=(q-p)^2=0$. Choosing the rest-frame of $q=k_1+k_2$,  we use the following asymmetric parametrization for the momenta, 
\bea
k_1 & = & k (+1,1, 0, \bo_7)
\no \\
k_2 & = & k (+1,-1, 0, \bo_7)
\no \\
k_3 & = & k (-1,\cos \theta, \sin \theta , \bo_7)
\no \\
k_4 & = & k (-1,-\cos \theta , -\sin \theta, \bo_7)
\no \\
p & = & k (1,\cos \alpha, \sin \alpha \cos \beta , \bn_7 \, \sin \alpha \sin \beta )
\eea
where $\bo_7$ is the null vector, and $\bn_7$ parametrizes the unit vectors in $\RR^7$. We have explicitly solved  the on-shell conditions $p^2=(q-p)^2=0$ which in particular imply that $p^0=k$. It will be convenient to express the kinematic variables in terms of $s=\alpha ' k^2$, and we have,
\bea
\label{tudef}
t =  -{s \over 2} (1-\cos \theta) 
\hskip 1in 
u =  -{s \over 2} (1+\cos \theta)  
\eea
as well as
\begin{align}
\label{t1def}
t_1  &=  -{ \alpha ' \over 4} (k_1-p)^2 =  -{s \over 2} (1- \cos \alpha) 
\no \\
 t_2  &= -{ \alpha ' \over 4} (k_4+ p)^2  =   - { s \over 2} (1-\cos \alpha \cos \theta -\sin \theta \sin \alpha \cos \beta )
\end{align}
along with $u_1=-s-t_1$ and $u_2=-s-t_2$. 

\subsection{Calculation of the discontinuity}

We  may  now  determine the discontinuity of the genus-one amplitude across the two-particle branch cut by substituting the genus-zero expression, $\cA^{(0)}$ of (\ref{A0}),  into the right-hand side of the discontinuity equation (\ref{stripped}).  The first few terms in the low-energy  expansion of the discontinuity were obtained in \cite{Green:2008uj} by substituting the terms in (\ref{A0b}) into  the right-hand side of (\ref{stripped}).    The lowest-order term in this expansion arises from the exchange of the massless supergravity states in both  factors of  $\cA^{(0)}$.  This gives the discontinuity of the supergravity one-loop amplitude that is of order $s \log(-2 \pi s)$.  The next terms, which are of the form $\zeta(3) s^4 \log(-2 \pi s)$ and $\zeta(5) s^6 \log(-2 \pi s)$, arise from the configuration in which the supergravity term is substituted for  one factor of $\cA^{(0)}$  and the  $\zeta(3)$ or $\zeta(5)$ terms in (\ref{A0b})  are substituted in the other $\cA^{(0)}$ factor.  The term of the form $\zeta(3)^2\, s^7 \log(-2 \pi s)$ gets two types of contributions.  One is  from  the  supergravity term in one  $\cA^{(0)}$  factor and the $\zeta(3)^2$ term in  (\ref{A0b}) in the other.  There is also a contribution that arises by substituting the $\zeta(3)$ term in (\ref{A0b}) into both factors of   $\cA^{(0)}$.

\sm

Rather than extend these calculations to a limited number of  higher-order terms our aim  here is to make contact with the all orders results of Theorem~\ref{theorem2}.   We will concentrate on the sub-class of contributions to the discontinuity in which one of the  $\cA^{(0)}$ factors in  (\ref{stripped})  is restricted to the lowest-order, supergravity, amplitude, and the other is general.    In this subsection only, we shall concentrate on the functional dependence of the discontinuity but not keep track of the factors of 2 and $\pi$ which enter at various stages. To show this clearly, we shall  use the notation $\approx$ instead of  the equal sign.   For this  purpose  it  is  convenient to separate  the massless particle exchange from  the  massive  exchanges by writing, 
\bea
\cA^{(0)}(k_1,k_2, -p,p-q) =  V(k_1,k_2,-p,p-q) + W(s,t_1)
\label{onesep}
\eea
where the massless  exchange tree amplitude is  given  by,  
\bea
V(k_1,k_2,-p,p-q) = { 1 \over s t_1 u_1} = - {1 \over s^2 t_1} -{1 \over s^2 u_1}
\eea
and the massive exchanges are contained in the function $W$ which was defined in (\ref{VS}).
The integration measure in (\ref{stripped}) reduces as follows,
\bea
d^{10} p \, \delta ^+ (p^2) \delta ^+  ( (q-p)^2  ) \approx  k^6 \, d^6 \bn \, d \alpha \, d \beta \, (\sin \alpha )^7 \, (\sin \beta )^6
\eea
and the volume of $S^6$ is given by $16 \pi^3/15$. 
Substituting (\ref{onesep}) for each of the  $\cA^{(0)}$ factors in the discontinuity equation (\ref{stripped}) gives the sum of four contributions,
\bea
{\rm Disc}_s \, \cA^{(1)} =  {\rm Disc}_s \, \left(  \cA^{(1)}\big|_{VV}+2 \cA^{(1)}\big|_{WV} + \cA^{(1)}\big|_{WW} \right)
\label{discsum}
\eea
where the notation  indicates the contribution that arises when each of the $\cA^{(0)}$ factors in (\ref{stripped}) is restricted to $V$ to $W$ (and we have used  the fact that $ \cA^{(1)}\big|_{WV} = \cA^{(1)}\big|_{VW} $). 
We are  particularly interested in analyzing the contribution,
\bea
i \, {\rm Disc}_s \, \cA^{(1)}\big |_{VW} \approx { s^4 \over 15} \int _0 ^\pi d \alpha \, 
  (\sin \alpha )^7  W(s,t_1)  \int _0 ^\pi d \beta \,    (\sin \beta )^6 \left ( {1 \over  t_2} + {1 \over u_2}  \right )
  \label{avw}
\eea
since ${\rm Disc}_s \, \cA^{(1)}\big|_{VV}$ is the well-known  discontinuity of the supergravity box diagram and ${\rm Disc}_s \, \cA^{(1)}\big|_{WW}$ is much more complicated.

\sm

Under the combined transformations $\alpha \to \pi - \alpha $ and $\beta \to \pi - \beta$ in (\ref{avw}), the variables $t_1$ is swapped with $u_1$, and $t_2$ is swapped with $u_2$, while the measure is invariant. Since the integrand  is even in $\beta$, we may extend the integration region from 0 to $2 \pi$ upon including a factor of $\thalf$. In summary, we get the following equivalent expression, 
\bea
i \, {\rm Disc}_s \, \cA^{(1)}  \big |_{VW} \approx  { s^4 \over 15 \, \sin \theta } \int _0 ^\pi d \alpha \, 
 (\sin \alpha )^6 W(s, t_1) 
\int _0 ^{\pi}  d \beta \, {   (\sin \beta )^6  \over  -a + \cos \beta}  
\eea
where we have defined the variable $a>1$ to be given by,
\bea
\label{adef}
a = { 1 - \cos \theta \cos \alpha \over \sin \theta \sin \alpha}
\eea
The integral over $\beta$ is readily calculated and we find, 
\bea
\label{DiscF}
i \, {\rm Disc}_s \, \cA^{(1)} \big |_{VW}\approx  {s^4 \over 15 \, \sin \theta} \int _0 ^\pi d \alpha \, 
(\sin \alpha )^6 W(s,t_1) 
\left ( (a^2-1)^{{5 \over 2}} -a^5 +{5 \over 2} a^3 -{15 \over 8} a \right )  
\eea
where the variables $a,b$ are defined  above.

\subsection{Comparison of unitarity with Theorem~\ref{theorem2}  }

To compare the expression of the previous subsection with the result from string theory, we express the discontinuity in $s$ of $\mA_*^{(1)}(s,t,u)$ defined in (\ref{thm2a}) and (\ref{thm2b})  in terms of an integral over the variable $t_1$ (defined in (\ref{t1def}). This gives, 
\bea
\label{disc}
i \, {\rm Disc}_s \cA_{{\rm non-an}}  
=  16 \pi^3 s^2 \, \mA(s,t,u;0)
\eea
where a factor of $-i \pi$ arises from the discontinuity of the logarithm and the second factor from the relation between $\mA$ and $\cA_{{\rm non-an}}$.   The latter may be evaluated as follows, 
\bea
\mA (s,t,u;0)  & = &  s^2  \int _0 ^1 \!\! dx \int _0 ^1 \!\! dy \,  
 x^5 y^2 (1-y)^2 \, W(s,-sxy+t(1-x))\no\\
& =& s^2 \int _0^1 dx \int _0^x dw \, w^2 (x-w)^2 W\left(s,-\frac{s}{2} (1+z)\right) 
\label{discma}
\eea
where $w=xy$ and we have simplified the argument of $W$ in this integral by defining,
\bea
z =  -\frac{2}{s} \left(-sxy  +t(1-x)\right) = 2w + (1-\cos\theta)(1-x)
\label{zdef}
\eea
The range of the variable $z$ is the interval $[-1,+1]$. We now change variables from $w$ to $z$,
\bea
\mA (s,t,u;0)  & = & {  s^2 \over 32}   \int _0^1 dx \int _{-1}^1 dz \,\,W(s,t_1) \theta (x+z+c-xc)
\no\\ && 
\hskip 0.6in \times 
\theta(x-z-c+cx)  \Big (x^2-(z+c-cx)^2 \Big )^2
\eea
If $z+c >0$ then the argument of the first $\theta$-function is automatically positive for all $x\in [0,1]$, while if $z+c<0$ it is the the argument of the second $\theta$-function which is automatically positive. Shifting $x$ in each sector so that its range starts at $x=0$, and  rescaling $x$ by $1-c^2$, one observes that both integrals may be combined as follows, 
\bea
\mA (s,t,u;0)  = { s^2 \over 32 \sin^6 \theta }  \int _{-1}^1 dz \, W(s,t_1) \int _{|z+c| } ^{1+cz} dx  \,   \left (x^2-(z+c)^2  \right ) ^2
\eea
Parametrizing $z$ by $\alpha$ with $z=-\cos \alpha$, and using the variable $a$ from (\ref{adef}),
\bea
1+cz & = &  a \, \sin \theta \, \sin \alpha 
\no \\
|z+c| & = &  \sqrt{a^2-1} \, \sin \theta \, \sin \alpha 
\eea
we obtain the integral representation for $\mA$. Including the normalization factor in (\ref{disc}) provided by Theorem \ref{theorem2} to obtain $\cA_{{\rm non-an}}$, we find, 
\bea
i \, {\rm Disc}_s\,\cA _{{\rm non-an}}    =     {  8 \pi^3 s^4 \over 15 \, \sin \theta}  
\int _{-{\pi \over 2} }^{{\pi \over 2}}  d\alpha \, ( \sin \alpha)^6  \,  W(s,t_1) 
 \left (  (a^2-1)^{{ 5 \over 2}} -  a^5 + { 5 \over 2}  a^3 -{ 15 \over 8} a \right )
\eea
One may easily verify the integral of the leading contribution from $W$, which is $2 \zeta (3)$,  and compare it with the results from the earlier calculation in (\ref{Anonan}) and (\ref{Ahatfour}). The functional form of the result including the prefactor of $s^4/15$, agrees with the result obtained from unitarity in (\ref{DiscF}) up to factors of 2 and $\pi$.

\end{document}